\documentclass{emulateapj}
\usepackage{graphicx}
\usepackage{hyperref}
\usepackage{color}

\newcommand{\ergs}{${\rm erg \ cm^{-2} \ s^{-1}}$ }

\newcommand{\todo}{\ifmmode {\Huge \bullet} \else {\Huge$\bullet$}\fi}

\newcommand{\vFWHM}{\ifmmode V_{\mbox{\tiny FWHM}} \else $V_{\mbox{\tiny FWHM}}$ \fi}

\newcommand{\kms}{\ifmmode {\rm km\,s}^{-1} \else km\,s$^{-1}$ \fi}

\newcommand{\ergcms}{\ifmmode {\rm ergs\,cm}^{-2}\,{\rm s}^{-1} \else ergs\,cm$^{-2}$\,s$^{-1}$\fi}
\newcommand{\ergcmsA}{\ifmmode{\rm ergs}\, {\rm cm}^{-2}\,{\rm s}^{-1}\,{\rm\AA}^{-1} \else ergs\, cm$^{-2}$\, s$^{-1}$\, \AA$^{-1}$\fi}
\newcommand{\ergcmsHz}{\ifmmode{\rm ergs\,cm}^{-2}\,{\rm s}^{-1}\,{\rm Hz}^{-1} \else ergs\,cm$^{-2}$\,s$^{-1}$\,Hz$^{-1}$\fi}
\newcommand{\phcms}{\ifmmode {\rm ph\,cm}^{-2}\,{\rm s}^{-1} \else ,ph\,cm$^{-2}$\,s$^{-1}$\fi}
\newcommand{\phcmsA}{\ifmmode {\rm ph\,cm}^{-2}\,{\rm s}^{-1}\,{\rm\AA}^{-1} \else ph\,cm$^{-2}$\,s$^{-1}$\,\AA$^{-1}$\fi}

\newcommand\Msun{\ifmmode M_{\odot} \else $M_{\odot}$\fi}
\newcommand\msun{\ifmmode M_{\odot} \else $M_{\odot}$\fi}
\newcommand\Lsun{\ifmmode L_{\odot} \else $L_{\odot}$\fi}

\newcommand\mpyr{\ifmmode \Msun\,{\rm yr}^{-1} \else $\Msun\,{\rm yr}^{-1}$ \fi}

\newcommand{\Luv}{\ifmmode L_{1450} \else $L_{1450}$\fi}
\newcommand{\Lop}{\ifmmode L_{5100} \else $L_{5100}$\fi}
\newcommand{\Lthree}{\ifmmode L_{3000} \else $L_{3000}$\fi}
\newcommand{\lledd}{\ifmmode L/L_{\rm Edd} \else $L/L_{\rm Edd}$\fi}
\newcommand{\ledd}{\ifmmode L_{\rm Edd} \else $L_{\rm Edd}$\fi}
\newcommand{\lamLlam}{\ifmmode \lambda L_{\lambda} \else $\lambda L_{\lambda}$\fi}
\newcommand{\lbol} {\ifmmode L_{\rm Bol} \else $L_{\rm Bol}$\fi}
\newcommand{\llbol}{\ifmmode \log\left(\lbol/\ergs\right) \else $\log\left(\lbol/\ergs\right)$\fi}

\newcommand{\fuv}{\ifmmode f_{\lambda}\left(1450\AA\right) \else $f_{\lambda}\left(1450 {\rm \AA}\right)$\fi}
\newcommand{\fthree}{\ifmmode f_{\lambda}\left(3000\AA\right) \else $f_{\lambda}\left(3000{\rm \AA}\right)$\fi}
\newcommand{\fH}{\ifmmode f_{\lambda}\left(1.65\micron\right) \else
$f_{\lambda}\left(1.65\micron\right)$\fi}

\newcommand{\mbh}{\ifmmode M_{\rm BH} \else $M_{\rm BH}$\fi}
\newcommand{\lmbh}{\ifmmode \log\left(\mbh/\Msun\right) \else $\log\left(\mbh/\Msun\right)$\fi}

\newcommand \Hbeta {\ifmmode {\rm H}\beta \else H$\beta$\fi}
\newcommand \hb    {\ifmmode {\rm H}\beta \else H$\beta$\fi}
\newcommand  \mgii  {\ifmmode {\rm Mg}{\textsc{ii}} \else Mg\,{\sc ii}\fi}
\newcommand  \MGII  {\ifmmode {\rm Mg}\,{\sc ii}\,\lambda2798 \else Mg\,{\sc ii}\,$\lambda2798$\fi}
\newcommand  \siiv  {\ifmmode {\rm Si}\, {\sc iv}\ \else Si\,{\sc iv}\fi}
\newcommand  \SIIV  {\ifmmode {\rm Si}\,{\sc iv}\,\lambda1399 \else Si\,{\sc iv}\,$\lambda1399$\fi}
\newcommand  \civ  {\ifmmode {\rm C}\, {\sc iv}\ \else C\,{\sc iv}\fi}
\newcommand  \CIV  {\ifmmode {\rm C}\,{\sc iv}\,\lambda1549 \else C\,{\sc iv}\,$\lambda1549$\fi}
\newcommand  \NV  {\ifmmode {\rm N}\,{\sc v}\,\lambda1240 \else N\,{\sc v}\,$\lambda1240$\fi}
\newcommand  \nv  {\ifmmode {\rm N}\,{\sc v}\ \else N\,{\sc v}\fi}
\newcommand  \pv  {\ifmmode {\rm P}\,{\sc v}\ \else P\,{\sc v}\fi}
\newcommand  \LyA  {\ifmmode {\rm Lyman}\,{\sc $\alpha$}\,\lambda1216 \else Lyman\,{\sc $\alpha$}\,$\lambda1216$\fi}
\newcommand  \lya {\ifmmode {\rm Lyman}\,{\sc $\alpha$}\ \else Lyman\,{\sc $\alpha$}\fi}
\newcommand  \feii     {Fe\,{\sc ii}}
\newcommand  \feiii     {Fe\,{\sc iii}}
\newcommand  \aliii  {\ifmmode {\rm Al}{\textsc{iii}} \else Al\,{\sc iii}\fi}
\newcommand  \ALIII  {\ifmmode {\rm Al}\,{\sc iii}\,\lambda1857 \else Al\,{\sc iii}\,$\lambda1857$\fi}

\newcommand  \CIII  {\ifmmode {\rm C}\,{\sc iii]}\,\lambda1909 \else C\,{\sc iii]}\,$\lambda1909$\fi}
\newcommand  \oi    {\ifmmode \left[{\rm O}\,{\textsc i}\right] \else [O\,{\sc i}]\fi}
\newcommand  \OI    {\ifmmode \left[{\rm O}\,{\textsc i}\right]\,\lambda6300 \else [O\,{\sc i}]$\,\lambda6300$ \fi}

\newcommand  \oii   {\ifmmode \left[{\rm O}\,{\textsc ii}\right] \else [O\,{\sc ii}]\fi}
\newcommand  \OII   {\ifmmode \left[{\rm O}\,{\textsc ii}\right]\,\lambda3727 \else [O\,{\sc ii}]\,$\lambda3727$ \fi}
\newcommand  \oiii  {\ifmmode \left[{\rm O}\,{\textsc iii}\right] \else [O\,{\sc iii}]\fi}
\newcommand  \OIII  {\ifmmode \left[{\rm O}\,{\textsc iii}\right]\,\lambda5007 \else [O\,{\sc iii}]\,$\lambda5007$\fi}
\newcommand  \ovi    {\ifmmode \left[{\rm O}\,{\textsc vi}\right] \else O\,{\sc vi}\fi}
\newcommand  \neiii   {\ifmmode \left[{\rm Ne}\,{\textsc iii}\right] \else [Ne\,{\sc iii}]\fi}
\newcommand  \nev   {\ifmmode \left[{\rm Ne}\,{\textsc v}\right] \else [Ne\,{\sc v}]\fi}

\newcommand{\lmg}{\ifmmode L\left(\mgii\right) \else $L\left(\mgii\right)$\fi}
\newcommand{\fwmg}{\ifmmode {\rm FWHM}\left(\mgii\right) \else FWHM(\mgii)\fi}
\newcommand{\fwciv}{\ifmmode {\rm FWHM}\left(\civ\right) \else FWHM(\civ)\fi}
\newcommand{\fwhm}{\ifmmode {\rm FWHM} \else FWHM\fi}

\begin{document}

\title{Variability of Low-ionization Broad Absorption Line Quasars Based on Multi-epoch Spectra from The Sloan Digital Sky Survey}
\author{W. Yi\altaffilmark{1,2,9}, W. N. Brandt\altaffilmark{1,4,5}, P. B. Hall\altaffilmark{3}, M. Vivek, \altaffilmark{1},   C. J. Grier\altaffilmark{1,5,6}, N. Filiz Ak\altaffilmark{7,8}, D.~P. Schneider\altaffilmark{1,4}, AND S. M. McGraw\altaffilmark{1} }

\altaffiltext{1}{Department of Astronomy \& Astrophysics, The Pennsylvania State University, 525 Davey Lab, University Park, PA 16802, USA}  
\altaffiltext{2}{Yunnan Observatories, Kunming, 650216, China} 
\altaffiltext{3}{Department of Physics and Astronomy, York University, Toronto, ON M3J 1P3, Canada}
\altaffiltext{4}{Department of Physics, The Pennsylvania State University, University Park, PA 16802, USA}
\altaffiltext{5}{Institute for Gravitation and the Cosmos, The Pennsylvania State University, University Park, PA 16802, USA}  
\altaffiltext{6}{Steward Observatory, The University of Arizona, 933 North Cherry Avenue, Tucson AZ 85721, USA }
\altaffiltext{7}{Faculty of Sciences, Department of Astronomy and Space Sciences, Erciyes University, 38039, Kayseri, Turkey}
\altaffiltext{8}{Astronomy and Space Sciences Observatory and Research Center, Erciyes University, 38039, Kayseri, Turkey}
\altaffiltext{9}{Key Laboratory for the Structure and Evolution of Celestial Objects, Chinese Academy of Sciences, Kunming 650216, China}

\begin{abstract} 
We present absorption variability results for 134 bona fide \mgii\ broad absorption line (BAL) quasars at  0.46~$\lesssim z \lesssim$~2.3 covering days to $\sim$ 10 yr in the rest frame. We use multiple-epoch spectra from the Sloan Digital Sky Survey, which has delivered the largest such BAL-variability sample ever studied. \mgii-BAL identifications and related measurements are compiled and presented in a catalog. 
We find a remarkable time-dependent asymmetry in EW variation from the sample, such that weakening troughs outnumber strengthening troughs, the first report of such a phenomenon in BAL variability. 
Our investigations of the sample further reveal that (i) the frequency of BAL variability is significantly lower (typically by a factor of 2) than that from high-ionization BALQSO samples; (ii) \mgii\ BAL absorbers tend to have relatively high optical depths and small covering factors along our line of sight; (iii) there is no significant EW-variability correlation between \mgii\ troughs at different velocities in the same quasar; and (iv) the EW-variability correlation between \mgii\ and \aliii\ BALs is significantly stronger than that between \mgii\ and \civ\ BALs at the same velocities. 
These observational results can be explained by a combined transverse-motion/ionization-change scenario, where transverse motions likely dominate the strengthening BALs while ionization changes and/or other mechanisms dominate the weakening BALs. 
\end{abstract}


\section{Introduction}
\label{sec:intro}

Intrinsic absorption lines in quasars, indicative of outflowing material from circumnuclear regions of quasars, are mainly manifested as broad absorption lines (BALs, e.g., \citealp{Weymann91}) and mini-BALs (e.g., \citealp{Hamann04}) in quasar spectra that are blueshifted with respect to systemic redshift. 
BAL quasars, which make up  $\sim$20\% of all quasars (e.g., \citealp{Gibson09,Paris17}), are usually  classified into three subtypes depending on the ionization levels of the transitions present in their absorption troughs. The majority are classified as high-ionization BAL quasars (HiBALs), which  contain deep and wide absorption troughs caused by species such as \civ, \nv, and \ovi. Low-ionization BAL quasars (LoBALs, $\sim$10\% of the BAL-quasar population, e.g., \citealp{Trump06}) are characterized by the presence of additional low-ionization species (e.g., \mgii, \aliii) in their spectra. Fe low-ionization BAL quasars (FeLoBALs, an even more rare subtype)  show prominent \feii\ and/or \feiii\ in addition to other low-ionization species.

Accumulating observational evidence indicates that BAL QSOs are likely associated with feedback from active galactic nuclei (AGN), an effective process of controlling the co-evolution between supermassive black holes (SMBHs) and host galaxies (e.g., \citealp{Di05,Fabian12}). 
The inferred kinetic luminosities of AGN/QSO outflows from low to high redshifts suggest that they could potentially play a critical role in  AGN/QSO feedback (e.g., \citealp{Crenshaw12,Borguet13,Yi17,Arav18}). 
BAL outflows thus have been proposed as important contributors in the regulation of the growth of both SMBHs and their hosts, and may  even dominate their coevolution history. 
Signatures of these outflows are seen in spectra across a wide range of wavelengths, and they can provide abundant and powerful diagnostics to explore or at least constrain geometrical structure, dynamics, chemical enrichment, and inner physics, greatly improving our understanding of AGNs. 

\begin{deluxetable*}{lcrccccc}
\tabletypesize{\scriptsize}
\tablecolumns{7}
\tablewidth{0pc}
\tablecaption{Individual/sample-based BAL variability studies of (\feii)LoBAL QSOs}
\tablehead{\colhead{Reference} & \colhead{ No. of BAL}\tablenotemark{a}  & \colhead{$\Delta t$ (yr)}\tablenotemark{b} & \colhead{No. of epochs} & \colhead{No. of Var}\tablenotemark{c}
& \colhead{ Type}\tablenotemark{d} & \colhead{Redshift}   }
\startdata
\citet{Junkkarinen2001} & 1/1 & $\sim$5.9 & 2 & 1 & L & 0.848  &  \\
\citet{Hall2011} & 1/1 & 0.6 -- 5 & 4 & 1 & F &  0.848 & \\

\citet{Vivek2012a} & 1/1 & $\sim$10.9 & 7 & 1 & L & 0.92 &  \\

\citet{Vivek2012b} & 4/5 & 0.3 -- 9.9 & 4 -- 14 & 0 & F  &  0.5 -- 2.0 & \\

\citet{Vivek14} & 17/22 & 0.3 -- 9.9 & 3 -- 7 & 8 & L/F  & 0.2 -- 2.1  & \\

\citet{Zhang2015} & 8/28 & 1 -- 10 & 2 -- 3 & 5 & H/L/F  &  0.55 -- 2.92 & \\

\citet{McGraw2015} & 11/12 & 0.03 -- 7.6 & 2 -- 10 & 3 & F  &  0.69 -- 1.93 & \\

\citet{Rafiee2016} & 3/3 & 0.03 -- 7.6 & 2 -- 4 & 3 & L/F  & 0.83 -- 2.34  & \\

\citet{Stern2017} & 1/1 & $\sim$9 & 3 & 1 & L & 2.345  & \\
This work & 134/134 & 0.005 -- 14 & 2 -- 47 & 45 & L/F & 0.46 -- 2.3  & 
\enddata
\tablecomments{  In our study, 134 \mgii-BAL QSOs have been selected from the main list (2109 BALQSOs) by a criterion of having at least two-epoch spectra. Among these \mgii-BAL QSOs, 80 have two-epoch spectra,  32  have three-epoch spectra, and 22  have $\ge$4 spectra. (a: Number of \mgii-BAL QSOs/ the whole sample size. b: Time span in the observed frame. c: Number of variable objects. d: L for LoBAL, F for FeLoBAL, H for HiBAL )}
\label{tab1}
\end{deluxetable*}

The inner regions of quasars cannot be spatially resolved with current technology, leading to a poor understanding of the nature of BALs. 
Fortunately, BAL variability provides one of the most powerful diagnostics for exploring the origin of outflows. 
Variations of HiBAL troughs, in depth or in width (or both), are relatively common based on results from previous studies (e.g., \citealp{Gibson08,Cap12,Filizak13}). In some cases, dramatic changes either from individual-object (e.g., \citealp{Hamann08,Grier15}) or large-sample studies, have been reported (e.g., \citealp{Gibson10,Filizak12}). 
BALs showing monolithic velocity shifts over the entire BAL trough, which may be associated with the acceleration of outflows, have also been reported, though rarely (e.g., \citealp{Grier16}). 
The absolute amplitude of BAL variability does not increase either with higher radio-loudness or radio luminosity among  HiBAL QSOs (e.g., \citealp{Welling14,Vivek16}), and similar results have been reported for samples of radio-selected (Fe)LoBAL \footnote{Hereafter, we will use the term (Fe)LoBAL QSOs when a sample under study includes both LoBAL and FeLoBAL QSOs} QSOs \citep{Zhang2015}.

Systematic studies of BAL variability based on large samples of quasars have become increasingly feasible using data from large sky surveys carried out by dedicated facilities, such as the Sloan Digital Sky Survey (SDSS, \citealt{York00}), which have greatly advanced our understanding of the properties of BALs. Indeed, previous sample-based studies have demonstrated that HiBALs are typically variable on timescales from days to years in the quasar rest frame (e.g., \citealp{Gibson08,Gibson10,Cap11, Cap12, Filizak12, Filizak13, Welling14, Wang15,He17}). These investigations are almost exclusively  based on high-ionization BALs. 
LoBALs, however, have thus far been studied sparsely, usually only including individual objects or  small samples (see Table \ref{tab1}).

\citet{Vivek14} reported that highly variable low-ionization BALs tend to occur at higher detached velocities, low equivalent widths, and low redshifts in a sample of 22 (Fe)LoBAL QSOs. 
This study also found that LoBAL variability is more frequently detected in QSOs showing strong Fe emission lines but found no clear correlation between continuum flux and LoBAL variability. 
Additionally, some studies have indicated that BAL variability in FeLoBAL QSOs is less common than for non-Fe LoBAL QSOs (e.g., \citealp{Vivek14,McGraw2015}), while other studies have found a high fraction of BAL variability from a radio-selected sample (e.g., \citealp{Zhang2015}). 
However, inferences on FeLoBAL variability from previous studies can yield only tentative results since their sample sizes are inadequate to draw firm statistical conclusions.

Currently, the origins of BAL variability are poorly understood; however, transverse motions across our line of sight (LOS) and changes in the ionization state of outflowing gas are two of the most widely accepted explanations accounting for BAL variability. 
The emergence or disappearance of BALs is often interpreted as gas moving into or out of the LOS (e.g., \citealp{Hamann08,Hall2011,Zhang2015,Rafiee2016}), although some studies did find observational evidence of ionization changes among small-to-moderate equivalent width (EW) BAL troughs (e.g., \citealp{McGraw2017,Stern2017}).  
Variability behaviors such as correlated variations with continuum fluctuations or coordinated EW variations for BALs of the same species but at different velocities are usually attributed to changes in the ionization state of the gas (e.g., \citealp{Hamann11,Filizak13,Wang15,DeCicco18}).  
Observed coordinated EW variability in troughs at different velocities indicates that at least $56\% \pm 7\%$ of BAL variability is due to variability in the ionization state of the absorbing gas \citep{Filizak13,He17}, as no obvious mechanism exists to explain such coordinated variability using transverse gas motion. However, because different-velocity absorbers associated with different densities in the same object do not always exhibit coordinated BAL variability in response to the same ionizing continuum variations \citep{Arav12}, it is difficult to determine definitively which mechanism is dominant in individual cases. 
Constructing a large sample to investigate LoBAL variability systematically can improve our understanding of BAL variability from low- to high-ionization transitions as a whole. Although LoBAL QSOs are not the dominant population, their intrinsic properties may provide unique and valuable clues about nuclear outflows and hence shed light on the origin of BAL variability. 

 Section~\ref{sec:sample} describes the sample compilation, data reduction, and related analyses/methods adopted in this work. 
Section~\ref{mgii_identification} describes the identification of \mgii-BAL troughs and introduces the definition of ``BAL complex'' before quantifying their properties. The quantitative measurements are illustrated and demonstrated in Section \ref{Quantifying}. Statistical results are presented in detail in Section~\ref{stat_view}. In Section \ref{var_timescales}, we discuss the effect of saturation in our results and propose a model to account for the observed BAL variability. 
The paper concludes with a summary of the main conclusions in Section \ref{sum_sec}.  Throughout this work, we use a cosmology in which $H_0 = 70$~km~s$^{-1}$~Mpc$^{-1}$, $\Omega_M = 0.3,$ and $\Omega_{\Lambda} = 0.7$.

\section{Sample compilation and data analysis}
\label{sec:sample}
We use spectroscopic data from the Sloan Digital Sky Survey-I/II (hereafter SDSS; \citealp{York00}) and the Baryon Oscillation Spectroscopic Survey of SDSS-III (hereafter BOSS; \citealp{Eisenstein11}). 
SDSS and BOSS are wide-field, large-sky survey projects using the dedicated 2.5 m telescope at the Apache Point Observatory, New Mexico \citep{Gunn06, Smee13}. 
There are 2109 BAL QSOs in our main SDSS-based sample (see \citealt{Filizak13} for further discussion), among which 2005 are classified as ``regular'' BAL QSOs, and 104 are classified as ``special'' BAL QSOs that have characteristics such as long-time coverages in spectroscopy, overlapping BAL troughs, or other unusual features.  HiBAL (e.g., \civ\ and \siiv) variability of the sample has been intensively analyzed in previous studies (e.g., \citealp{Gibson09,Gibson10,Filizak12,Filizak13,McGraw2017}), but there is an obvious deficiency of investigations of LoBAL variability (\mgii-BAL variablity in this sample has not yet been explored).  
Objects in the main sample are luminous BAL QSOs whose SDSS spectra have relatively high signal-to-noise ratios and strong BALs in their SDSS spectra.   
To ensure the detection of major absorption lines, we only consider quasars with at least one spectrum with an average S/N $>6$ from 3020 to 3100 \AA~ or 2000 to 2150 \AA\ in the rest frame depending on their redshifts. 
Since this work is focused on the variability of \mgii\ BALs, a redshift constraint is applied at 0.46 $\leq z \leq$ 2.3 to ensure an adequate coverage of the \mgii\ region, within which 1526 BAL quasars are included. 
We choose to use the modified Balnicity index values for \mgii\ BALs (BI$_{M,0}$) from  \citet{Gibson09} to search for \mgii\ BALs, as the traditional BI metric does not include low-velocity ($v<3000$ km s$^{-1}$)  BALs, which are frequently seen among LoBALs. 
We require the formal presence of one or more unambiguous \mgii-BAL troughs (BI$_{M,0} > 10$ km s$^{-1}$) in the SDSS-I/II spectra and that there exist at least two spectroscopic epochs for each object. The average number of spectroscopic epochs in the sample is three, corresponding to six unique spectroscopic pairs.  

\begin{figure} [h]
  \begin{center}
      \includegraphics[width=8.5cm, height=6cm, angle=0]{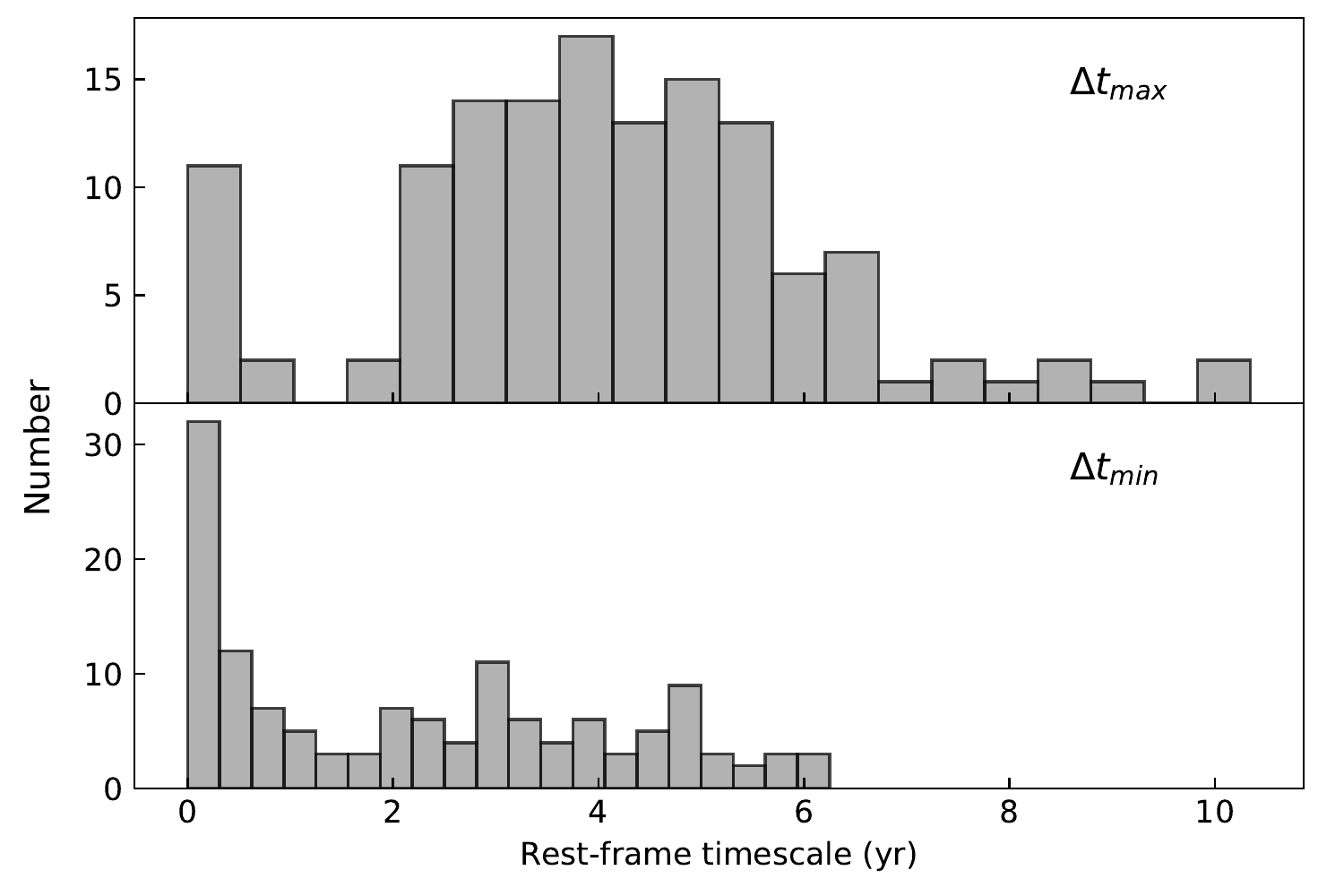}
       \caption{Distribution of the maximum (upper panel) and minimum (lower panel) sampled rest-frame timescales of the 134 \mgii-BAL QSOs in our sample.} 
   \end{center}
   \label{fig1}
\end{figure}

Based on visual inspection, we exclude seven extreme overlapping-trough \mgii-BAL quasars, such as SDSS J030000.57+004828.0 \citep{Hall02}, which have no available continuum windows below 2800 \AA. We also exclude two objects that have \mgii~ absorption too shallow to formally qualify as a BAL and a spectral turnover near 2800 \AA\ to which it is difficult to fit a continuum; namely, SDSS J010540.75$-$003313.9 \citep{Hall02} and J140800.43+345124.7. Such objects cannot be reliably studied using the continuum normalization approach adopted here. These exclusions leave 134 quasars, which comprise the \mgii-BAL sample used throughout this study.  In addition, we obtained follow-up spectra using the Lijiang 2.4m Telescope (LJT; \citealp{Fan15}) and the Hobby-Eberly Telescope (HET; \citealp{Ramsey98,Chonis16}) for $\sim$6\%  BAL QSOs in the sample. The spectral observations from SDSS-I/II and SDSS-III provide a long time baseline for most objects, typically longer than two years in the rest frame (see Figure 1). The maximum/minimum time spans (see Figure \ref{fig1}) for each object are obtained from the two spectra that give the maximum/minimum sampled rest-frame timescales.

\begin{deluxetable}{lccccc}
\tabletypesize{\scriptsize}
\tablecolumns{6}
\tablewidth{0pc}
\tablecaption{Properties of \mgii-BAL QSOs in the sample}
\tablehead{\colhead{Name (SDSS)} & $z$  & N$_e$& $R^*$ & $M_i$ } 
\startdata
J010352.46+003739.7  &  0.703  &    5  &  6.25   &  $-$25.73\\
 \bf{.} & \bf{.} &\bf{.} & \bf{.}  & \bf{.}  \\
 \bf{.} & \bf{.} &\bf{.} & \bf{.}  & \bf{.}  \\
J235843.48+134200.2  &  1.134  &   3 &  $-$1.0   &  $-$25.67 
\enddata
\tablecomments{ Note: N$_e$ is the number of epochs; $M_i$ is the absolute $i$-band magnitude \citep{Schneider10}; $R^*$ is the radio loudness \citep{Shen11}, where $R^*=-1$ means not available in that catalog. The full table is available in the online version.} 
\label{tab2}
\end{deluxetable}

In this work, we adopt redshift values from \citet{Hewett10} for most objects in our sample. However, about 8\% of our quasars have relatively large redshift discrepancies ($\Delta z\geq$ 0.01) from different catalogs (e.g., \citealp{Gibson09,Schneider10,Paris17}),  as these measurements are affected by strong absorption.  
If a discrepancy was found in at least two other quasar catalogs, we re-determine their redshifts based on the following criteria: 

\begin{enumerate}

\item
If typical narrow emission lines (e.g., \oii, \oiii) are present in the spectra, we choose the redshift that best matches the positions of these lines.

\item
If narrow emission lines are absent but the \mgii\ broad emission line is present with a symmetric profile, we choose the redshift that  best matches the \mgii\ line profile using the SDSS non-BAL quasar composite template \citep{VandenBerk01}. 

\item
For a few objects with available near-IR spectra (see Section \ref{Eddington_dependence}), we determine their redshift using the \oiii\ or H$\alpha$ emission line. 

\item
For objects that do not meet any of the three conditions above, we choose the redshift that best matches with the positions of the variety of emission features present.  During this process, high-ionization emission lines have a low priority since they are often blueshifted with respect to the systemic velocities  of quasars (e.g., \citealp{Shen16}). 

\end{enumerate}

In conjunction with other quasar catalogs \citep{Schneider10,Shen11}, we tabulate several basic properties of our quasar sample in Table \ref{tab2}.

\begin{figure*}
\center{}
 \includegraphics[width=16cm, height=8cm, angle=0]{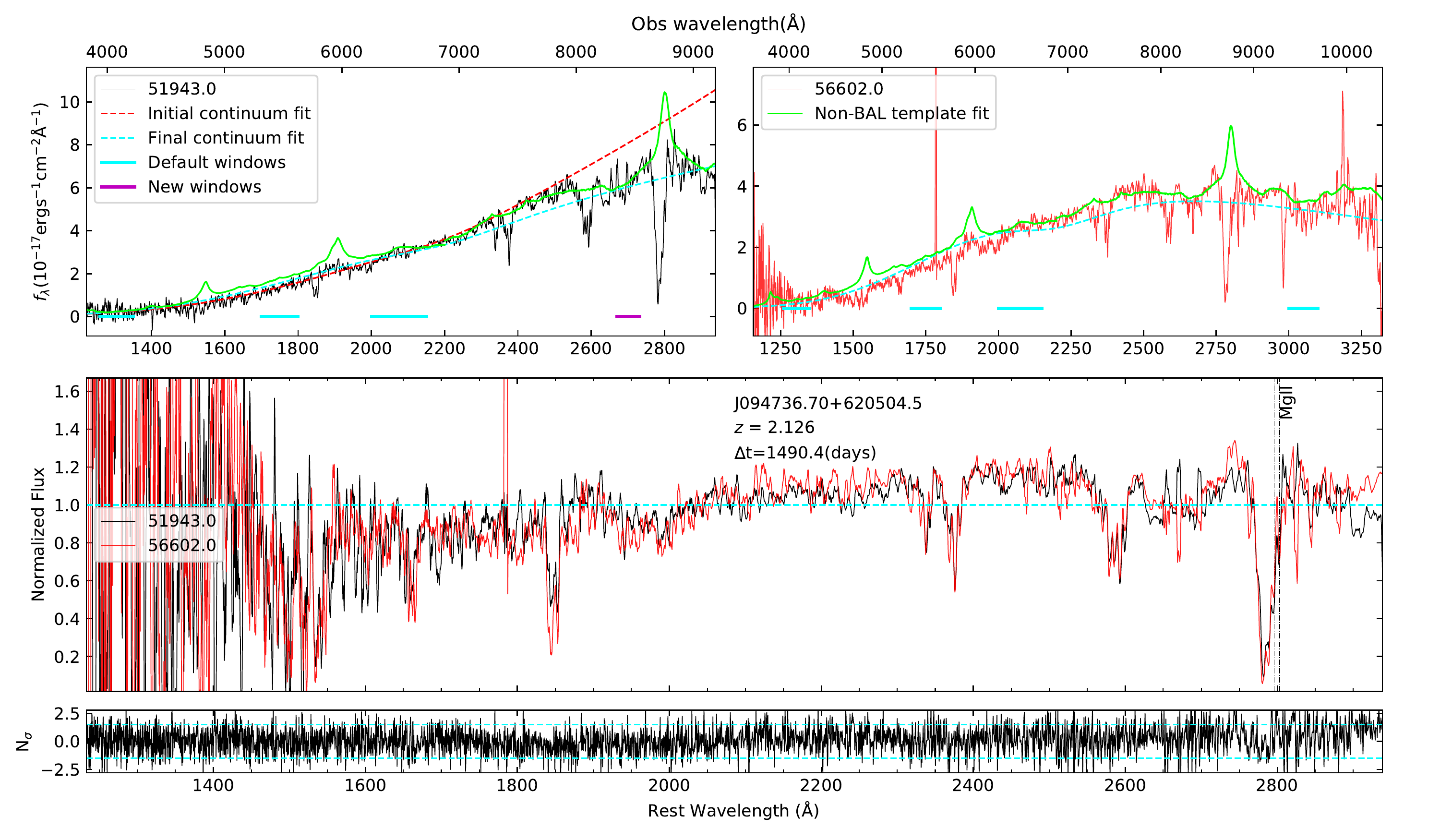}
      \caption{ Demonstration of the continuum and non-BAL template fits: The top-left panel displays the differences between initial (red dashed line) and final (cyan dashed line) fits for the first (SDSS I/II) epoch at MJD=51943 spectrum with a narrower wavelength coverage. The top-right panel shows the BOSS spectrum at MJD=56602, which covers a wider wavelength range. The non-BAL template fits are plotted above the two-epoch spectra (green lines). Comparisons between the two continuum-normalized spectra are presented in the middle and bottom panels ($N_\sigma$; see Section \ref{variable_region_define}). All spectra are smoothed by an 8-pixel boxcar filter for display. }
\label{normspec2com}
\end{figure*}

\subsection{Continuum fits}

Our spectroscopic fitting can be separated into two processes: the continuum fit and the non-BAL template fit. The continuum fit is executed first to determine the continuum level, which in turn is adopted as a benchmark for the non-BAL template fit.  

In previous studies of BAL variability, a variety of models have been adopted to fit the continuum, such as power-law models  (e.g., \citealp{Cap11,Cap12}), reddened power-law models (e.g., \citealp{Gibson08, Gibson09,Filizak13}), and polynomial models (e.g., \citealp{lundgren07}).  
Following previous studies  (e.g., \citealp{Gibson09,Filizak13,Grier16}), we adopt a reddened power-law model to fit the continuum emission with a nonlinear least-squares fitting algorithm. We modeled the continuum in each spectrum using a SMC-like reddened power-law function from \citet{Pei92} with free parameters including the amplitude, power-law index, and extinction coefficient. 
The initial continuum-fit windows are composed of relatively line-free (RLF) regions defined by \cite{Gibson09}: 1250--1350~\AA, 1700--1800 \AA, 1950--2200 \AA, 2650--2710 \AA, 2950--3700 \AA, 3950--4050 \AA, and 4140--4270 \AA. 
However, these default windows do not always suitably sample the continuum for all objects, so they are adjusted where necessary to reach an acceptable final fit. In particular, obvious emission/absorption lines present in the default windows among all spectra have been excluded.

A number of operations were performed before and during the continuum-fit process, including pixel masking for each continuum-fit window  and setting additional constraints (details are given in the following paragraph).  
All spectra are then converted to the rest frame using redshifts as discussed above. 
Emission and/or absorption features, in some cases, appear to be frequently present in  the same fitting regions. Therefore, we adopt a sigma-clipping algorithm that consists of fitting the continuum function to the RLF windows for each spectrum, and then  rejecting data points that deviate by more than 3$\sigma$ from the continuum fit for each pixel in each window.  The final continuum fit is obtained by re-fitting the remaining data points. We do not attach physical meaning to the values of $E(B - V )$ obtained from our fits since there is an obvious degeneracy between the UV spectral slope and the magnitude of intrinsic reddening. The uncertainty of the continuum fit is obtained via a Monte Carlo approach through randomizations of spectral errors in the RLF windows.

In cases where BAL QSOs have spectra from both SDSS and BOSS, we choose the highest S/N spectrum from BOSS as a reference to set additional constraints to fit SDSS-I/II spectra, as BOSS spectra have a wider wavelength coverage.  This aspect is particularly important in the scenario where the default number of RLF windows is different between the two spectra because of their difference in wavelength coverage. 
Using the additional constraints, the final continuum fit of the SDSS spectrum can be improved to be consistent with its BOSS counterpart that has a wider wavelength coverage. 
An example is shown in the top-left panel of Figure \ref{normspec2com}, where the initial SDSS-I/II fit (red dashed line) would be significantly inconsistent with that from the BOSS spectrum displayed in the top-right panel of Figure \ref{normspec2com}.

The continuum-fit process works well for most LoBAL QSOs in our sample. However, for $\sim$ 20\% objects in the sample, heavy absorption are often seen, or the $2200$--$3000$ \AA\ regions are contaminated by complex \feii\ and \mgii\ emission/absorption lines, or the \mgii-BAL troughs lie in particularly noisy regions of the spectrum.  In such cases, automatic fits may be unreliable, as the RLF regions may be unsuitable for use in these spectra. We identified these cases by visual inspection and manually re-fit them either by re-identifying RLF windows, increasing/decreasing weights to different windows, or adopting polynomial functions.

\subsection{Non-BAL template fit}

\begin{figure*}
\center{}
 \includegraphics[width=18cm, angle=0]{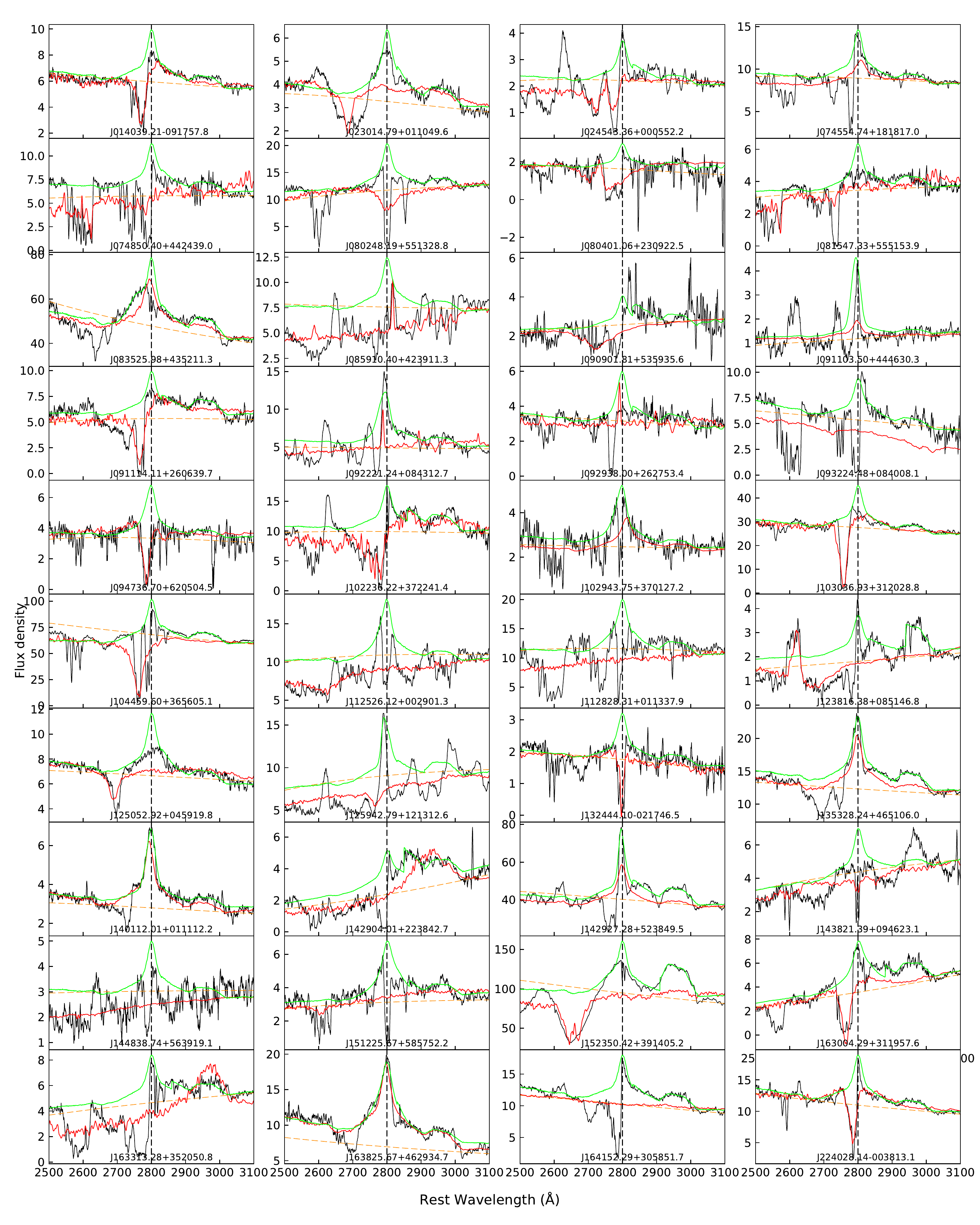}
      \caption{  Demonstration of the diversity of various typical profiles around the \mgii~  emission line in our sample. Vertical dashed lines mark the center of the \mgii~ doublet. For each object, the actual spectrum (black line), best-fit spectral model (red line) from the SDSS pipeline, our reddened power-law fit (dashed orange line), and our non-BAL template fit (green line) are indicated. A few quasars show prominent bumps at $\lambda_{\rm rest}\sim 2600$ \AA, likely due to \feii\ emission; they did not affect our measurements of relative BAL variability between different epochs (see text for details).}
\label{mgiiBAL_diversity}
\end{figure*}

While our reddened power-law fits provide us a reasonable estimate of the continuum level for our quasars, these fits do 
not take emission features into account, and thus they are insufficient for BAL studies in spectra with absorption troughs at low velocities that superimpose them on the \mgii~ emission line; the continuum fits also does not take into account the underlying blended contamination from \feii\ (see examples in Figure~\ref{mgiiBAL_diversity}), etc. 
Based on an investigation of a sample of 68 \mgii-BAL profiles, \citet{Zhang10} reported that the \mgii-BAL composite spectrum is similar to the non-BAL composite spectrum of \citet{VandenBerk01} except for some subtle differences regarding narrow emission lines and \feii~ features. 
Therefore, we adopt a model combining the non-BAL composite spectrum \citep{VandenBerk01} and an SMC-like reddening function \citep{Pei92} as a template to fit the spectra in our sample.

During the initial template-fitting process, we only include the local region between 2200 and 3400 \AA, where most spectra can be matched by the same non-BAL template with different reddening. The reddened non-BAL template can be simply expressed as the following function:

\begin{equation}
f(\lambda) = a f_t(\lambda)  e^{-\tau_{\lambda}}
\label{eq1}
\end{equation}
where $f_t$ is the non-BAL quasar template and $\tau_{\lambda}$ is a reddening factor that depends on the wavelength $\lambda$ \citep{Pei92}.  
Because our reddened power-law fits can reproduce the underlying continua to a higher precision, they were adopted as benchmarks to guide the non-BAL template fits. Only data points above 95\% of the continuum-normalized spectra were taken into account as the default, which  effectively clips most absorption troughs. 
The template fits have independent $E(B-V)$ values from those derived from the corresponding continuum fits, though the two values are  approximately equivalent in most cases. 

Generally, the initial template fit  works well for most cases where the \mgii\ emission lines are unabsorbed or slightly absorbed. 
In cases where the flux in the initial template fits around the \mgii\ emission-line is underestimated, additional Gaussian components (no more than two) are adopted to fit the \mgii\ emission-line. 
During this process, we use un-smoothed spectra in the wavelength range between 2790 and 2810 \AA\  to capture the peak of the \mgii\  line. 
For intermediately and heavily absorbed \mgii\ emission lines, however, no adjustments have been made after the initial template fits even if the fits appear to overestimate the apparent peak strength of \mgii\ emission lines (e.g., J0947+6205 and J1030+3120 in Figure \ref{mgiiBAL_diversity}) as long as they are consistent among different spectroscopic epochs for the same object. The over fitted parts of the \mgii\ emission lines likely have little effect on the measurement of relative BAL variability since no dramatic \mgii\ emission-line variability was found in the sample, in agreement with that from \citet{Sun15}.  
The over-fitted regions are consistent with each other over different epochs. 
Uncertainties may be introduced when reconstructing absorbed \mgii\ emission lines, especially in cases where the emission line is very heavily absorbed by a low-velocity trough. We visually inspect all cases where the \mgii-BAL troughs are attached, or superimposed, on the \mgii\ emission lines.

In most cases, our initial template fits match the spectrum well on the redward side of \mgii, at $\lambda>2800$~\AA. However, a few  objects showing significant \feii\ and \mgii\ absorption troughs or other unusual features may have large uncertainties due to difficulty in reconstructing their intrinsic spectra. Therefore, additional broadened \feii\ components were also taken into account during the fitting process for such objects. 
We use a combination of UV \feii\ templates from \citet{Salviander07} for the  2200--3090 \AA\ region and from \citet{Tsuzuki06} for the 3090--3500 \AA\ region.  Because some \feii\ broad bumps ubiquitously appeared among these spectra at $\lambda > 2820$~\AA, the original \feii\ template was split into three main regions: $2830 < \lambda < 2910$ \AA, $2910 < \lambda < 3050$ \AA, and $3050 < \lambda < 3400$ \AA. Before fitting, these \feii\ components were linearly interpolated and then broadened by 5-pixel Gaussian kernel smoothing. 
Due to the absorption, we did not use additional \feii\ components to fit \feii\ emission features at $\lambda < 2800$ \AA; thus the fits in this region may deviate from the spectra due to strong \feii\ emission features (e.g., J0245+0005 and J0911+4446 in Figure \ref{mgiiBAL_diversity}). 

We use the same fitting procedures to fit the spectra around \civ\ and \aliii\ ($\lambda < 2200$\AA) in our \civ-BAL subsample, 
which  allows an investigation of BAL variability from low- to high-ionization species simultaneously (see Section \ref{EWvar_mgii_civ} for details). 
Many objects in our sample have two apparently different spectral slopes at blue ($\lambda < 2200$\AA) and red ($\lambda > 2200$\AA) wavelengths, which leads to an inability to fit the whole spectrum with a single reddened non-BAL quasar template. 
To alleviate this problem, we opted not to tie the amplitude and reddening of the regions around the other emission lines to those from the local fit around \mgii. Instead, they are treated as independent parameters during the fitting process. Thus, in some cases, a break at 2200 \AA\ caused by the independent fits can be seen. However, such a break does not affect our measurements of BAL variability, as the break is located far from the central positions of the two lines (at 1860 \AA\ and 2800 \AA\ for \aliii\ and \mgii, respectively).

\subsection{Consistency of the two methods }
To compare the results derived from the reddened power-law fits and non-BAL template fits, we measure the quantity $\Delta$EW (the difference in equivalent width for the same BAL trough at two different epochs) for \mgii\ and \civ\ BALs using the two methods.  

\begin{figure}[h]
\center{}
 \includegraphics[height=5cm,width=6cm,  angle=0]{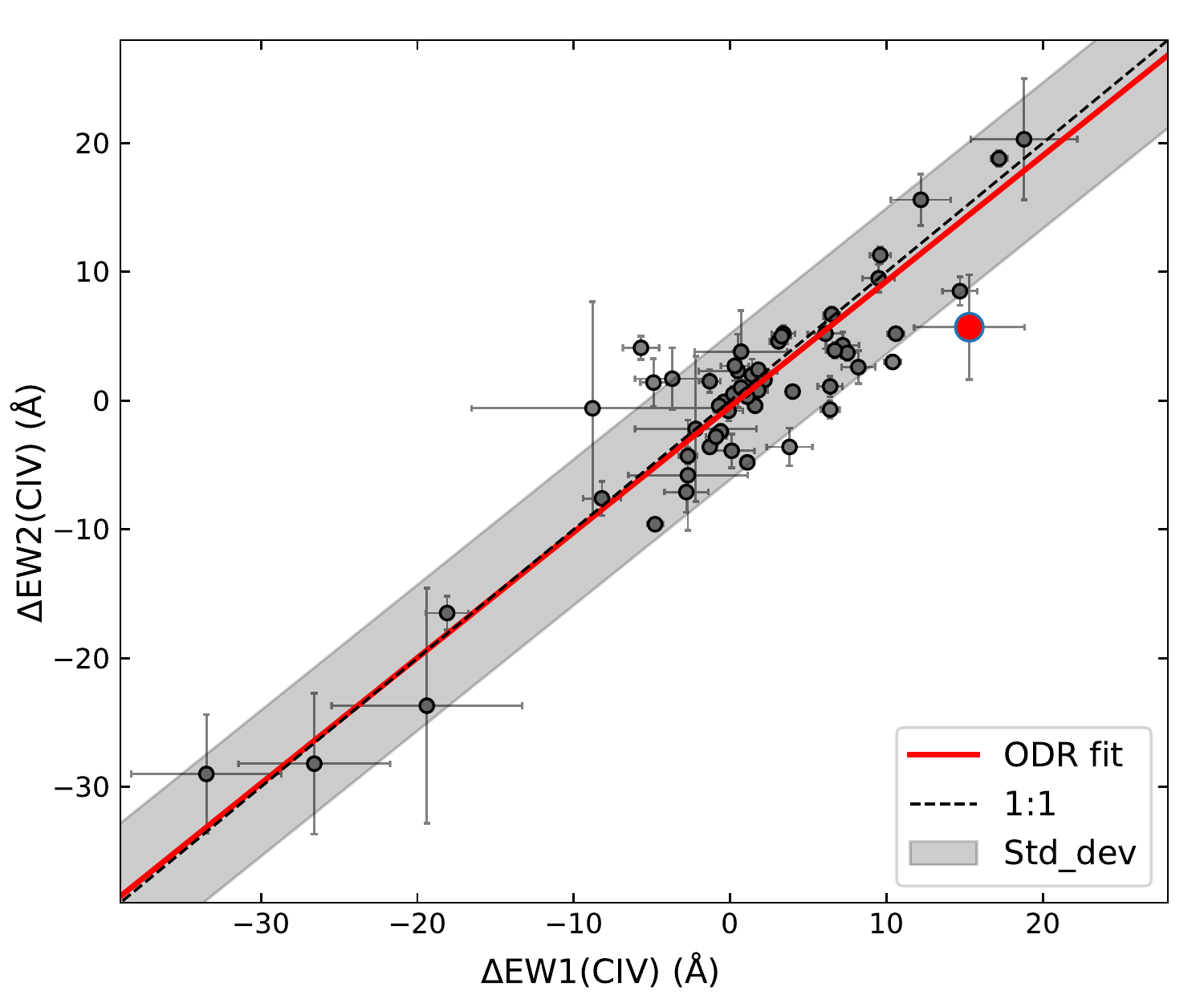} 
 \includegraphics[height=5cm,width=6cm,  angle=0]{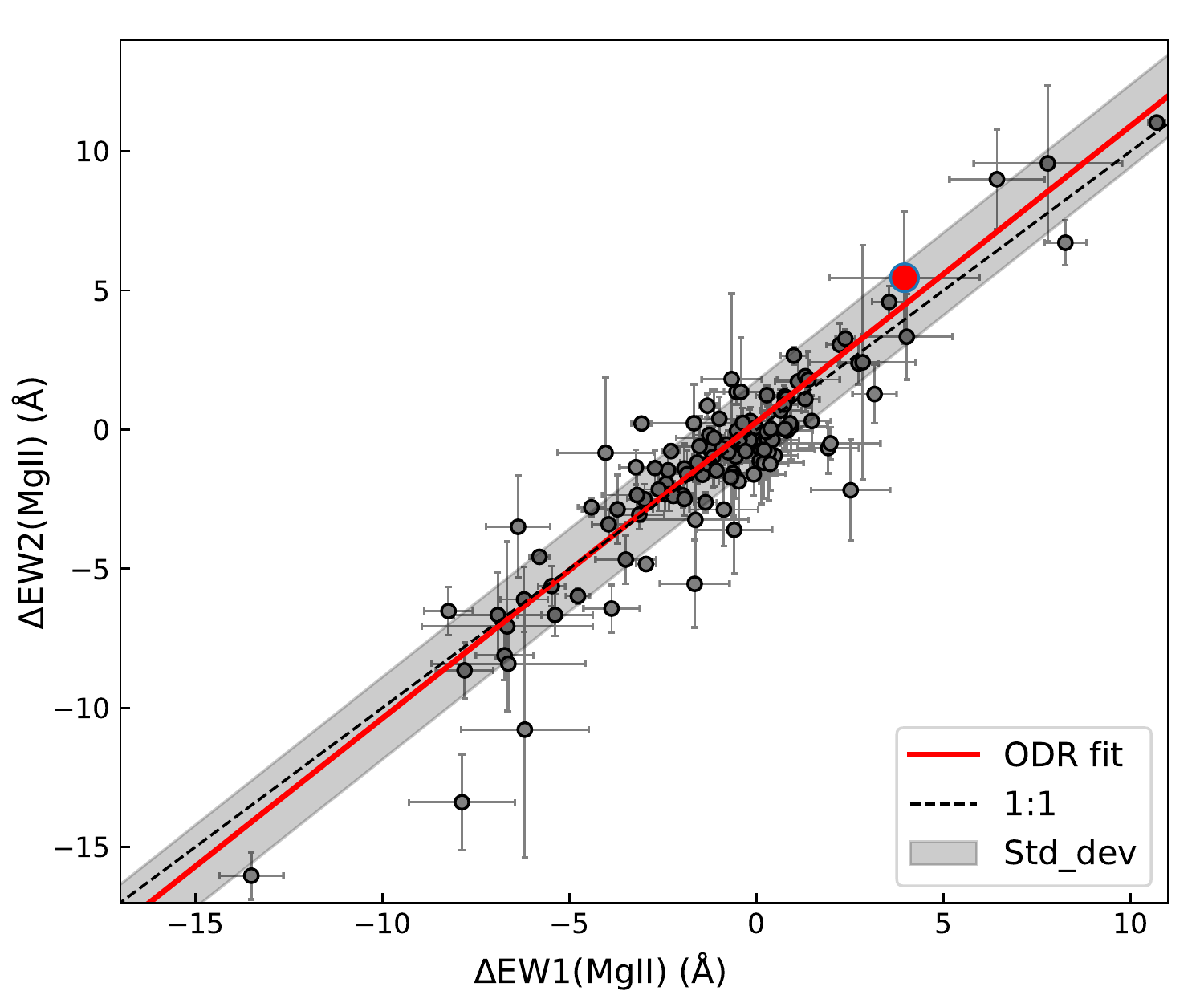} 
      \caption{Comparison in EW variations from non-BAL template fits ($\Delta$EW1) and reddened power-law fits ($\Delta$EW2). The black dashed lines shows the boundary where $\Delta$EW1 equals $\Delta$EW2. Grey regions present linear fits based on orthogonal distance regression within a standard deviation. The big red circles refer to measurements in J0947+6205 as shown in Figure \ref{normspec2com}.}
\label{civEWvar_byTPfits}
\end{figure}

In general, the \civ\ BALs have much deeper and wider absorption troughs than those of \mgii\ BALs, although the actual optical depths are often hidden due to reddening and partial covering. In cases where objects have been intermediately absorbed, the reddened non-BAL template can model their unabsorbed spectra in a reasonable sense over different epochs for the same object, which may provide a more reliable normalization than that derived from a reddened power-law fit. For heavily absorbed objects, particularly for the template fit around the \civ\ emission line at the blue end of the spectra, the uncertainty in EW of a trough may increase significantly due to the effects of reddening, low S/N, abundance of heavy absorption features, and most importantly, a lack of identifiable continuum regions, leading to a larger scatter (more than one standard deviation in Figure \ref{civEWvar_byTPfits}) and discrepancy in $\Delta$EW derived from the two methods.

We found strong linear correlations of $\Delta$EW (see Figure \ref{civEWvar_byTPfits}) using the two methods based on Pearson tests ($r>0.8$ and $p<10^{-23}$), suggesting that our non-BAL template fits do not introduce false variability signals, and are thus unlikely to adversely affect our BAL variability analysis. Because the non-BAL template fits are likely to yield more accurate EWs that better characterize BALs, we adopt these as our standard method for normalizing spectra for analyses throughout.  A large fraction of best-fit models from the SDSS pipeline failed to match the spectra in our sample (as shown in Figure \ref{mgiiBAL_diversity}), highlighting the importance of a new fitting method for this work. Compared with best-fit models from the SDSS pipeline, our non-BAL template fits provide much more reasonable fits to these spectra.

\section{\mgii-BAL  identification and measurement} 
\label{mgii_identification}
As we imposed a requirement of BI$_{M,0} \ge 10$ \kms\ when compiling the sample, each object is expected to show at least one obvious trough blueward of the \mgii\ emission line. However, the identification and investigation of \mgii\ BALs can be somewhat difficult, as these troughs are often contaminated by \feii\ emission/absorption lines, the Balmer continuum, and dust reddening. 
Indeed, the spectral shape between 2200 and 3400 \AA\ in our sample differs from object to object,  and sometimes even changes  from epoch to epoch for an individual object (see Section \ref{section_bal_complex}). 

Following previous studies (e.g. \citealp{Trump06,Filizak13}), we define velocities associated with blueward absorption troughs to be negative. Positive velocities indicate features that are at longer wavelengths than the rest-frame emission-line center. Wavelengths are converted to LOS velocities  using the red component ($\lambda = 2803$ \AA) of the \mgii\ doublet.

\subsection{Identification of \mgii-BAL troughs}
 \label{ew_definition}
\mgii~ broad emission and absorption lines have a wide variety of characteristics (see Figure~\ref{mgiiBAL_diversity}). \mgii-BAL troughs are typically weaker and narrower than those from \civ\ in the same object. The \mgii\ emission lines in the sample usually show slight  asymmetry and blueshift with respect to the \oii\ or \oiii\ emission lines, but most of them are blueshifted within the range of intrinsic uncertainties ($<$ 250 \kms, see \citealt{Shen16}). 
In addition, some objects' profiles are obviously characterized by deep, wide troughs around the \mgii~ emission line plus significant reddening and apparent Fe emission/absorption, making the identification of ``clean'' \mgii-BAL components difficult.

After performing the non-BAL template fits, we smoothed each spectrum using a boxcar filter over five pixels ($\sim$ 69 \kms per pixel) to aid in identifying BALs and mini-BALs (for definition see \citealp{Hall02, Hamann04}). 
We then made  measurements of equivalent widths, minimum/maximum velocities, velocity widths, trough depths, and variable regions (defined below), which all can be used to quantify BAL properties. 

Both BAL and mini-BAL troughs signal the presence of circumnuclear outflows along our LOS. 
We include both BAL/mini-BAL troughs in our analysis and do not distinguish them throughout this work. Note that the \mgii-BAL doublets with a separation of 769 km s$^{-1}$ cannot be distinguished if the corresponding absorption lines are saturated and wider than the doublet separation. We thus  choose a velocity-width threshold of 1250 \kms\ to search for all BAL/mini-BAL troughs after considering both the separation of 769 \kms\ and minimum velocity width of 500 \kms\ for mini-BAL troughs. 

For our study, we consider only BAL troughs in the wavelength range from 2560 \AA\ to 2820 \AA\ ($-25000$ to 1800 \kms). In general, most objects only show a single prominent \mgii-BAL trough, but this is a conservative estimate since $\sim$ 40 objects exhibit apparent \feii\ absorption troughs, which in turn may contaminate \mgii-BAL troughs and render them indistinguishable in the spectra.  
For simplicity, we exclude \mgii-BAL troughs blended with other species such as \feii\ unless \aliii\ and/or \civ\ show BAL troughs in a similar velocity range (for details see Section \ref{EWvar_mgii_civ}). 
 \feii\ troughs are identified by checking whether UV62/63 \feii* absorption troughs are present in a similar LOS velocity range as the \mgii\ troughs if UV1 and UV2/3 \feii\ absorption troughs are clearly seen in the spectrum.

Although the actual errors in EW measurements are dominated by the continuum placement and reconstruction of the \mgii\ emission line, the propagated errors that are derived from statistical errors of spectra and Monte Carlo simulations of the continuum fit  often well reflect systematic uncertainties, particularly for quantitative analyses of BAL variability. In cases where absorption troughs are superimposed on the \mgii\ emission-line profile, the uncertainties in EW may increase significantly if the \mgii\ emission line cannot be well reconstructed. However, this work is mainly focused on relative BAL variability, which is generally not sensitive to the reconstruction of the emission line provided the continuum placements are close to the true continuum level at each individual epoch.

\subsection{BAL troughs and BAL complexes}
\label{section_bal_complex}

\begin{figure}[h]
\center{}
 \includegraphics[height=9cm,width=8.5cm,  angle=0]{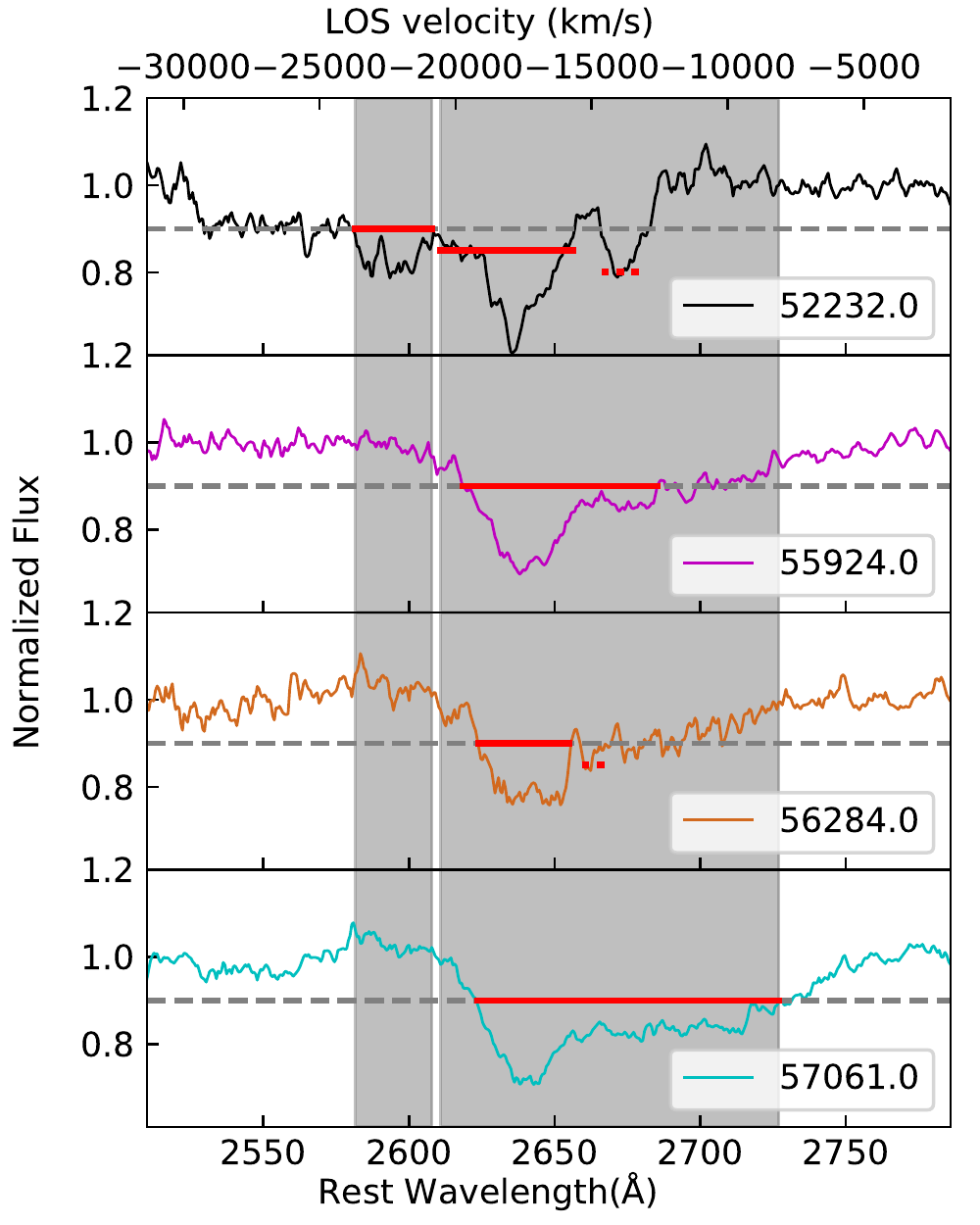} 
      \caption{A demonstration of our procedure for identifying BAL complexes (see text for details) in four spectra of J0835+4352. Red solid lines and dashed lines show  BALs and mini-BALs, respectively. The two grey regions show the final ranges for two independent BAL complexes identified by our algorithm.}
\label{bal_complex} 
\end{figure}

Previous canonical definitions of BAL troughs associated with BI were mainly suitable to find BAL troughs in a single-epoch spectrum. Our primary purpose in this work, however, is focused on BAL variability among multi-epoch spectra. Therefore, we use a slightly modified BAL-trough definition, which is derived from both spectra for each epoch pair. 
Following a similar approach to \citet{Filizak13}, we define the ``BAL complex'' to deal with the issue. 
Defining a BAL complex is often necessary, as some absorption troughs may appear in one spectroscopic epoch with isolated components, and later disappear or show quite different shapes.

As shown in Figure \ref{bal_complex}, spectral profiles and features of the \mgii~ emission/absorption lines sometimes show different profiles in different epochs.  The procedure for dealing with this issue during spectral analyses is the following: 

\begin{enumerate}

\item
After dividing by the non-BAL template fits, all absorption troughs with velocity width larger than 1250 \kms\ are sorted  by order of LOS velocities in each  spectrum, and each one is flagged and marked with an identifier and its velocity ranges (minimum and maximum velocities) for the purpose of cross comparisons among all unique pairs of epochs for the same object. 

\item
We then compare all results from every spectrum for that object. 
A BAL trough is identified as an overlapped pair if its velocity range partially or totally overlaps the range of another BAL from different epochs for the same object. 

\item
Finally, we consider all pairs of overlapping troughs as a single ``BAL complex'', with a velocity range that spans all of the overlapping troughs. 

\end{enumerate}

We demonstrate this procedure for identifying BAL complexes in the QSO  J0835+4352 (Figure \ref{bal_complex}).  
In the top panel in the first spectrum, there are two BALs (thick red line) and one mini-BAL (dotted line) initially meeting our definition for the two types. Then, we gathered all unique troughs by examining regions where the velocity ranges of the troughs overlap or lack thereof in subsequent epochs, which yields two BAL complexes as shown in the two shaded regions in Figure \ref{bal_complex}.

\section{Quantifying BAL Variability}
\label{Quantifying}
Adopting a single indicator of variability can yield questionable results due to the difficulty of quantifying the systematic uncertainties inherent in the spectra; some visible variations can be caused by photon noise or systematic uncertainties. 
In addition, using a single metric to quantify variability may face  challenges due to the large variety of line profiles and shapes (see Figure~\ref{mgiiBAL_diversity})  and the sometimes complicated variability patterns of BALs from epoch to epoch, as demonstrated by Figure \ref{bal_complex}. 
To alleviate these issues, we used three different metrics to quantitatively measure BAL variability. Our three chosen metrics ($\sigma_{\rm \Delta EW}$, $N_\sigma$, and $\chi_{1-2}^2$) are complementary to one another and when combined can gauge BAL variability more reliably.

\begin{figure*}
\center{}
 \includegraphics[width=18.6cm, height=6cm, angle=0]{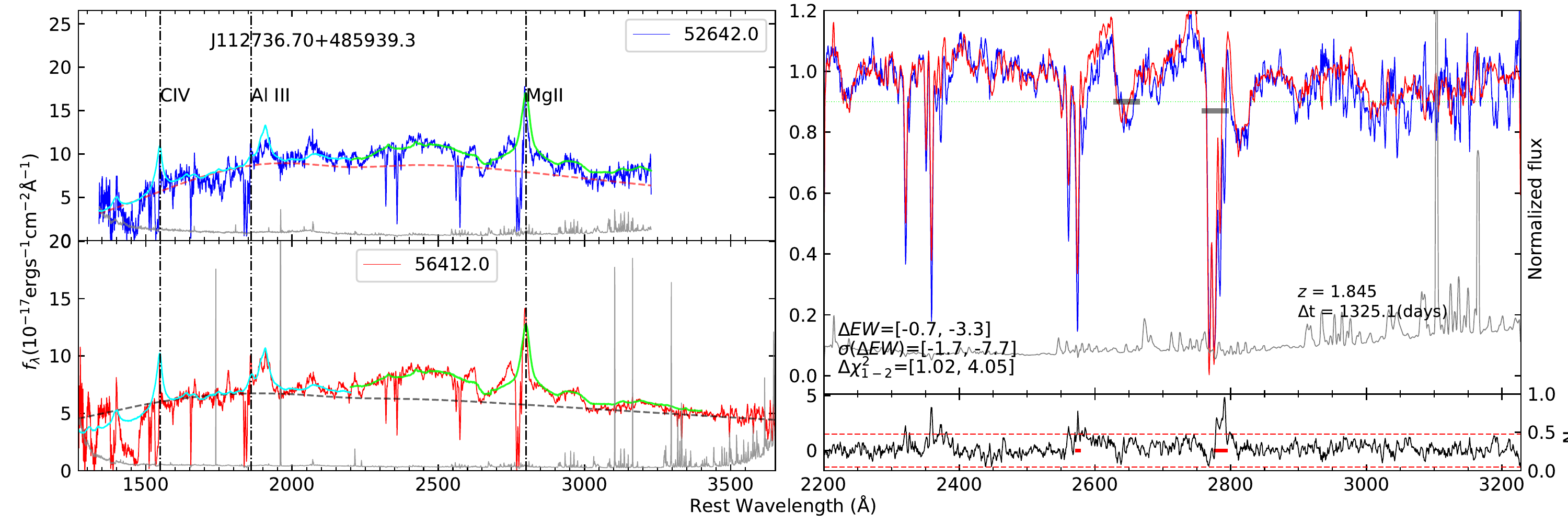} 
      \caption{Demonstration of pair comparisons between two epoch spectra for a given quasar. Left panels display different spectroscopic  epochs (blue and red lines), errors (grey lines) and non-BAL template fits (green/cyan lines stand for fits around \mgii~ and \civ) for the same object. The top right panel shows the normalized spectra by non-BAL template fits, in which grey horizontal lines indicate all BAL troughs found in the pair. The bottom values embraced by square brackets are $\Delta$EW, $\sigma_{\rm \Delta EW}$, and $\chi^2_{1-2}$ for each trough pair ordered from short to long wavelengths. The bottom panel presents the distribution of $N_{\sigma}$ for each pixel, in which variable regions are marked by thick red lines between 2560 and 2820 \AA.  }
\label{vr_demo}
\end{figure*}

\subsection{BAL variability metric  - I: $\sigma_{\rm{\Delta EW}}$  }
\label{sigma_delta_ew}

For each spectroscopic pair with times MJD1 and MJD2, the related quantities utilized in this work are defined as follows:

\begin{eqnarray}
\Delta\rm{EW} = \rm{EW}_2 - \rm{EW}_1 ; ~~  Err_{\rm{\Delta EW}} = \sqrt{ \sigma_{\rm EW_1}^2+\sigma_{\rm EW_2}^2  }  
\label{eqv1}
\end{eqnarray}
where EW$_1$ and EW$_2$ are the EWs measured from an epoch pair at epochs MJD1 and MJD2. $\rm \sigma_{EW_1}$ and $\rm \sigma_{EW_2}$ are the corresponding errors, which are propagated from the uncertainty in the continuum fit and the spectral statistical error.  $\rm{Err}_{\rm{\Delta EW}}$  is the final  uncertainty of $\rm{\Delta EW}$. 

For an individual BAL QSO, the fractional EW measurements of all BALs and mini-BALs in each epoch pair are used to measure the overall BAL variability. The fractional EW variation and its uncertainty are determined as follows: 
\begin{eqnarray}
\frac {\Delta\rm{EW}}  {\langle \rm{EW} \rangle} = {\frac {(\rm{EW}_2 - 
\rm{EW}_1)} {(\rm{EW}_2 + \rm{EW}_1) \times 0.5}},
\nonumber \\ \nonumber \\
\rm{Err}_{\frac {\Delta\rm{EW}}  {\langle \rm{EW} \rangle}} = \frac 
{4 \times (\rm{EW}_2 \sigma_{\rm{EW}_1} + \rm{EW}_1 
\sigma_{\rm{EW}_2} )} {(\rm{EW}_2 + \rm{EW}_1)^2}
\label{eqv3}
\end{eqnarray}
where $\rm{Err}_{\frac {\Delta\rm{EW}}  {\langle \rm{EW} \rangle}}$ is the final uncertainty of $\frac {\Delta\rm{EW}}  {\langle \rm{EW} \rangle}$. 
Both variations in the EW and fractional EW, as introduced above, can be chosen to quantify BAL variability regarding relative changes in the amplitude and proportion, respectively. Based on these quantities, we define the significance of amplitude/fractional EW variations   as follows: 
\begin{eqnarray}
\sigma_{\Delta\rm{EW}} = \Delta\rm{EW}/Err_{\rm{\Delta EW}}, 
\nonumber \\ \nonumber \\
\sigma_{\frac {\Delta\rm{EW}}  {\langle \rm{EW} \rangle}} = \frac {\Delta\rm{EW}}  {\langle \rm{EW} \rangle}/\rm{Err}_{\frac {\Delta\rm{EW}}  {\langle \rm{EW} \rangle}} 
\label{eqnarray_sigma}
\end{eqnarray}

\subsection{BAL variability metric - II: $N_\sigma$}
\label{variable_region_define}

In some cases, variability is present across the entire BAL (or mini-BAL) trough, while in other cases only a portion of the trough varies. 
EW may not necessarily be a good metric of variability in cases where only a small portion of the trough varies, or both increasing and decreasing portions appear in a BAL trough.  As an alternate approach, bona fide variable troughs can be defined as a trough containing at least one significantly variable region.  To identify these variable regions, we measure deviations between two spectra for each pixel in units of sigma, which is defined by the following equation:

\begin{equation}
N_{\sigma}(\lambda) =  {\frac { F_2 - F_1 } 
{\sqrt {\sigma_2^2  + \sigma_1^2 }}}
\label{eqv2}
\end{equation}
$F_1$  and $F_2$ are the normalized flux densities.  $\sigma_1$ and  $\sigma_2$ are corresponding (normalized) flux-density uncertainties at wavelength $\lambda$. Both $\sigma_1$ and $\sigma_2$ include spectral errors and quantitative uncertainties from the  fitting procedures mentioned above. 
Initially, the regions that were significantly variable were defined as those where $|N_\sigma| \geq 1$ in a consistent direction for at least five consecutive pixels (e.g., \citealp{Gibson08, Filizak13}),  which allows the detection of variable regions wider than $\approx275~\rm{km\,s^{-1}}$ in our spectra.  This requirement was imposed to minimize the possibility of random fluctuations being categorized as variable regions. 

Due to the diversity of \mgii-BAL troughs and contamination by \feii\ emission/absorption, 
we use a more conservative threshold ($|N_{\sigma}|>1.5$) to define our variable regions compared to that ($|N_{\sigma}|>1$) from \citet{Filizak13}. Adopting this more conservative threshold is particularly helpful for quantitative analyses in the blue/red ends of spectra due to flux-calibration uncertainties based on our visual inspection (see Figure \ref{vr_demo} for an example).

\subsection{ BAL variability  metric - III: $\chi_{1-2}^2$}
\label{chi2_trough}

In cases where an absorption trough contains some regions of increasing EW and other regions of decreasing EW, using the variance  (mean square deviation) to quantify variability can  better reflect true variability. 
We thus calculate the reduced $\chi^2$ for each distinct trough pair, which combines features both from the overall EW and individual variable regions in a quantitative way. This quantity is defined as: 
\begin{equation}
\chi_{1-2}^2 =  \sum \frac{N_{\sigma}(\lambda)^2}{N}
\label{eq11}
\end{equation}
where $N$ is the number of pixels spanning the BAL trough. 
The greater the value of $\chi_{1-2}^2$, the more likely that the observed variations are not produced by systematic noise.

\subsection{Classification of BAL variability}
\label{variability_classification}

\begin{figure}[h]
\center{}
 \includegraphics[height=6cm,width=8cm, angle=0]{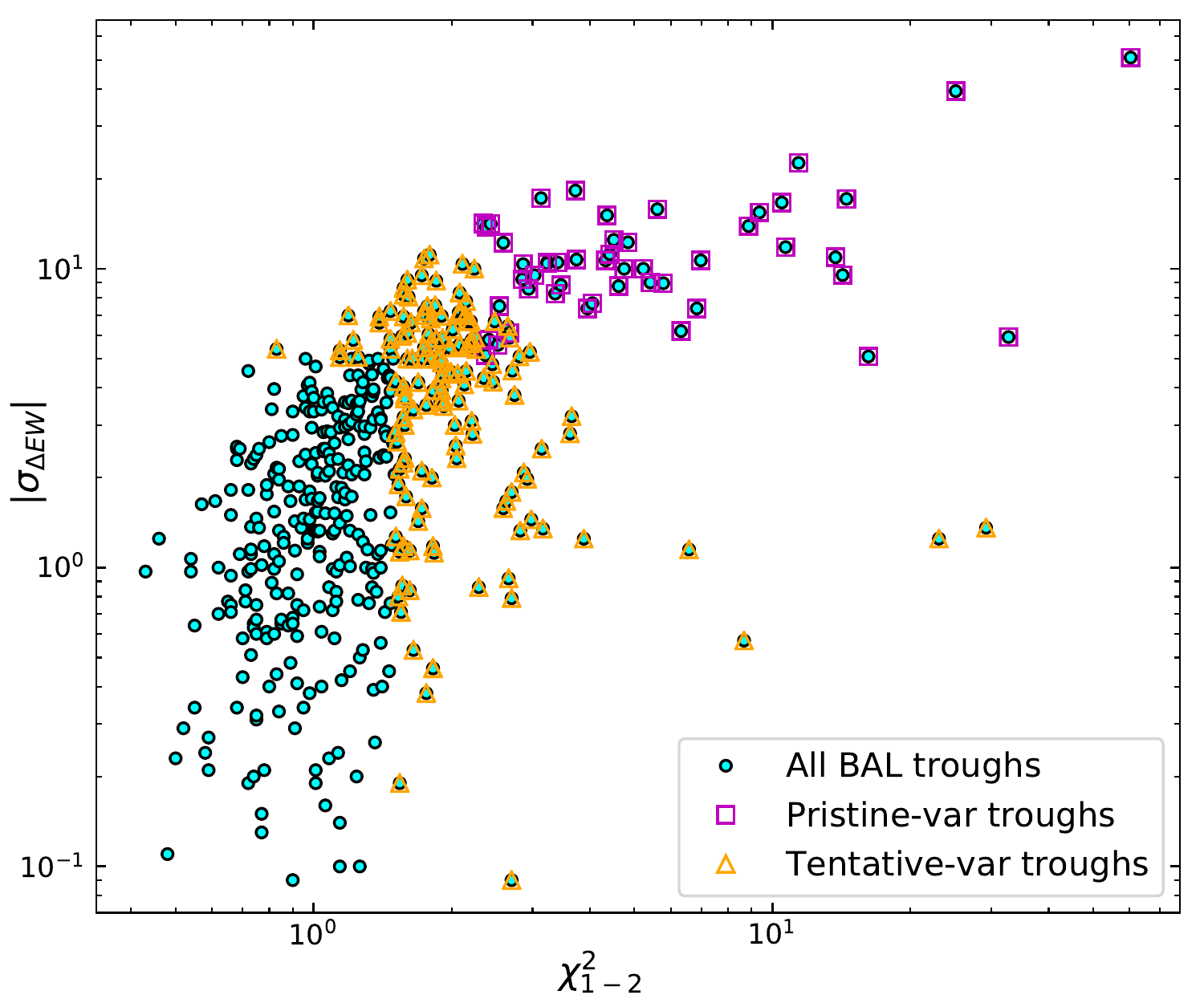}
      \caption{Distribution of the absolute $\sigma_{\rm\Delta EW}$ versus $\chi_{1-2}^2$ from all BAL troughs (cyan), in which pristine (magenta squares) and tentative (orange triangles) variable troughs are indicated.} 
    \label{sigEW_vs_chi2}
\end{figure}

In addition to utilizing the three above-mentioned variability metrics, we conduct a visual inspection of  
at least one spectroscopic pair for each object in our sample, which guarantees that our results are not affected by issues such as an improper fit. 
These three metrics all have different strengths, and thus different weights are assigned into their reliability, depending on the specific properties of each object. 
For example, if a BAL trough  has low S/N, the variable region  may not be suitable for quantitative analyses and instead the significance of EW variations in the trough may better characterize BAL variability; if a wide trough varies only in a small portion, then $N_{\sigma}$ is more suitable for  BAL variability analyses.

We investigate the relation between the  $\sigma_{\rm\Delta EW}$ and  $\chi_{1-2}^2$ metrics sampled from all unique \mgii-BAL troughs. 
The  $\sigma_{\rm\Delta EW}$ and $\chi_{1-2}^2$ metrics are highly correlated (see Figure \ref{sigEW_vs_chi2}; a Pearson test yields $r=0.76$). However, some apparent scatter can be seen in the distribution, which is likely  linked to large deviations in extreme cases. It is therefore necessary to combine these metrics for a more reliable analysis of BAL variability.

To identify cases of true, significant variability, we thus adopt the following criteria: 
\begin{enumerate}
\item
To be considered a ``pristine'' variable trough, a spectral pair must 
(1) have $\sigma_{\rm\Delta EW}>5$, (2) have at least one variable region in the trough, and (3) have a $\chi_{1-2}^2$ value larger than 2.3. 

\item
We define a trough to have ``tentative" variability if it has  at least one variable region, or $\chi_{1-2}^2>1.5$, or $\sigma_{\rm\Delta EW}>5$.

\item
Non-variable troughs are defined as those not meeting either of the two conditions above. 

\end{enumerate}

Two types of BAL variability are often seen in our sample. The first is that both increasing and decreasing portions can be present in the same BAL trough, leading to small $\sigma_{\rm\Delta EW}$ but large $\chi_{1-2}^2$. Another  is that  only a small portion of a wide BAL trough varies (either increasing or decreasing), causing large  $N_{\sigma}$  but small $\chi_{1-2}^2$.  
The three different quantitative metrics ($\sigma_{\rm\Delta EW}$, $N_{\sigma}$, and $\chi_{1-2}^2$) are therefore meant to complement one another in detecting different types of BAL variability in a sample with a variety of absorption-line profiles from object to object and significant profile variability from epoch to epoch accompanied by varying S/N ratios.

\section{Statistical view of the sample}
\label{stat_view}

In this section, we present the distributions of our measurements and our corresponding analyses.

\begin{deluxetable*}{cccccccccccccccc}
\tabletypesize{\tiny} 
\tablecaption{\mgii-BAL trough identifications and quantitative measurements}
\tablewidth{0pt}
\tablehead{
Name &RA & Dec &  MJD1 & MJD2 & $\chi_{1-2}^2$ & VR\tablenotemark{a}  & $v_{max}$\tablenotemark{b} & $v_{min}$\tablenotemark{c}
&  d1\tablenotemark{d} & d2\tablenotemark{d} &  EW1 &  EW2 	\\
(SDSS) & (J2000) & (J2000) & & & & & (\kms) & (\kms) & & & (\AA) & (\AA)
} 
\startdata
 & & & 51816 & 53726 & 1.9 & 0 & $-$11334 & $-$8665 & 0.22 & 0.15 & 5.4 $\pm$ 0.16 & 3.8 $\pm$ 0.16 \\
 & & & 51816 & 55214 & 3.23 & 1 & $-$11334 & $-$8665 & 0.22 & 0.14 & 5.4 $\pm$ 0.16 & 3.4 $\pm$ 0.11 \\
 & & & 51816 & 55475 & 2.09 & 0 & $-$11334 & $-$8665 & 0.22 & 0.16 & 5.4 $\pm$ 0.16 & 4.1 $\pm$ 0.09 \\
 & & & 51816 & 57279 & 4.51 & 2 & $-$11334 & $-$8665 & 0.22 & 0.12 & 5.4 $\pm$ 0.16 & 2.9 $\pm$ 0.12 \\
J0103+0037 & 15.96864 & 0.62772 & 53726 & 55214 & 0.68 & 0 & $-$11001 & $-$9134 & 0.18 & 0.16 & 3.2 $\pm$ 0.13 & 2.8 $\pm$ 0.09 \\
 & & & 53726 & 55475 & 0.79 & 0 & $-$11068 & $-$8732 & 0.16 & 0.17 & 3.6 $\pm$ 0.15 & 3.9 $\pm$ 0.08 \\
 & & & 53726 & 57279 & 0.95 & 0 & $-$11001 & $-$9134 & 0.18 & 0.15 & 3.2 $\pm$ 0.13 & 2.6 $\pm$ 0.1 \\
 & & & 55214 & 55475 & 1.6 & 0 & $-$11068 & $-$8732 & 0.14 & 0.17 & 3.2 $\pm$ 0.11 & 3.9 $\pm$ 0.08 \\
 & & & 55214 & 57279 & 0.65 & 0 & $-$10935 & $-$9267 & 0.17 & 0.16 & 2.6 $\pm$ 0.09 & 2.5 $\pm$ 0.09 \\
 & & & 55475 & 57279 & 2.94 & 1 & $-$11068 & $-$8732 & 0.17 & 0.12 & 3.9 $\pm$ 0.08 & 2.7 $\pm$ 0.11 \\
\hline
 \bf{.} & \bf{.} &\bf{.} & \bf{.} & \bf{.} & \bf{.} & \bf{.} & \bf{.} & \bf{.} & \bf{.} & \bf{.} & \bf{.}  \\
 \bf{.} & \bf{.} &\bf{.} & \bf{.} & \bf{.} & \bf{.} & \bf{.} & \bf{.} & \bf{.} & \bf{.} & \bf{.} & \bf{.}   \\
 \bf{.} & \bf{.} &\bf{.} & \bf{.} & \bf{.} & \bf{.} & \bf{.} & \bf{.} & \bf{.} & \bf{.} & \bf{.} & \bf{.}   
\enddata
\tablecomments{This table summarizes our identifications and measurements of all unique \mgii-BAL troughs in the sample.  a: the number of variable regions (VR) in the trough pair. b/c: the max/min velocity (\kms) limits of the trough. d: BAL depths at MJD1 and MJD2.  The full table is available in the online version. } 

\label{result_demo_table}
\end{deluxetable*}

\subsection{Distribution of EW variation versus timescale}
\label{overall_var}

\begin{figure*}
\center{}
 \includegraphics[height=9cm, width=18cm,  angle=0]{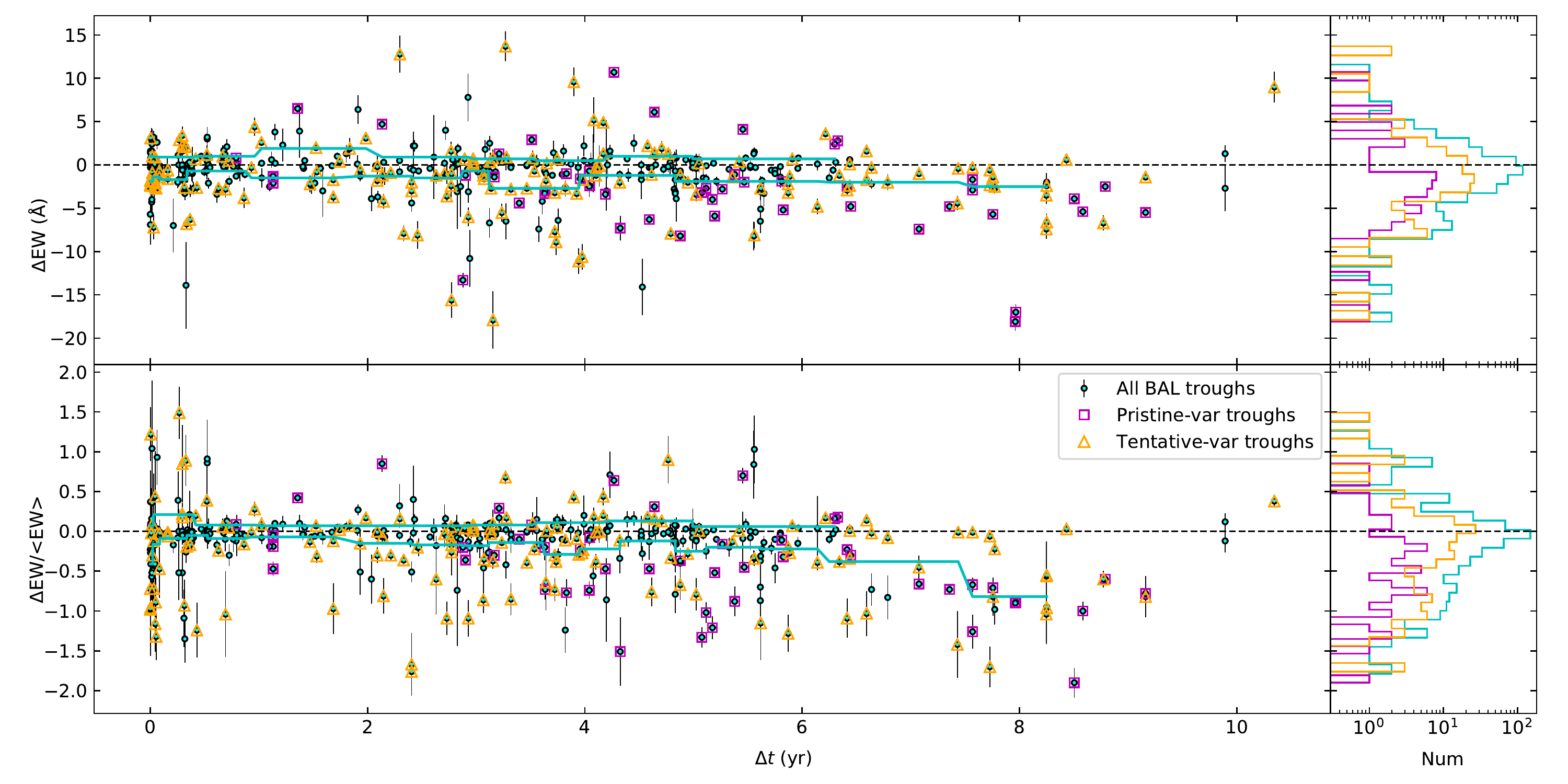}
      \caption{Distributions of $\Delta$EW and $\Delta \rm{EW}/ \langle \rm{EW} \rangle$ versus the rest-frame timescale ($\Delta t$), in which pristine (magenta squares) and tentative (orange triangles) variable troughs are over-plotted above all BAL troughs (cyan) between successive epochs. Solid cyan lines in the left panels represent the median $\Delta \rm {EW}$ within a sliding window of 15 time-ordered weakening ($\Delta$EW $<0$) or strengthening ($\Delta$EW $>0$) troughs. In the right panels, we show corresponding histograms of the distributions for the three groups.  An apparent feature of the distributions is that the number of weakening troughs is  larger than that of strengthening troughs.} 
      \label{ewave_vs_dew_tspan}
\end{figure*}

Measurements of all \mgii-BAL troughs are provided in Table \ref{result_demo_table} and related figures in this section.  
The quasar J1411+5328 has two unique \mgii-BAL troughs, and each of them has 1081 pair-wise combinations of spectra based on the available 47 epochs; this object thus contributes $\approx$ 76\% of the data points in the \mgii-trough distribution of the sample. 
Because we do not want a single object to dominate the distribution, particularly for small EW variations on short timescales, for this object we randomly choose three different-epoch spectra (corresponding to six combinations in searching for BAL troughs) for the statistical analysis.

The distributions of BAL variability versus timescale are displayed in Figure \ref{ewave_vs_dew_tspan}, in which the three groups (all, pristine-variable, and tentative-variable \mgii\ troughs) appear to show preferentially negative offsets in the distribution of EW variations, which indicate that strengthening troughs ($\Delta \rm {EW}>0$ \AA) are less common than weakening troughs ($\Delta \rm {EW}<0$ \AA). 
We adopted a non-parametric triples test \citep{Randles80} to examine the distribution of $\Delta \rm {EW}$ for all BAL troughs, and this asymmetry was found to be significant at $>$99\%. 
Note that the non-parametric triples test does not assume the distribution is centered about zero, so as a check, we applied a two-sample K-S test for the distributions of all weakening (based on absolute values) and strengthening BAL troughs, which also indicates a significantly asymmetric distribution around zero ($p=0.0002$).

To display the overall trend in $\Delta \rm {EW}$ versus timescale, we calculate the median of $\Delta \rm {EW}$ using a sliding window of 15 time-ordered data points for weakening and strengthening troughs, separately (see solid cyan lines in Figure \ref{ewave_vs_dew_tspan}). In general, both absolute and fractional EW variations become larger with the increase of timescale for weakening troughs; in addition, almost all troughs are from the pristine/tentative subsets at $\Delta t>6$~yr, indicating that \mgii-BAL variability occurs with greater frequency on long timescales. For the strengthening troughs, there is no clear tendency regarding the distribution of $\Delta \rm {EW}$ versus timescale.

Such a phenomenon of unequal weakening and strengthening rates of pre-existing BALs has not been reported previously. \citet{Rogerson18} note that when newly emerged BAL candidates are observed 6--12 months later, their $\Delta$EW values show a large range but correspond to an overall average decline in EW (the blue point in their Figure 20), but unlike us they studied a biased subset of the BAL population.

\subsection{BAL variability versus EW} \label{ew_variability}

\begin{figure}[h]
\center{}
 \includegraphics[height=6cm,width=8.7cm,  angle=0]{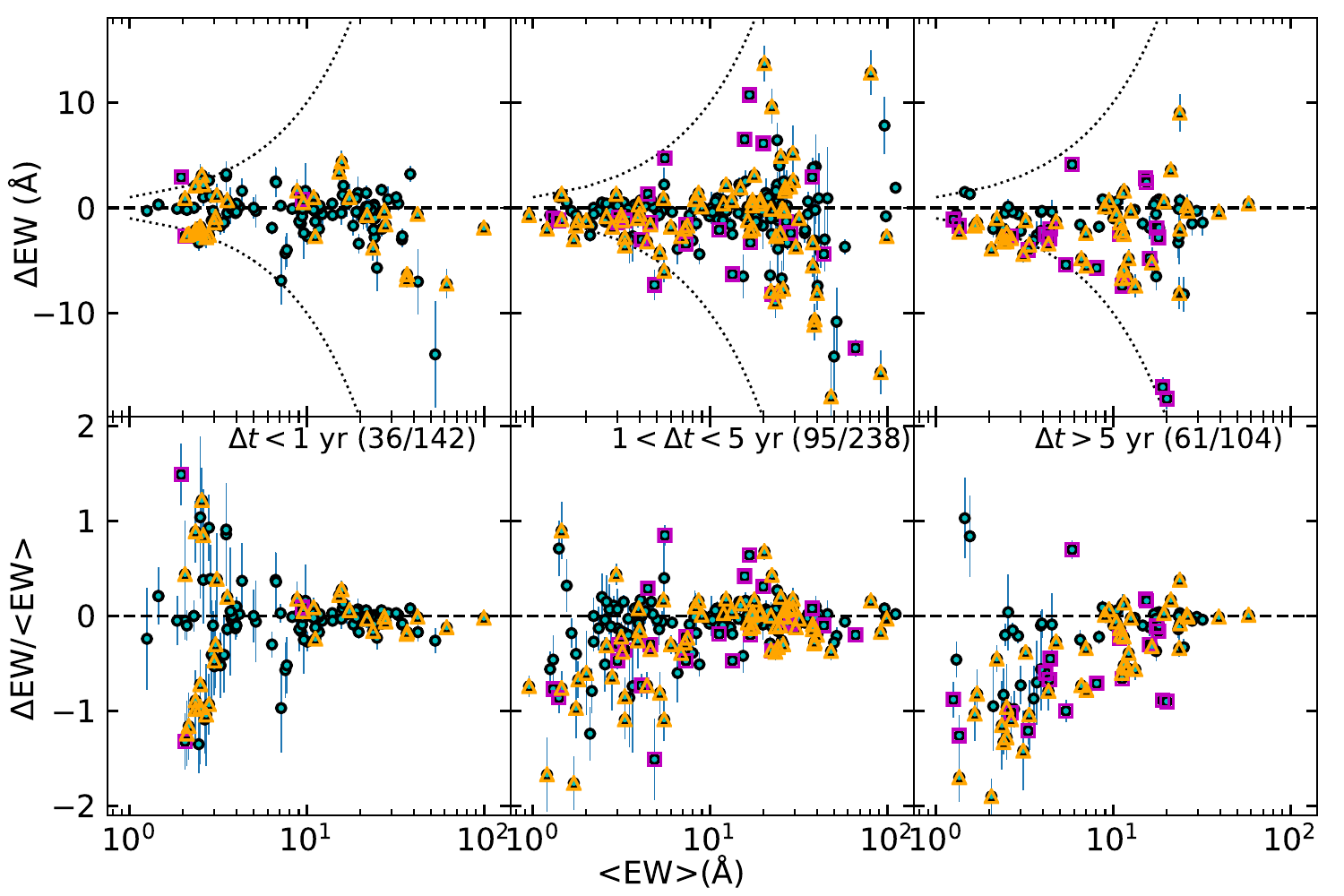} 
      \caption{Distributions of EW variability versus $\langle \rm{EW} \rangle$ sampled for \mgii\ BALs over short, intermediate, and long timescales, in which pristine (magenta squares) and tentative (orange triangles) variable troughs are over-plotted. Dotted curves denote the boundary where $|\Delta \rm{EW}|$ is equal to $\langle \rm{EW} \rangle$. The numbers of pristine+tentative and all troughs in each panel are given in parentheses. There is a clear trend that the longer the timescale sampled, the more asymmetric the distribution of EW variations appears.}  
      \label{dew_sml_tspan}
\end{figure}
Previous studies have found that HiBAL variability tends to become larger in samplitude with the increase of timescale both for absolute and fractional EW variations (e.g., \citealp{Gibson08,Filizak13}).  
We here examine the distributions of EW variability versus mean EW while first separating the LoBAL sample into short, intermediate, and long timescales.

The data in Figure \ref{dew_sml_tspan} show that the maximum absolute and fractional EW variations of \mgii-BAL troughs at the 90\% quantile level at $\Delta t <1$ yr ($-$6 \AA\ and $-$1; see the left two panels in Figure \ref{dew_sml_tspan}) are obviously larger than those from \citet{Filizak13}, where the maximum values of absolute and fractional EW variations of \civ-BAL troughs are approximately equal to 2 \AA\ and 0.3 at the 90\% quantile level,  respectively (see Figures 15 and 16 in their work).  
The fractional EW variations of weak \mgii-BAL troughs ($\langle \rm{EW} \rangle <10$~\AA) are dramatically larger than those from HiBALs on short timescales, although these weak \mgii-BAL troughs have large fractional uncertainties. 
On the other hand, the level of asymmetry for the EW variability shows an increasing trend as the sampling timescale increases (see the spread of EW variation versus timescale in Figure~\ref{ewave_vs_dew_tspan}). We also note that a similar trend for the HiBAL sample of \citet{Filizak13} for both \civ\ and \siiv\ BALs may start to appear at $\Delta t>2.5$ yr (see Figure 16 in their work), although they do not comment on it. 
However, the large differences in the timescale coverage, the average number of spectroscopic epochs, and the ionization potentials for the two samples in this work and \citet{Filizak13}, together make comparisons difficult.

Quantitatively, we analyze the correlations of $|\Delta \rm{EW}|$ versus $\langle \rm{EW} \rangle$ and $|\Delta \rm{EW}|/\langle \rm{EW} \rangle$ versus $\langle \rm{EW} \rangle$, based on the Spearman rank-correlation test. We find strong correlations between $|\Delta \rm{EW}|/\langle \rm{EW} \rangle$ and $\langle \rm{EW} \rangle$ ( $>$99.9\%) over all the three timescales. 
When considering the distribution of  $|\Delta \rm{EW}|$ versus $\langle \rm{EW} \rangle$, we find a strong correlation ($>$99.9\%) only on the intermediate timescale.  
No correlations of $|\Delta \rm{EW}|$ versus $\langle \rm{EW} \rangle$ were found over short and long timescales. 
The statistical results of the fractional EW variations for \mgii\ BALs are consistent with the HiBAL sample of \citet{Filizak13} and LoBAL sample of \citet{Vivek14}. They together confirm a similar tendency of larger fractional EW variations occurring at lower EWs in the whole BAL population.

As a comparison,  significant correlations, both in the absolute and fractional EW variations, were found for HiBAL variability by \citet{Filizak13}; however, they also reported that the significance from \siiv\ BALs ($>95$\%) is lower than that from \civ\ BALs  ($>99$\%). The weakening trend of correlations between EW variations and mean EWs from high- to low-ionization BALs suggests an underlying link between BAL variability and the ionization potential, which may contribute to differences between HiBAL and LoBAL variability.  Investigations of the relation between BAL variability and ionization potentials will be presented in Section \ref{EWvar_mgii_civ} .

\subsection{BAL variability versus BAL-profile properties} \label{v_width_depth}

\begin{figure}[h]
\center{}
 \includegraphics[height=10cm,width=8.7cm,  angle=0]{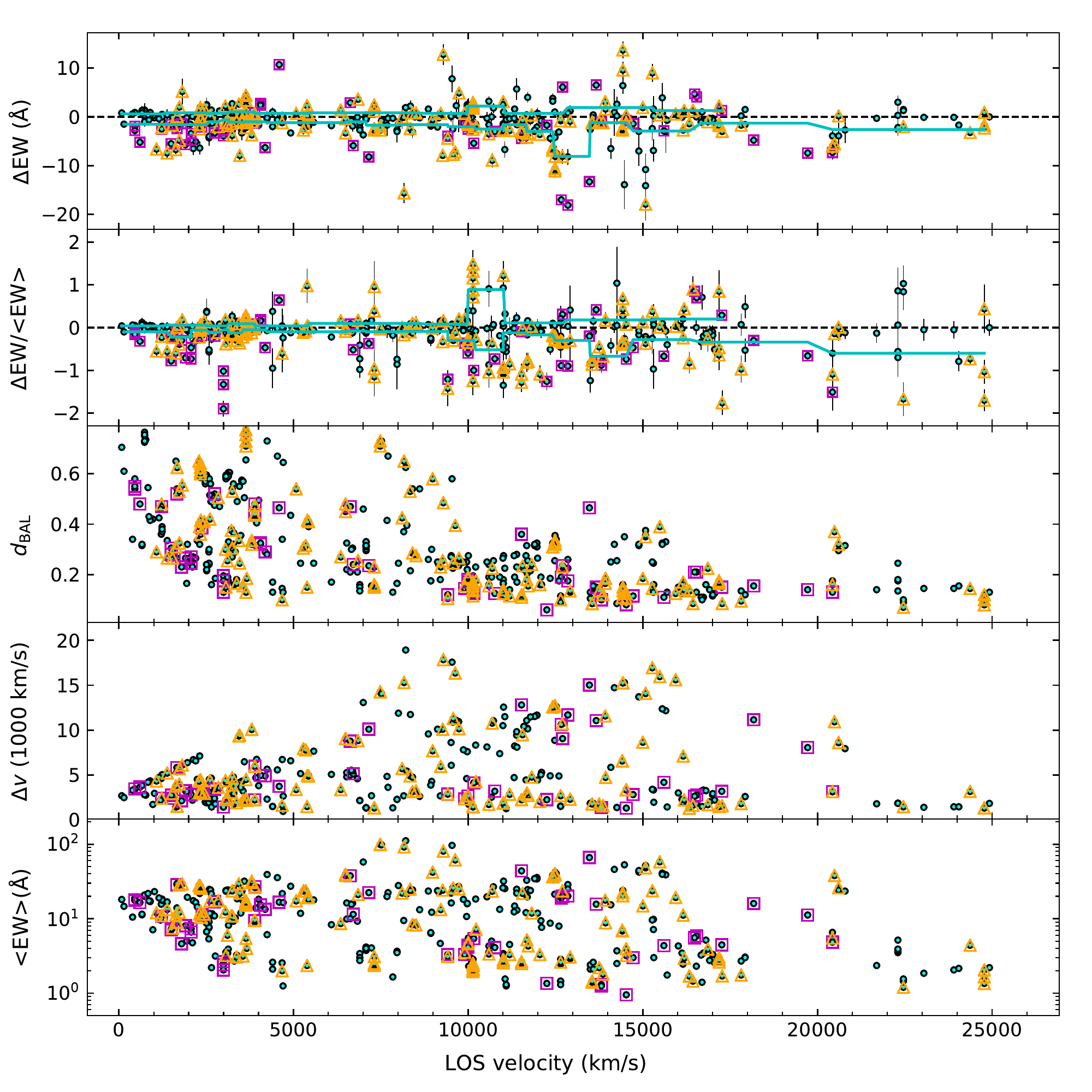} %
 \caption{Distributions of the mean velocity versus $\Delta$EW, $\Delta$EW/$\langle$EW$\rangle$, mean BAL depth, velocity width (1000 \kms), and mean EW based on all  \mgii-BAL troughs. Pristine (magenta) and tentative (orange) variable troughs are over-plotted. Cyan solid lines represent the median values of absolute and fractional EW variations within a sliding window of 15 velocity-ordered data points (all with $\Delta$EW $<0$ or $\Delta$EW $>0$). There are significant correlations of LOS velocity versus BAL depth and LOS velocity versus fractional EW variation.} 
      \label{fdew_ew_los_velocity}
\end{figure}
BAL-profile properties, such as trough width, trough depth, and LOS velocity provide useful diagnostics for the analysis of BAL variability. We investigate the relations of these properties in this subsection.

Consistent with  \citet{Vivek14}, the fractional EW variations are generally larger at higher velocities of  $v > 10000$ \kms\ than those at lower velocities of $v < 10000$ \kms\ (see Figure \ref{fdew_ew_los_velocity}), although our sample size is $\sim$ 8 times larger than their sample. 
The apparent deficiency of data points between $\sim$ 5$\times 10^3$ \kms\ and $\sim$ 7$\times 10^3$ \kms\ may be caused by the exclusion of troughs blended with \feii\ troughs among FeLoBAL QSOs (see Section \ref{mgii_identification}).  
We also found that there is a strong anti-correlation between LOS velocities and trough depths (see the third panel of Figure \ref{fdew_ew_los_velocity}). 
High-velocity ($>21000$ \kms) \mgii-BAL troughs tend to have both shallow depths and narrow trough widths, mainly associated with weak BAL troughs ($\langle \rm EW \rangle<5$ \AA). 
By contrast, low-velocity ($<4000$ \kms) \mgii\ BALs tend to have higher depths and  wider trough widths, as well as stronger BALs than those troughs at $>21000$ \kms. Similar observational results were reported by \citet{Filizak14} for their sample of HiBAL QSOs.

Quantitatively, we found a significant correlation between LOS velocity and $|\Delta \rm{EW}|/\langle \rm{EW} \rangle$ using the Spearman test ($r_s=0.39$, $p_s<10^{-10}$); there is also a weak correlation between LOS velocity and $|\Delta \rm{EW}|$ ($r_s=0.22$, $p_s=10^{-6}$).  
These results are different from \citet{Filizak13}, where no significant correlations were found in both distributions for \civ, supporting our speculation of potential differences between HiBAL and LoBAL variability. 
Larger fractional EW variability occurring at higher velocities can be explained by the fact that higher-velocity troughs are weaker, and trough weakness is the main driver behind the observed increasing fractional EW variations at high velocities.  
In addition, we found that the distributions of depth, width, LOS velocity, and $\langle \rm{EW} \rangle$ sampled by all \mgii\ troughs and pristine+tentative variable troughs are consistent with each other ($p=0.19,p=0.97,p=0.56,p=0.74$, respectively) based on the two-sample K-S tests.

\begin{figure}[h]
\center{}
 \includegraphics[height=6cm,width=8.7cm,  angle=0]{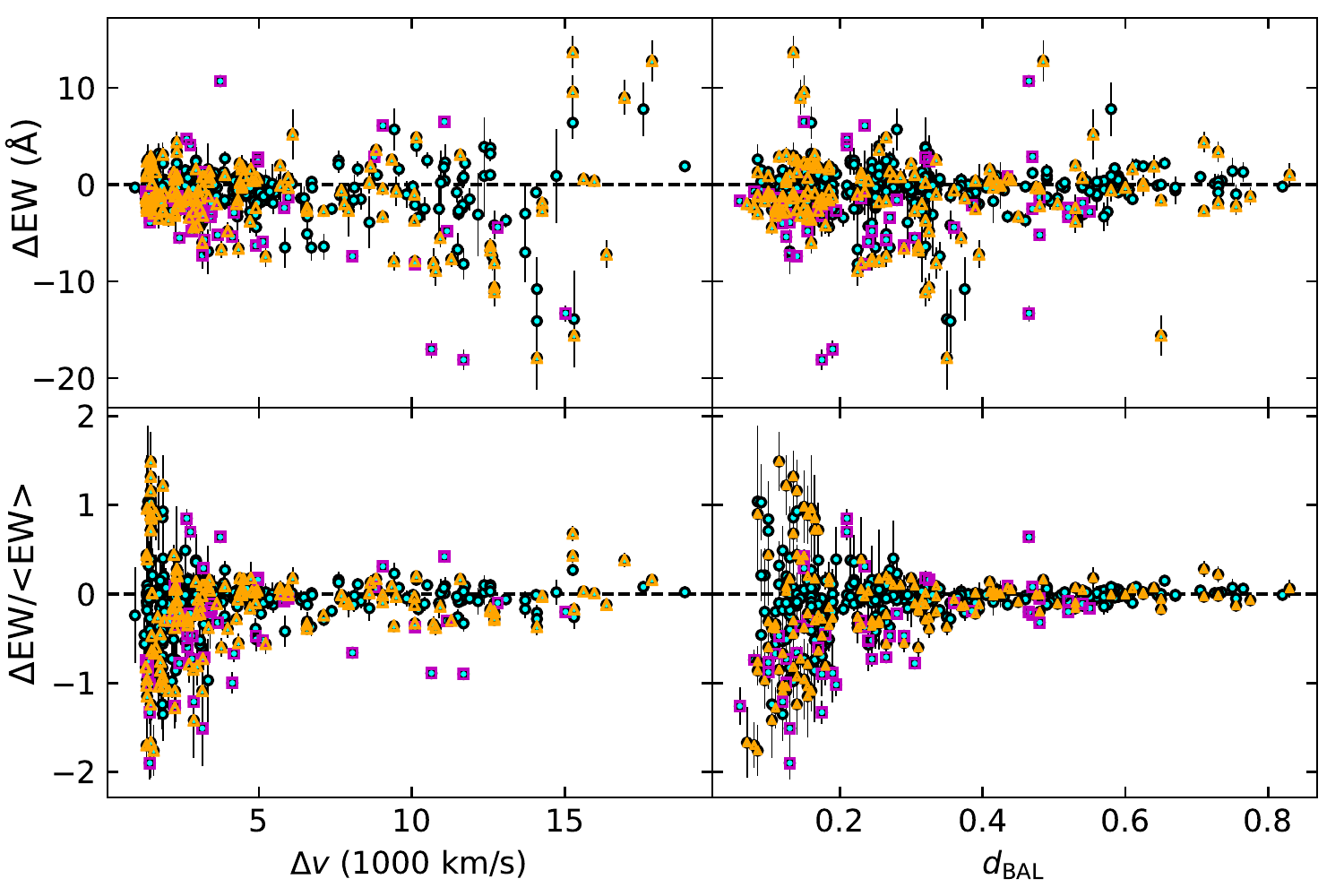} 
      \caption{Distributions of EW variability versus trough width ($\Delta v$) and depth ($d_{\rm BAL}$) sampled by all \mgii\ troughs (cyan), in which pristine (magenta) and tentative (orange) variable troughs are over-plotted. The distributions of $\Delta$EW/$\langle$EW$\rangle$ versus $\Delta v$ and $d_{\rm BAL}$ show a similar decreasing trend.}  
      \label{f_dew_width_depth}
\end{figure}
In addition, the distributions of EW variability versus width and depth are presented in Figure \ref{f_dew_width_depth}, in which we can see three  apparent features. First, larger absolute EW variations tend to occur in wider ($\Delta v>10000$ \kms) troughs. Second,  distributions of fractional EW variation versus width and  depth show a similar decreasing trend.  Third, there is no clear dependence for the distribution of  $\Delta$EW versus depth.  
Consistent with \citet{Filizak13}, the deepest BAL troughs (those with  $d_{\rm BAL}>0.6$) appear to show less variation than shallower ones, particularly for the distribution between fractional EW variations and depths, where a ``flat'' tendency appears at $d_{\rm BAL}>0.3$. However, we found that shallow \mgii\ BALs  ($d_{\rm BAL}<0.3$) on average decrease their EWs, opposite to the result for \civ\ BALs from \citet{Filizak13} (see Figure 17 in their work). This finding provides a critical clue regarding the difference of BAL variability between HiBAL and LoBAL QSOs, which will be systematically investigated in Section \ref{var_timescales}.

\subsection{ Variable BAL-trough and BAL-quasar fractions}
\label{variable_region_section}

To find out the highly variable portions over the whole BAL troughs in the sample, 
we analyze the variable regions (VRs) as defined in Section \ref{variable_region_define} for the sample.

\begin{figure}[h]
\center{}
 \includegraphics[height=7cm,width=8.6cm,  angle=0]{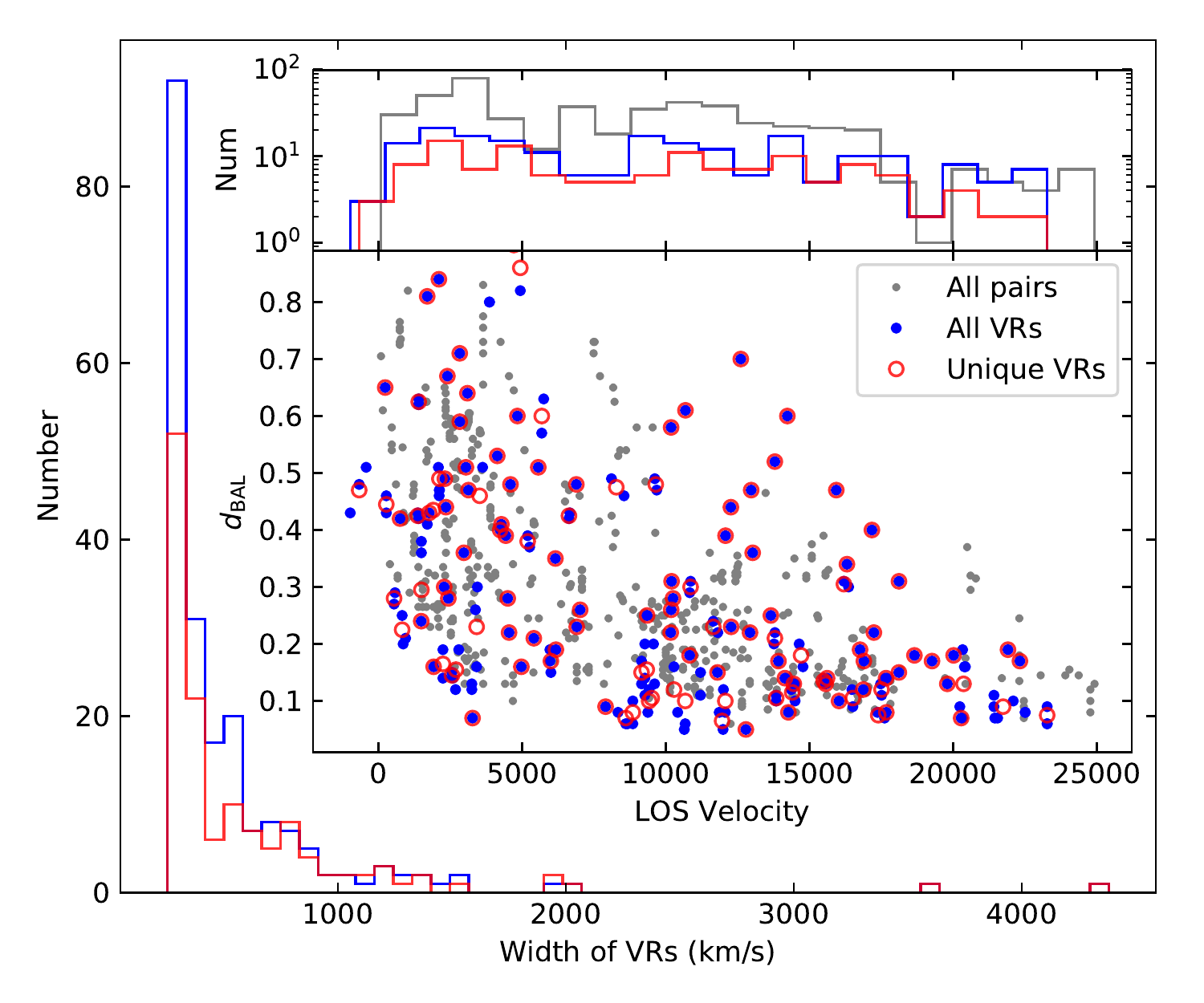} 
      \caption{The large panel shows the velocity width distributions of variable regions, in which the blue histogram represents all variable regions and the red histogram represents the results after the BAL-complex analysis. The inset panel displays the relation between mean velocities and depths from all VRs (blue dots) and unique VRs (red circles) compared with the distribution of all \mgii-BAL troughs (grey dots). The depth of \mgii-BAL troughs decreases with the increase of LOS velocities.} 
\label{width_VRs}
\end{figure}

We first search for the total number of identified VRs from all unique epoch pairs in our sample. 
To eliminate overlapping VRs from the same object over different-epoch pairs, the same analysis used to define BAL-trough complexes (see Section \ref{section_bal_complex}) is adopted to generate a set of independent VRs for each object. We found 101 independent \mgii~VRs (see red circles in the inset panel of Figure~\ref{width_VRs}) in our sample of 134 quasars over all spectroscopic epochs, and almost all of them have VR widths less than 2000 \kms~ (see the large panel of Figure \ref{width_VRs}). The peak in the distribution of VR widths is located at $\sim$ 270 to 370 \kms, which is close to the minimum threshold (5 consecutive pixels) set in searching for VRs. 

Using the VR as the criterion to search for variable troughs, we found that about 37$_{-4.5}^{+5.2}$\% \mgii-BAL troughs have at least one VR. 
As a comparison, the fractions of pristine and tentative variable \mgii-BAL troughs are about 10$_{-2.3}^{+2.9}$\% and 29$_{-3.7}^{+4.4}$\%, respectively (the 1$\sigma$ error bounds are calculated following \citealt{Gehrels1986}), according to the criteria defined in Section \ref{variability_classification}. The apparent low fraction of pristine variable troughs arises from the strict criteria for an unambiguous identification of BAL variability.  
Thus, the fraction either from VRs or pristine+tentative variable troughs in our sample is in agreement with that from \citet{Vivek14} with a fraction of $\sim$ 36\% from a sample of 22 (Fe)LoBAL QSOs.

We further examined those LoBAL QSOs with at least one variable region sampled by all epoch pairs.  As a result, the 101 independent \mgii~VRs are found in 54 quasars ($\sim$ 40\%); additionally, 28 of these quasars ($\sim$ 21\%) are classified as the pristine variable type. The above results further support the argument that the incidence of \mgii-BAL variability is significantly lower than that of \civ-BAL variability (typically $\sim$ 50\%--60\%; see \citealp{Cap12,Filizak13}), especially when considering the longer timescales covered in our sample. This is yet another piece of intrinsic differences may exist between HiBAL and LoBAL variability. We will test this hypothesis in the following sections.

\subsection{BAL variability dependence with quasar properties} \label{mag_rd_rl_Edd}

\citet{Filizak13} found that \civ-BAL variability does not show any strong dependences on quasar  bolometric luminosity, redshift, radio loudness, BH mass, or Eddington ratio. We here present the results of our investigation of these dependencies in our \mgii\ BAL sample.

\subsubsection{ Luminosity and redshift}
\begin{figure}[h]
\center{}
 \includegraphics[height=7cm,width=8.6cm,  angle=0]{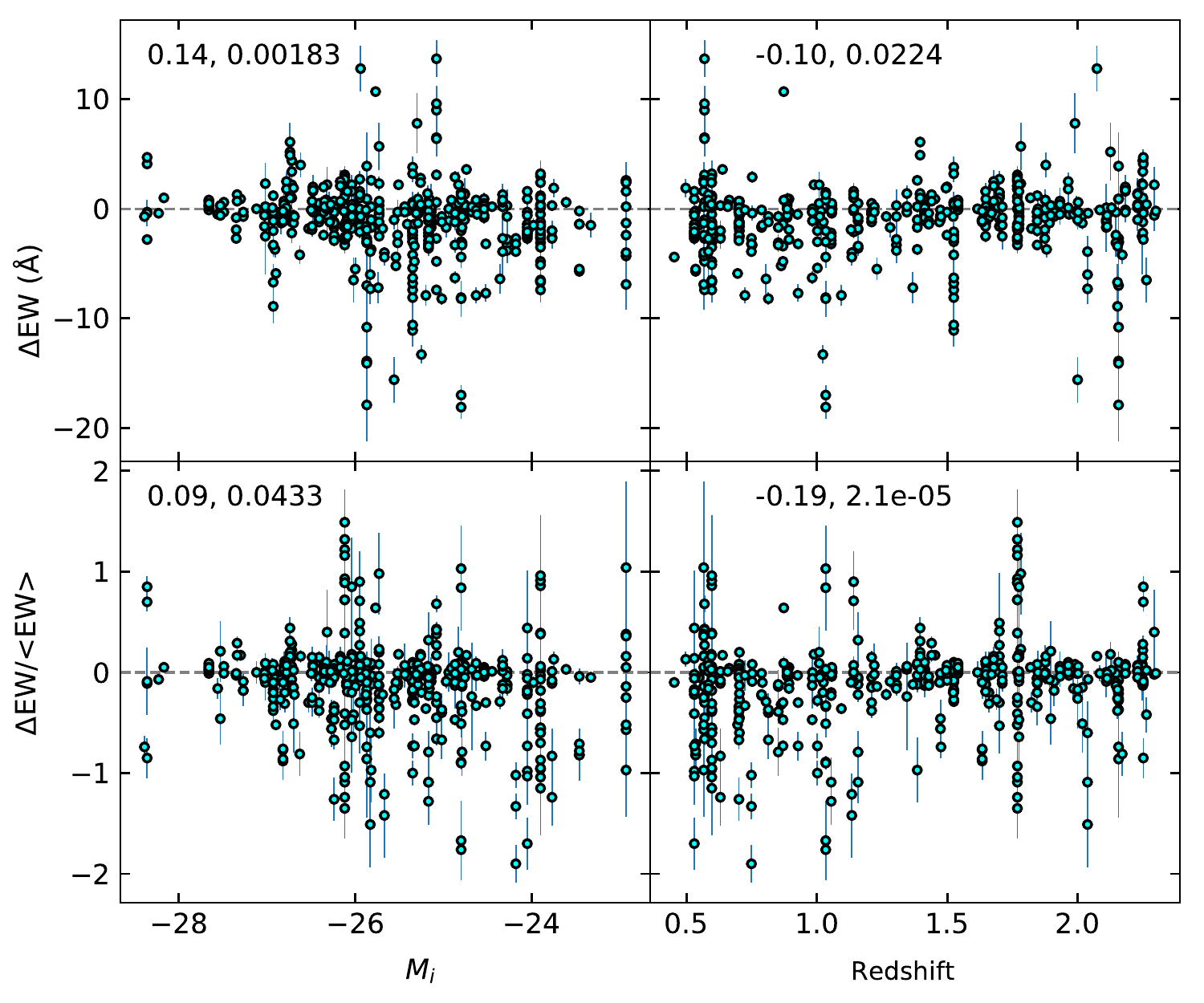} 
      \caption{Distributions of $\Delta$EW and $\Delta \rm{EW}/ \langle \rm{EW} \rangle$ versus the absolute $i$-band magnitude ($M_i$) and redshift, where no correlations are found. Spearman test results ($r_s, p_s$) using $|\Delta$EW$|$ are shown in each panel. }
\label{totvar_vs_imag}
\end{figure}

The results are displayed in Figure \ref{totvar_vs_imag}, in which we find no evidence that the systemic redshift has any correlation either with $\Delta$EW or $\Delta \rm{EW}/ \langle \rm{EW} \rangle$ in our sample, disfavoring the argument that highly variable LoBAL QSOs tend to cluster at low redshifts as proposed by \citet{Vivek14}.  
No correlation between BAL variability and redshift may imply that \mgii\ BALs  have little cosmological evolution in this redshift range ($0.46<z<2.3$). 
We also found no significant correlation between \mgii-BAL variability and luminosity, consistent with previous studies using HiBAL QSOs.

\subsubsection{BAL variability of radio-loud LoBAL QSOs}
Radio-loud BALQSOs are currently not well understood, as they are rare and it was once thought that strong radio emission prohibited the presence of BAL troughs (e.g., \citealp{Stocke92,Becker00}). In this section,  we examine BAL variability in the radio-loud subset of our sample.

Our sample consists of $\sim$ 23\% radio-loud quasars with radio loudness ($R^*$) larger than 10, and only three are among the objects reported by \citet{Becker00}. This fraction is anomalously high in our sample, as it is at least 23 times the radio-loud fraction in the BAL population ($<1\%$; \citealp{Shen11}); in addition, it is two times higher than the average radio-loud fraction ($\sim$~10\%) in the whole quasar population. Moreover, 6 of the 32 objects have $R^* > 100$. 
Note that these objects have significantly high reddening and 14 of them are selected from the 104 ``special'' list (see Section \ref{sec:sample}), so the high radio-loud fraction is therefore likely biased to a large extent.

\begin{figure}[h]
\center{}
 \includegraphics[height=7cm,width=8.6cm,  angle=0]{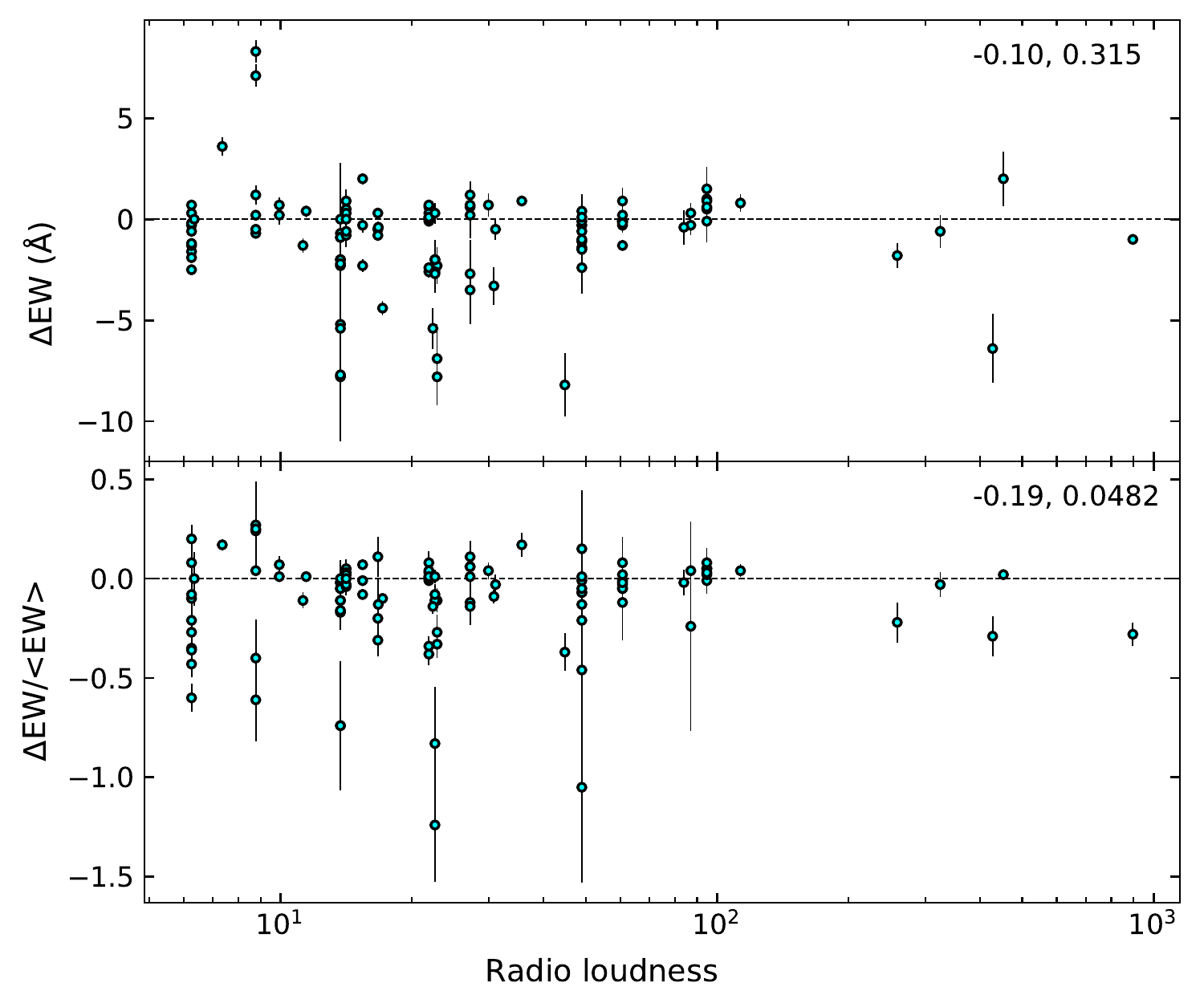} 
      \caption{Distributions of $\Delta$EW and $\Delta \rm{EW}/ \langle \rm{EW} \rangle$ versus radio loudness, in which there is no clear dependence. Spearman test results ($r_s, p_s$) using $|\Delta$EW$|$ are shown in each panel.}
\label{rloudness_fdew}
\end{figure}

The distributions of the absolute/fractional EW variations versus radio loudness are shown in Figure \ref{rloudness_fdew}. 
A Spearman test ($r_s \approx -0.09$, $p_s \approx 0.32$) reveals no correlation between $|\Delta \rm{EW}|$ and $R^*$. Similarly, no significant correlation between $|\Delta \rm{EW}/ \langle \rm{EW} \rangle|$ and $R^*$ has been found. 
Similar results were also reported for \civ-based BAL variability studies (e.g., \citealp{Filizak13}). 
The high fraction of radio-loud LoBAL QSOs in our sample and the lack of a significant difference in BAL variability from radio-loud to radio-quiet quasars (e.g., \citealp{Filizak13,Welling14,Vivek16}), in turn, may provide important insights into the inner structure of radio-loud LoBAL QSOs and the origin of jets and BAL outflows. 
Since powerful jets are naturally thought to interact with absorbing gas along our LOS and hence to play a role in BAL variability, it is thus worth exploring the underlying physics for radio-loud LoBAL QSOs in a dedicated work in the future.

\subsubsection{Central black hole mass and Eddington ratio}\label{Eddington_dependence}

BAL outflows are believed to be launched from the QSO nuclear or circumnuclear region, leading to the possibility that BAL variability may depend on the activity level of the central engine. 
Thus, we may expect to find a correlation or an anti-correlation between EW variations and Eddington ratio. 
However, the \mgii\ broad emission lines in our sample are usually affected by absorption. Therefore, we require alternate black hole (BH) mass estimators that do not rely on \mgii. In addition, reddening increases significantly at shorter wavelengths for some quasars in the sample. Therefore, a reliable BH mass estimate based on the single-epoch virial relation must avoid these complications.

To construct a ``clean'' subsample, a search for  H$\beta$-based BH masses has been done by cross-matching our sample with the SDSS DR7 quasar catalog \citep{Shen11}, in which 30 quasars are found. In addition, four LoBAL QSOs were found from a near-IR spectroscopic study using the H$\alpha$ emission line \citep{Schulze17}. Recently, four LoBAL QSOs (two of them from the 30 \hb-based objects) were observed at near-IR wavelengths by the TripleSpec instrument at the Palomar Hale 200-inch telescope (P200/TripleSpec; \citealt{Wilson04}). We estimate the BH mass and Eddington ratio from the high S/N near-IR  spectra using the same method as \citet{Schulze17}.  
For the two quasars with \hb-based BH masses from \citet{Shen11}, we refine their BH mass using the H$\alpha$ estimator from near-infrared spectroscopy, as their \hb\ lines are located at the red end of the SDSS spectra with low S/N. 
This process produces a sample of 36 quasars with BH masses and Eddington ratios derived from the hydrogen Balmer lines, allowing us to investigate the LoBAL variability dependence on the BH mass/Eddington ratio in a more reliable and uniform way, with the cost of reducing the sample size.  

 \begin{figure}[h]
 \includegraphics[width=8.6cm, height=7cm, angle=0]{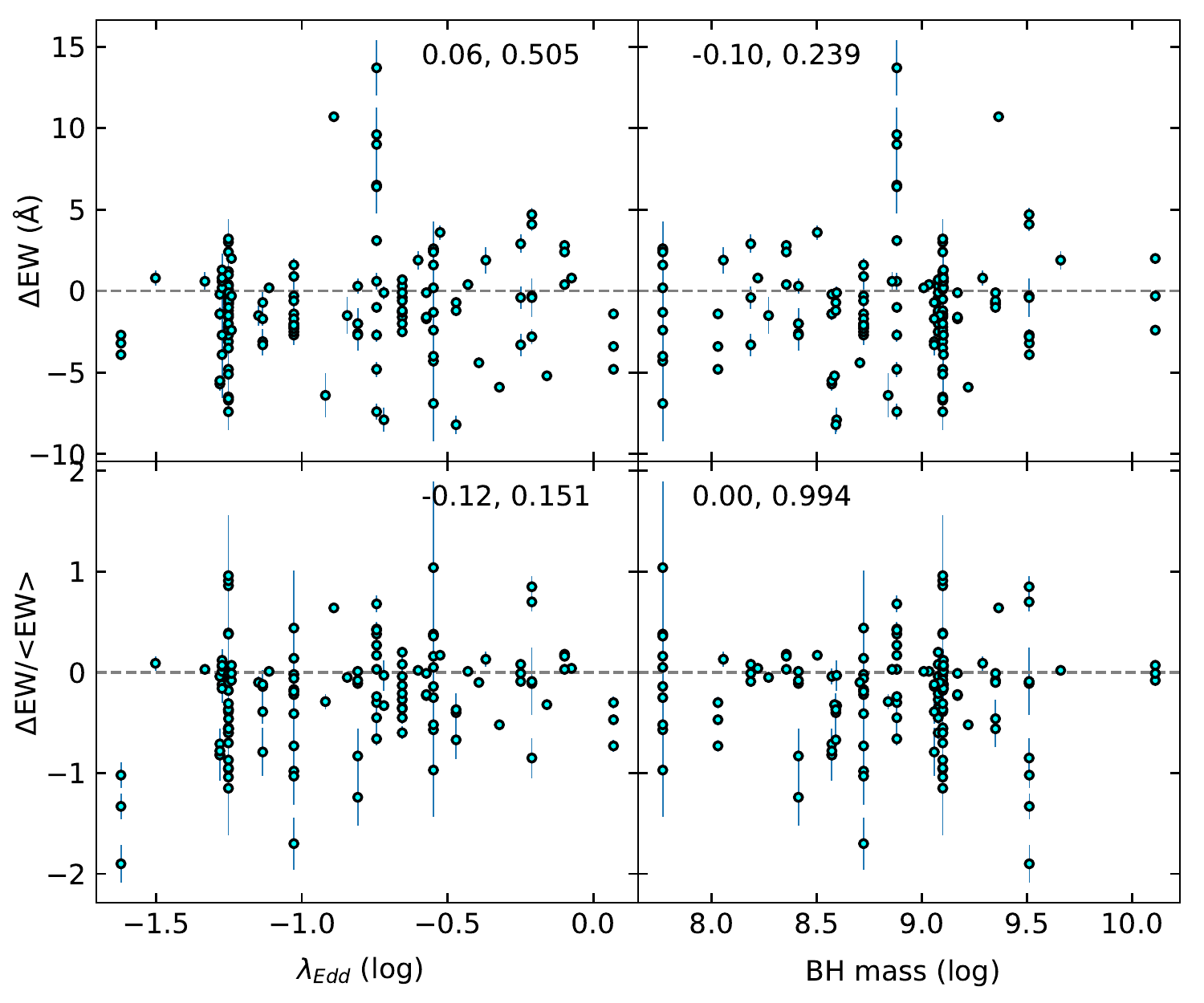}
      \caption{ Distributions of the absolute/fractional EW variations versus Eddington ratios and BH masses in the ``clean'' subsample of 36 \mgii-BAL QSOs with \hb- or H$\alpha$-based BH masses, in which no correlations are found. Spearman test results ($r_s, p_s$) using $|\Delta$EW$|$ are shown in each panel.}
\label{hb_BH_dew}
\end{figure}

We find no significant correlations ($|r_s|<0.15$,  $p_s>0.15$ based on Spearman tests) between EW variability and Eddington ratio, or between EW  variability and BH mass. The results, in combination with those from \citet{Filizak13}, probably indicate that Eddington ratio is not a dominant driver of BAL variability if we assume that the BH masses from their sample are not biased by the \civ\ estimator. 
However, limited by the subset size, we do not exclude the possibility of correlations entirely. A larger uniform sample is required to definitely determine the role of BH mass and Eddington ratio in LoBAL variability.

\subsection{Investigation of coordinated variability over different-velocity \mgii-BAL troughs}
\label{coordinate_var}
Previous studies based on high-ionization BAL QSOs with multiple troughs from the same species  have demonstrated that variations of multiple \civ\ troughs are strongly correlated (e.g., \citealp{Cap12,Filizak12,Filizak13,Wang15}). 
In our sample, some objects have multiple  \mgii-BAL troughs. Quasars with multiple BAL troughs from the same ion provide unique and powerful diagnostics to investigate differences/similarities between separated BAL troughs, allowing further understanding with respect to the origin of outflows and BAL variability.

\begin{figure}[h]
\center{}
 \includegraphics[height=9cm, width=8cm,  angle=0]{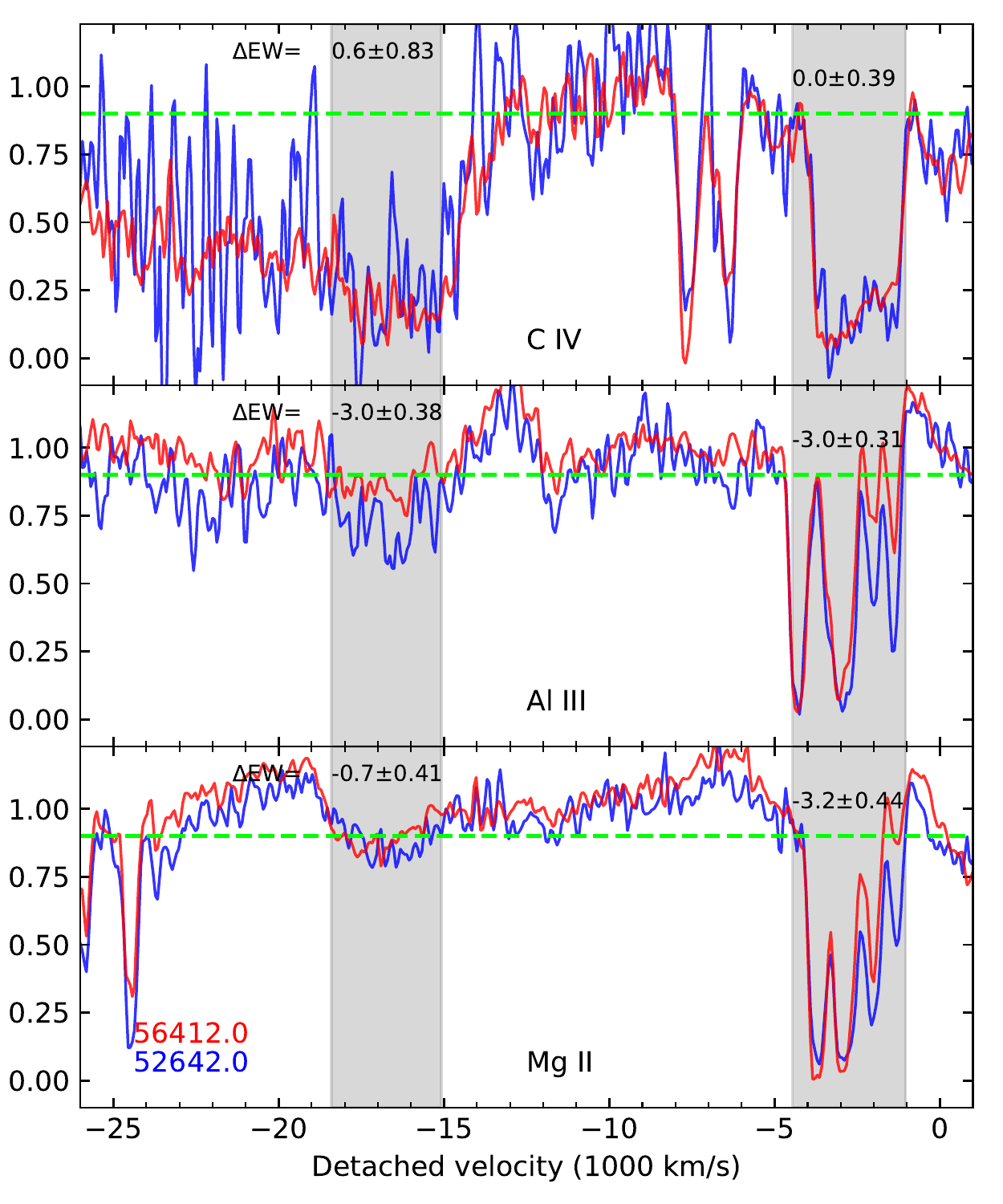} 
      \caption{Demonstration of coordinated BAL variability from the same ion and the comparison between EW variations from BALs caused by different ions (\civ, \aliii, and \mgii; see Section \ref{EWvar_mgii_civ}) in J1127+4859. Horizontal green lines represent the level at 90\% of the normalized continuum. Vertical grey zones represent different velocity troughs with corresponding measurements of $\Delta$EW.} 
\label{vcomJ112736}
\end{figure}

\begin{figure}[h]
\center{}
 \includegraphics[height=6cm,width=7cm,  angle=0]{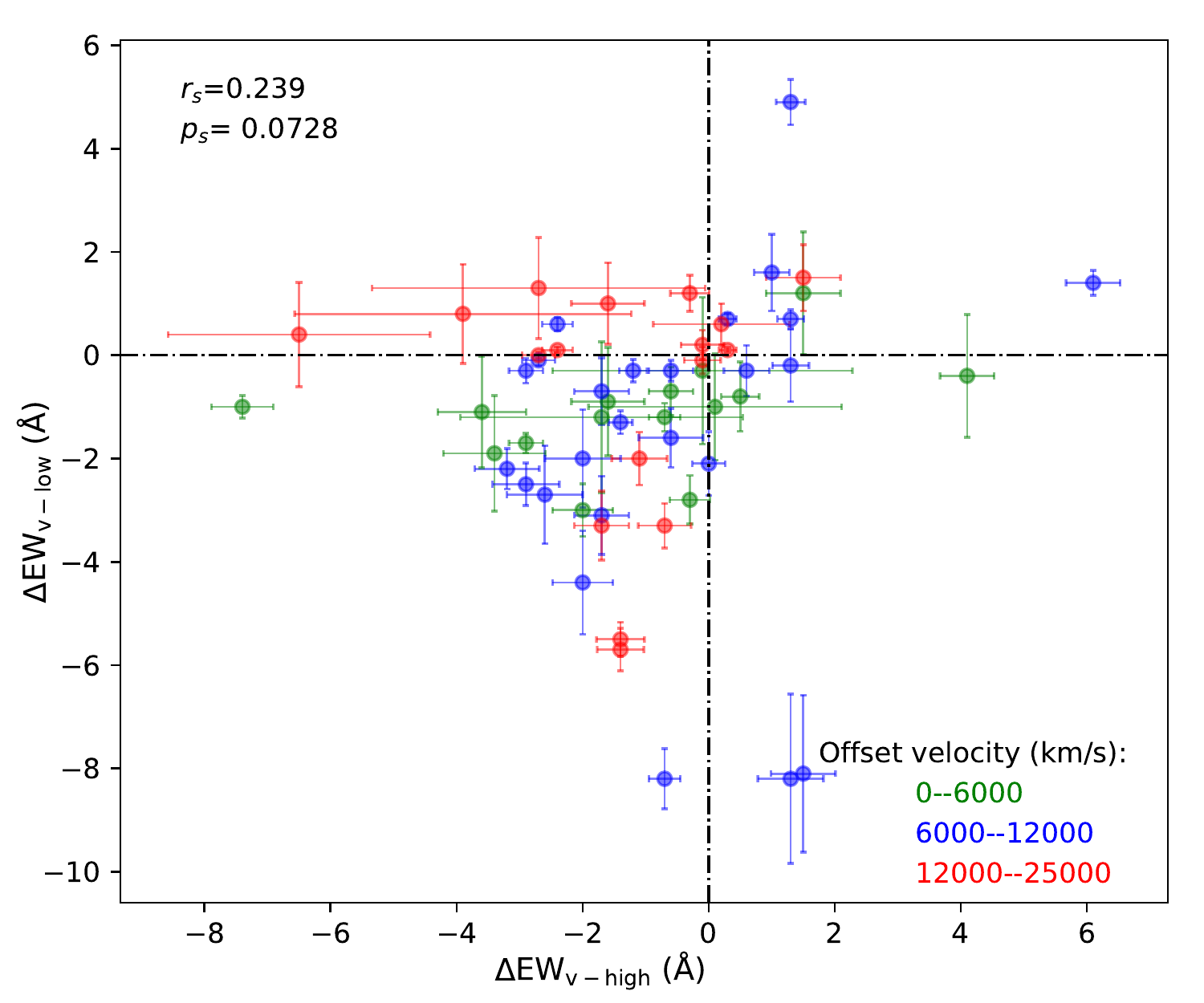}
  \includegraphics[height=6cm,width=7cm,  angle=0]{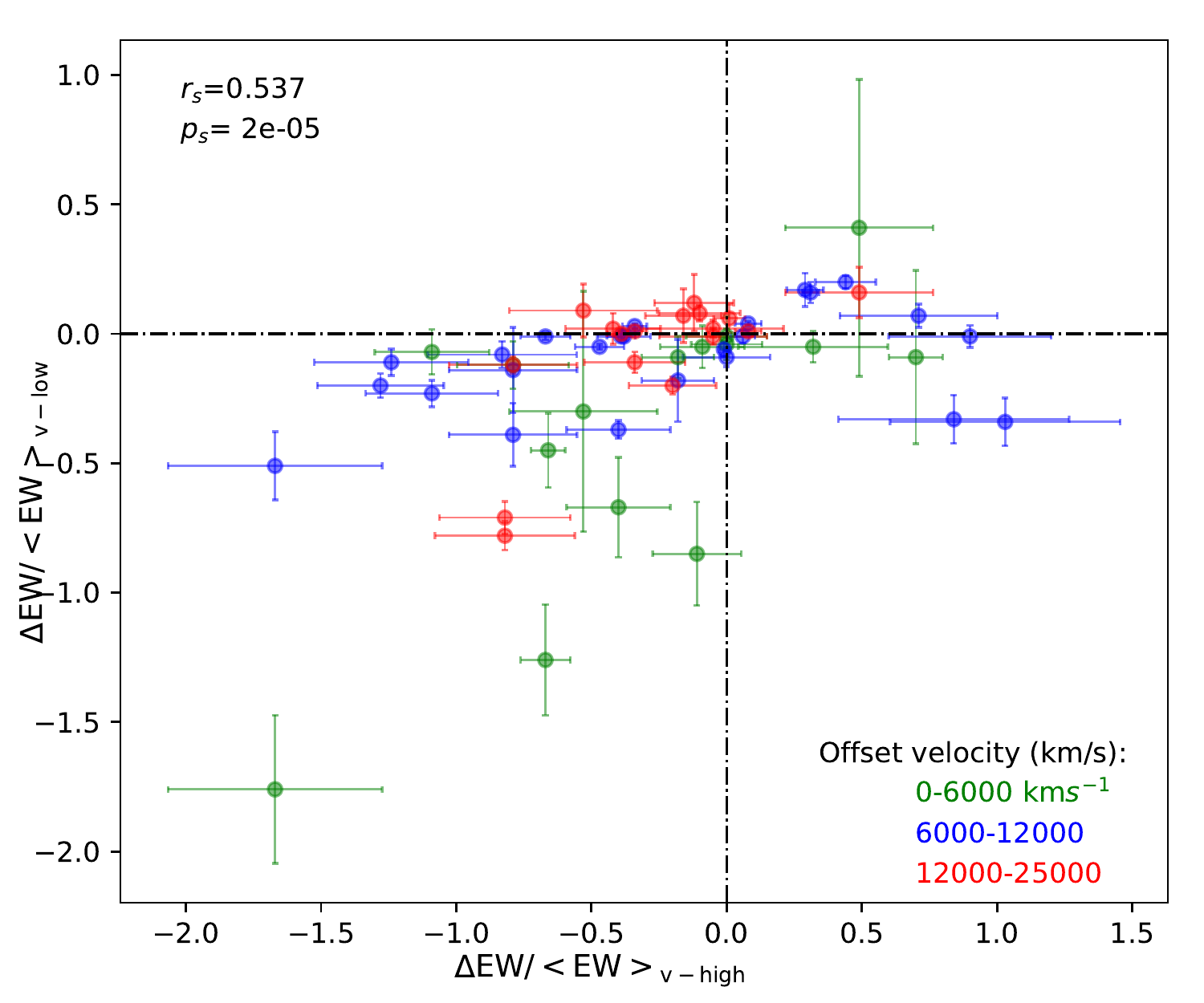}
      \caption{Distributions of absolute EW variations (upper panel) and fractional EW variations (lower panel) among BAL quasars with multiple \mgii\ troughs. Colors indicate the mean velocity separation between BAL-complex troughs. $r_s$ and $p_s$ values are derived from Spearman tests, where no correlation is found for absolute EW variations.} 
\label{coordination_ew}
\end{figure}

The subset for investigating coordinated BAL variability is constructed from an automatic search of all objects having two or more independent \mgii-BAL troughs. One of the difficulties is that approximately 40 FeLoBAL QSOs exhibit apparent \feii\ absorption troughs,  among which high-velocity \mgii-BAL troughs are likely blended with \feii\ components. 
Since this work is focused on \mgii-BAL variability, we did not account for multiple BAL components in most FeLoBAL QSOs. However, additional \mgii-BAL troughs were re-identified in the FeLoBAL subsample with the aid of  \aliii- and/or \civ-BAL troughs appearing in a similar velocity range  (see Figure \ref{vcomJ112736}). These features are added to the subsample for investigations of coordinated variability since these outflowing absorbers associated with different ions at same velocities are likely formed in the same region. This procedure, in turn,  provides an effective way to assess the reliability of weak \mgii-BAL troughs. 
To exclude further any multiple troughs not caused by \mgii, all objects in the subsample are  visually inspected. As a result,  23 objects hosting multiple \mgii-BAL troughs are identified, and we separate them into three groups according to the offset velocity (see Figure \ref{coordination_ew}). 

Based on our measurements of absolute/fractional EW variations, we perform a Spearman test to examine whether possible correlations of BAL variability  exist among different-velocity \mgii-BAL troughs.
We found that there is a significant correlation between the fractional EW variations. As a comparison, the correlation between absolute EW variations in our sample is not highly significant (see Figure \ref{coordination_ew}), obviously different from those based on HiBAL variability studies (e.g., \citealp{Filizak13,Wang15}), where highly significant correlations have been found using \civ-BAL troughs. 
Specifically, the absolute EW variations from the lowest offset-velocity group (green) shows a significant correlation while the other two groups (blue and red) do not show any correlations, probably implying an offset-velocity dependence of the coordinated BAL variability. This finding is consistent with the results of \citet{Rogerson18} (their section 4.7). 
However, the size of this subset is rather small, and thus we require a larger sample to draw a firm conclusion.  

For coordinated \mgii-BAL variability, a typical example is shown in Figure \ref{vcomJ112736} (see Figure~\ref{vr_demo} for unnormalized spectra). The object (J1127+4859) has two \mgii-BAL troughs with an offset velocity of 13000 \kms. One of them has a VR, $\sigma_{\rm \Delta EW}>5$, and $\chi_{1-2}^2>2.2$, which is identified as a pristine variable LoBAL QSO. 
It is apparent that the high-velocity trough nearly disappeared while the low-velocity trough depth decreased in EW in the later epoch.  
In addition, BAL variations occur across the whole high-velocity trough while only a portion of the low-velocity trough varied, likely due to saturation of the low-velocity trough.  
This object shows unambiguous coordinated LoBAL variability between the two different-velocity \mgii\ troughs along with similar \aliii-BAL variability in a corresponding velocity range.

For uncoordinated \mgii-BAL variability, a typical example is shown in Figure \ref{bal_complex}, where one of the two \mgii-BAL troughs varies from epoch to epoch. Noticeably, after combining $\sigma_{\rm \Delta EW}$ and $\chi_{1-2}^2$, we found that the low-velocity BAL trough composed by the first (MJD = 52232) and last (MJD = 57061) epoch spectra has a small $\sigma_{\rm \Delta EW}$ but large $\chi_{1-2}^2$, which is caused by approximately equal weakening/strengthening portions occurring in the same \mgii\ trough ($\Delta$EW = 0.4 \AA), and thus is hard to be explained by ionization changes.

Coordinated variability between widely separated troughs in the same ion can be most naturally explained by changes in the ionization state, 
although we cannot rule out the possibility that coordinated variability of absorption troughs is a combination of other effects.  
The fraction of uncoordinated \mgii-BAL variability in this subset can be roughly represented by these data points with opposite EW variations ($\sim$ 26\%). However, this fraction is a lower limit since there is a dramatically high fraction ($\sim$ 42\%) of \mgii-BAL troughs in this subset having $\sigma_{\rm \Delta EW}<5\sigma$ but $\chi_{1-2}^2>1.1$, which is potentially associated with the phenomenon where both increasing and decreasing portions are present for the same BAL trough (see Figure \ref{sigEW_vs_chi2} for an overall distribution in the sample). Therefore, we argue that this phenomenon, as well as partial covering for \mgii-BAL absorbers with relatively high optical depths  (see Section \ref{var_timescales}),  could dominate the behavior of \mgii-BAL variability over different-velocity troughs in this subset and hence lead to significantly weaker correlations in coordinated variability found in this subset compared to HiBAL QSOs.

\subsection{Investigation of BAL variability from low- to high-ionization species}
\label{EWvar_mgii_civ}
In addition to coordinated variability of troughs of the same ion at different velocities, correlations between EW variations of BAL troughs caused by \civ\ and \siiv\ have been reported in several sample-based studies. 
\citet{Gibson10} and \citet{Cap12} found no clear correlations in the absolute EW variations between  \civ\ and \siiv\ BAL troughs, although tentative correlations in fractional EW variations. 
However, highly significant correlations in EW variations of different HiBALs were found from larger samples (e.g., \citealp{Filizak13,Wang15}). In addition, \citet{Filizak14} reported that only a marginally significant correlation in EW variations was found between \civ\ and \aliii. 
To date, no such work has been done for \mgii-BAL troughs, as previous \mgii-BAL studies did not meet requirements both for the sample size and redshift range. 

In this section, we further investigate EW variations of BAL troughs caused by different ions at similar velocities. 
This subset includes 58 LoBAL QSOs with \mgii, \aliii,  and \civ\ BALs simultaneously present in the spectra for each quasar.  
The main advantages of studying BAL variability among different-ionization species are the following: 

\begin{enumerate}
\item
The \civ\ and \aliii\ troughs are usually deeper than \mgii\ in LoBAL QSOs, which  is particularly helpful to set boundaries  or identify shallow multi-velocity components from \mgii-BAL troughs. 
\item
Different-ion BALs are quite useful for selecting ``clean'' \mgii-BAL components in FeLoBALs, where high-velocity \mgii\ BALs tend to be contaminated by \feii\ absorption features. They can also be used as additional diagnostics to identify \mgii\ BALs attached on the \mgii\  emission line. 
\item
BAL variability in different ions at similar velocities also helps to probe changes in ionization state, if their ionization potentials are comparable, such as \mgii\ and \aliii, or \civ\ and \siiv. 

\item
\mgii, \aliii, and \civ\ BALs appearing in the same object provide a unique and powerful probe to study BAL variability simultaneously from low- to high-ionization transitions (the ratio of ionization potentials is $1:2:4$). 

\end{enumerate}

\subsubsection{BAL variability between \mgii~ and \aliii }
\label{dEW_mgii_vs_aliii}

\begin{figure}[h]
\center{}
 \includegraphics[height=12cm,width=6.5cm,  angle=0]{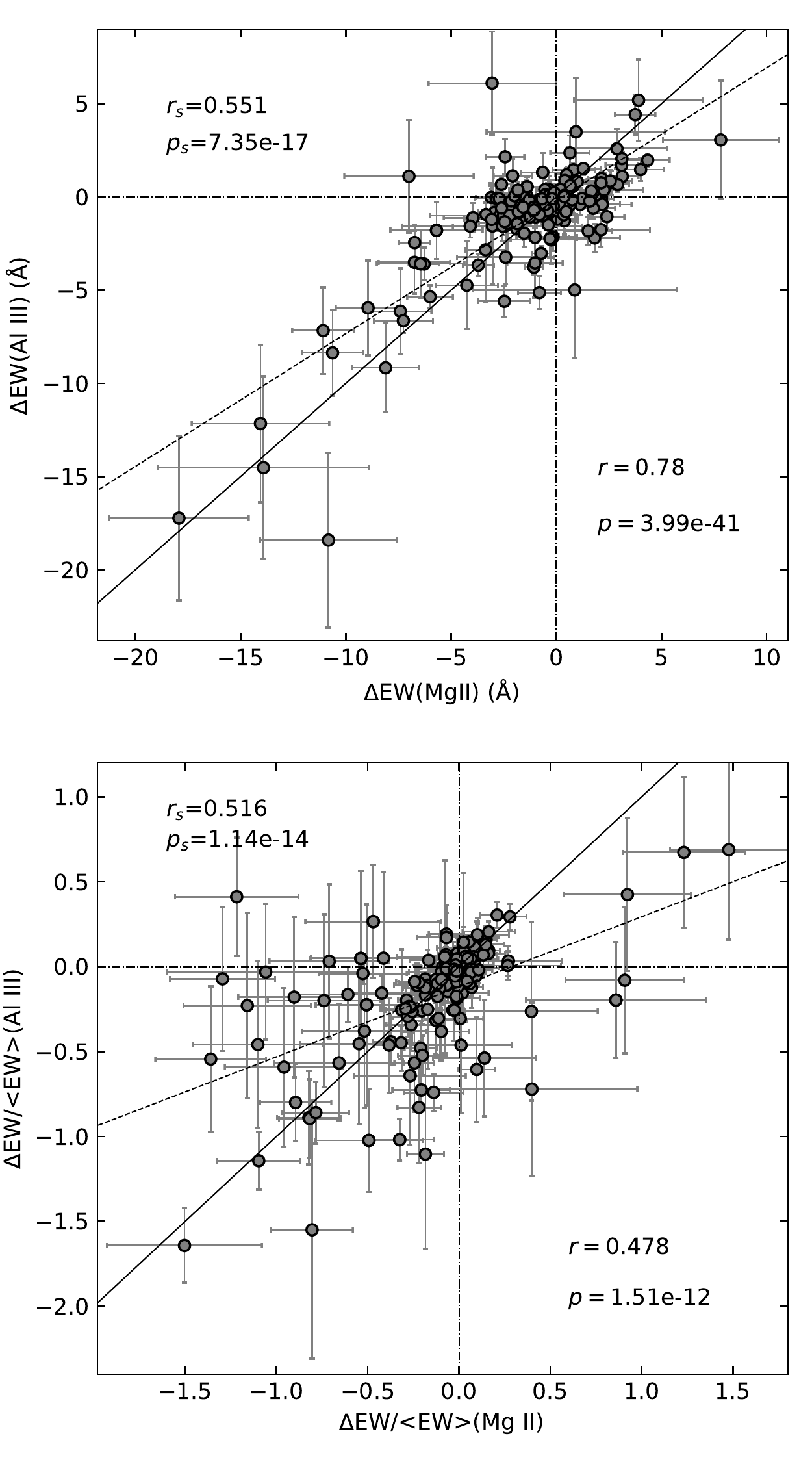}
      \caption{Comparison of EW variations (top panel) and fractional EW variations (bottom panel) between \mgii\ and \aliii\ troughs. Spearman test results are shown in the upper right of each panel; Pearson test results are shown in the bottom right of each panel. The solid black lines show the same absolute/fractional EW variations. The dotted lines represent the best linear fits. There is a strong linear correlation in the distribution of $\Delta$EW between \mgii\ and \aliii.} 
\label{dEW_mgiialiii} 
\end{figure}

Due to the differences in ionization potentials (28.5 eV for \aliii\ and 15.0 eV for \mgii) and the minimum photon energies needed to ionize Al II (18.8 eV) and Mg I (7.65 eV), \aliii\ troughs are expected to be  more common than \mgii\ troughs, as higher ionization troughs appear more frequently than lower ionization troughs in BAL quasars (e.g., \citealp{Hall02,Filizak14}). 
The presence of \aliii\ also places a quasar in the LoBAL family; therefore, \aliii\ can be used to compare different LoBAL subtypes with 
different-ionization potentials.

We restrict our subset to the 58 quasars showing both \civ\ and \mgii\ BALs as well as \aliii\ to enable an investigation of all three species simultaneously. 
The widths and depths of \aliii-BAL troughs are wider and deeper than \mgii\ troughs in general. To make appropriate comparisons,  
we use the same velocity ranges identified from the \mgii\ troughs to exactly match \aliii\ troughs for each object (see Figure \ref{vcomJ112736} for an example).

Our measurements are shown in Figure \ref{dEW_mgiialiii}. After performing Spearman tests, we found highly significant correlations between \mgii\ and \aliii\ BAL variability, both in absolute and fractional EW variations. Furthermore, the Pearson test reveals a strong linear correlation ($r=0.78$) in the distribution of absolute EW variations between \mgii\ and \aliii, implying that the two different-ion BALs are strongly associated. 
The EW variations of \mgii\ BALs tend to be stronger than those for \aliii, as we measure the slope of the best linear fit (dashed line) to be smaller than the one-to-one (solid black) line. This tendency is similar to that from \citet{Filizak14}, where they found that EW variations of \aliii\ BALs are stronger than \civ\ BALs on average. 

Compared with HiBAL variability studies, highly significant correlations, either in the absolute or fractional EW variations, were also reported between \civ\ and \siiv\ BALs (e.g., \citealp{Filizak13}). We compare BAL variability correlations over different species from low- to high-ionization transitions in Section \ref{conclusion_mgii_civ}.

\subsubsection{BAL variability between \mgii~ and \civ }
\label{dew_civ_mgii}

\begin{figure}[h]
\center{}
\includegraphics[height=6.5cm,width=8.7cm,  angle=0]{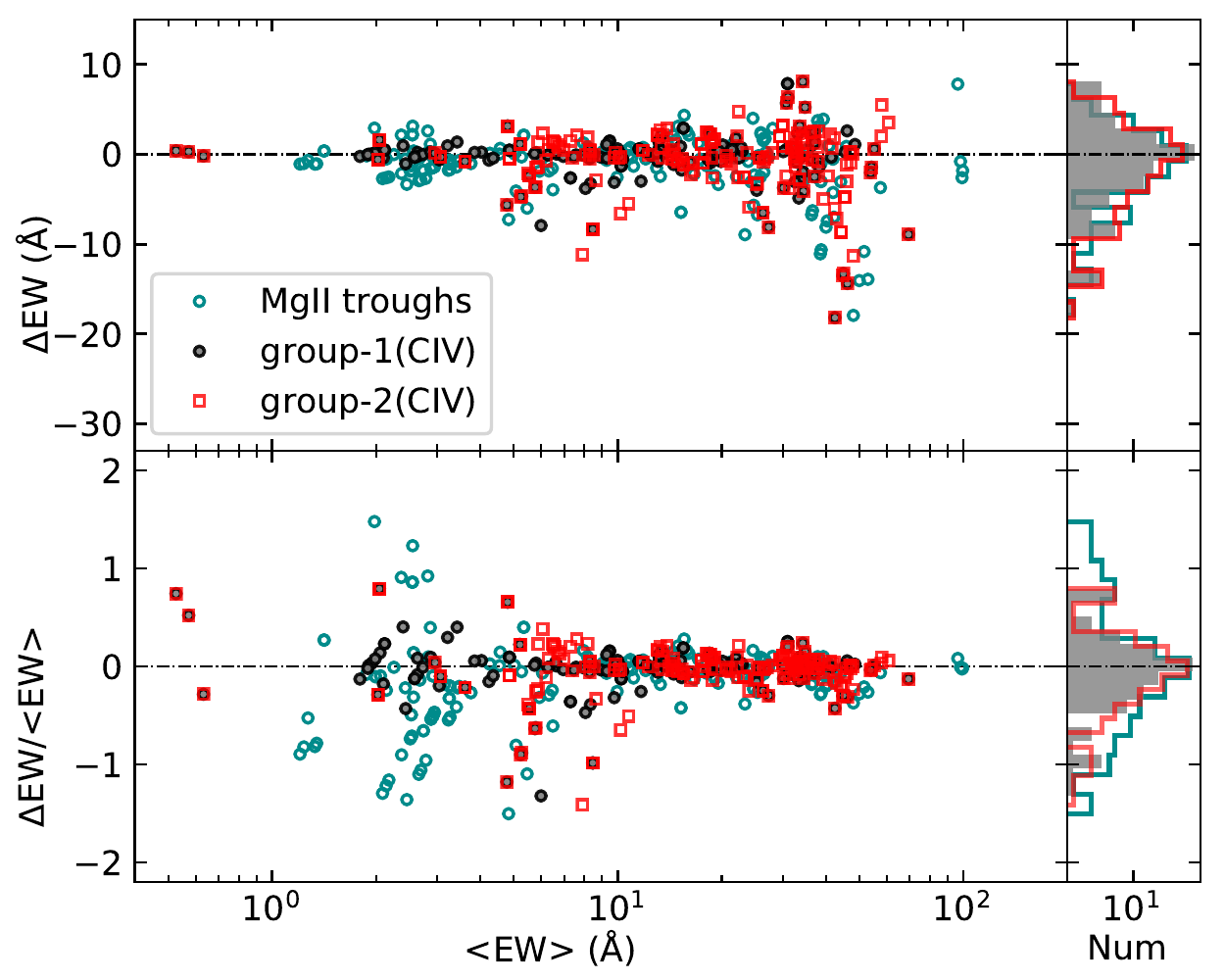}   
      \caption{Left panels show distributions of EW variations as a function of mean EWs in \civ\ BALs measured by the \mgii-BAL matched (group-1) and entire \civ~ (group-2) troughs, respectively; right panels are corresponding histogram distributions.  Cyan circles are \mgii\  troughs used for matching \civ\ BALs. In general, \civ\ BALs tend to have higher EWs and show larger absolute EW variations but smaller fractional EW variations than those of \mgii\ BALs.} 
\label{EW_matchciv_dEW} 
\end{figure}

Investigating the relationships between \mgii\ and \civ\ is complicated by wavelength-dependent reddening and larger differences in ionization potentials (64.5 eV for \civ\ and 15.0 eV for \mgii)  and the minimum photon energies needed to ionize C III (47.9 eV) and Mg I (7.65 eV).  
\civ-BAL troughs also tend to exhibit lower S/N ratios and lower flux calibration accuracies (due to their location at the blue end of spectra), reducing the reliability of quantitative measurements. 
Consistent with \citet{Filizak14}, another common feature in the subset is that these \civ\ BALs usually exhibit much deeper and wider troughs than those from \mgii\ and \aliii, although the actual \civ-BAL depths are hidden among those objects that are heavily affected by intrinsic reddening. Therefore, only a lower limit on the real depth can be derived from normalized spectra in such cases.

 \begin{figure}[ht]
\center{}
 \includegraphics[width=6.5cm, height=12cm, angle=0]{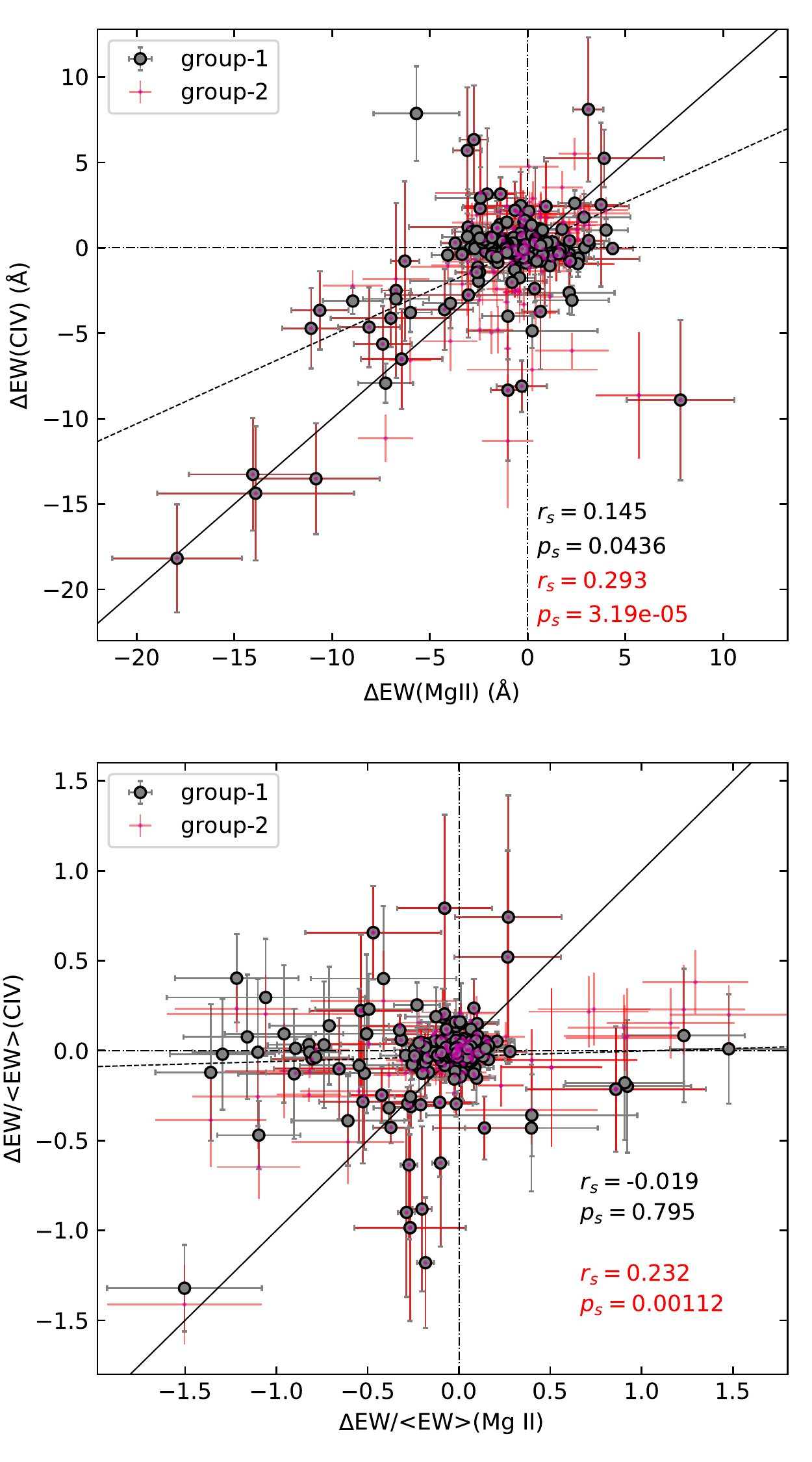}
      \caption{ EW variations from \mgii-BAL matched (group-1) and the whole \civ-BAL velocity ranges (group-2) against \mgii-BAL troughs. Spearman test results are shown in the bottom right of each panel. The solid black lines show the same absolute/fractional EW variations in group-1. The dotted line represents the best linear fit in group-1. This figure reveals a significantly weaker correlation between \mgii\ and \civ\  compared to that between \mgii\ and \aliii\ in $\Delta$EW.} 
\label{EWvar_mgiiciv_byT}
\end{figure}

Note that \civ-BAL troughs usually appear to be less variable in the velocity range exactly matched with a \mgii-BAL trough while being more variable in other portions for the same BAL trough (see the high-velocity \civ\ BAL in Figure \ref{vcomJ112736} as an example).  
We measure the EWs of the \civ\ BALs both from an exact match of the \mgii-BAL width that is defined as group-1 and the entire \civ-BAL width that is defined as group-2.  
Indeed, the absolute EW variation of \civ-BAL troughs derived from the former method (see black dots in Figure \ref{EW_matchciv_dEW}), on average, is   smaller than that derived from the latter (red squares in Figure \ref{EW_matchciv_dEW}).  Two reasons can explain this phenomenon. First, the \civ-BAL trough is ubiquitously stronger than the \mgii-BAL trough when both of them are present in the same spectrum. Second, \civ-BAL components corresponding to group-1 tend to be more saturated compared to the entire \civ-BAL troughs (group-2) and hence become less variable.  
Conversely, the absolute EW variations in \civ-BAL troughs from group-1 also appear to be smaller than those of corresponding \mgii-BAL troughs; a ``flat'' tendency in the distribution of fractional EW variations of group-1 \civ-BAL troughs starts to appear at $\langle \rm{EW} \rangle > 10$ \AA, which is 
apparently different from that of \citet{Filizak13} where the ``flat'' tendency appears at $\langle \rm{EW} \rangle  > 40$ \AA\ (see Figure 18 in their work).  Such differences are likely  attributed to a higher likelihood of saturated \civ-BAL troughs in our sample, as most of them exhibit one (or more) nonblack saturation shape(s) characterized by roughly constant but nonzero residual flux, a strong signature of saturation under a partial covering condition. Furthermore, 
the similar ``flat'' tendency of the fractional EW variability between \mgii\ and group-1 \civ\ troughs indicates that \mgii\ troughs are also saturated but at a lesser level.

Although EW variations of the \civ\ BALs measured by the two above methods are apparently different (particularly for the absolute EW variations), the results obtained by the two methods, from a statistical point of view, are consistent with each other regardless of which boundaries are used to measure the \civ-BAL EW (see the Pearson test results in Figure \ref{EWvar_mgiiciv_byT}).

Obviously, statistical results indicate that the correlations of BAL variability between \mgii\ and \civ\ are significantly weaker than those between \mgii\  and \aliii, implying that \mgii-BAL absorbers have stronger connections with \aliii-BAL absorbers than that with \civ-BAL absorbers.  
The significantly weaker correlations compared to previous HiBAL variability studies, to a large extent, can be explained by saturation effects and/or large differences in ionization potential and density. 
For an individual example, Figure \ref{vcomJ112736} demonstrates that the \mgii-BAL variability behavior highly resembles \aliii, while BAL-variability is uncertain in \civ.

In combination with the investigations of coordinated BAL variability for \mgii\ (see Section \ref{coordinate_var}), our results suggest that \mgii\ and \aliii\ BAL absorbers have strong connections while \mgii\ and \civ\ BAL absorbers, as well as different-velocity \mgii-BAL absorbers for the same quasar, appear to be more independent of each other.

\subsection{Comparison of BAL variability from \mgii, \aliii\ and \civ} \label{conclusion_mgii_civ}
As \civ\ and \mgii\ possess ionization potentials of 64.5 eV and 15.0 eV , in combination with the minimum photon energies of 47.9 eV and 7.65 eV to form them, respectively, we do not expect to find strong correlations in BAL variability such as those observed between \aliii\ and \mgii. 
The presence of \aliii\ (28.5 eV and 18.8 eV for the ionization potential and minimum photon energy, respectively) BAL troughs provides a natural bridge over the gap from low to high ionization transitions (with the ionization-potential ratio among \mgii, \aliii, and \civ\ $\approx 1:2:4$). 

Our statistical analyses (see Section \ref{EWvar_mgii_civ}) indicate that the correlation strength of BAL variability between \mgii\ and \aliii\ is significantly stronger than that between \mgii\ and \civ\ (see Table \ref{tab_3disap_trans3}). The result that species with closer ionization potentials are more correlated is consistent with \citet{Filizak14}, where they reported that the correlation of BAL variability between \civ\ and \siiv\ is highly significant  compared to a marginal correlation between \civ\ and \aliii.

These finding are consistent with the case where a gas cloud contains both \civ\ and \mgii\ BAL absorbers (with a large difference in the ionization potential) at the same velocity and the \civ\ BALs are often saturated due to its higher abundance. Consequently, saturation effects must play an important role in determining these correlations mentioned above, given a large difference in abundance and ionization potential between \mgii\ and \civ.

For an overview of various correlations measured throughout the whole section, we tabulate Spearman test results in Table \ref{tab_3disap_trans}, \ref{tab_3disap_trans2}, \ref{tab_3disap_trans3}.

\begin{deluxetable*}{lcccccc}
\tabletypesize{\scriptsize}
\tablewidth{0pc}
\tablecaption{}
\tablehead{ & $\Delta t$ &  \colhead{$\langle \rm EW \rangle$-S/I/L} & $\langle \rm EW \rangle$ & $v_{\rm LOS}$ & $d_{\rm BAL}$ & $\Delta v$  }
\startdata
$|\Delta \rm EW|$ & $0.14, 2\times 10^{-3}$  & N/Y/N & $0.18, 4\times 10^{-5}$ & $0.22, 9\times 10^{-7}$ & $0.15, 6\times 10^{-4}$ & $0.33, 3.6\times 10^{-14}$  \\ 
$|\Delta \rm EW|$/$\langle \rm EW \rangle$ & $0.16, 4\times 10^{-4}$  & Y/Y/Y & $-\textbf{0.53}, 5\times 10^{-36}$ & $0.4, 1\times 10^{-19}$  & $-\textbf{0.67}, 6\times 10^{-64}$ & $-0.31, 2.1\times 10^{-12}$ \\

$v_{\rm LOS}$ &- & -& $-0.26, 4\times 10^{-9}$ & - & $-\textbf{0.63}, 1\times 10^{-60}$ & $0.0, 0.92$  
\enddata
\tablecomments{ Spearman test results ($r_s,p_s$) of \mgii-BAL EW variability versus various properties measured from the entire sample. $\langle \rm EW \rangle$ and $\langle \rm EW \rangle$-S/I/L represent mean EWs on the whole and short/intermediate/long timescales, respectively, in which Y and N refer to correlation and non-correlation. Highly significant correlations ($r_s>0.5,p_s>10^{-3}$) are highlighted by bold fonts.}
\label{tab_3disap_trans}
\end{deluxetable*}

\begin{deluxetable*}{lccccccccccccccc}
\tabletypesize{\scriptsize}
\tablewidth{0pc}
\tablecaption{}
\tablehead{ & $M_i$ & $z$ & $R^*$ & $M_{BH}$ & $\lambda_{\rm Edd}$   }
\startdata
$|\Delta \rm EW|_{\mgii}$  & $0.14, 2\times 10^{-3}$ & $-0.1, 0.02$ & $-0.1, 0.31$ & $-0.1, 0.24$ & $0.06, 0.5$    \\
$|\Delta \rm EW|$/$\langle \rm EW \rangle_{\mgii}$  & $0.09, 0.04$ & $-0.2, 2\times 10^{-4}$ & $-0.19, 0.05$ & $0.0, 0.99$ & $-0.12, 0.15$     \\ 
$|\Delta \rm EW|_{CIV}$  &- &- &- &- &-    \\
$|\Delta \rm EW|$/$\langle \rm EW \rangle_{CIV}$  &- &- &- &- &- 
\enddata
\tablecomments{ Spearman test results ($r_s,p_s$) of EW variability versus different physical quantities from different subsets. }
\label{tab_3disap_trans2}
\end{deluxetable*}

\begin{deluxetable*}{lccccccccccccccc}
\tabletypesize{\scriptsize}
\tablewidth{0pc}
\tablecaption{}
\tablehead{ & $|\Delta \rm EW|_{\mgii}$2 & $|\Delta \rm EW|$/$\langle \rm EW \rangle_{\mgii}$2  & $|\Delta \rm EW|_{\aliii}$ & $|\Delta \rm EW|$/$\langle \rm EW \rangle_{\aliii}$  & $|\Delta \rm EW|_{CIV}$ & $|\Delta \rm EW|$/$\langle \rm EW \rangle_{CIV}$ }
\startdata
$|\Delta \rm EW|_{\mgii}$  & $0.24, 0.07$ & -& $\textbf{0.55}, 7\times 10^{-17}$ & - & $0.14, 0.04$  & - \\
$|\Delta \rm EW|$/$\langle \rm EW \rangle_{\mgii}$  & - & $ \textbf{0.5}, 2\times 10^{-5}$ & -  & $\textbf{0.52}, 1\times 10^{-12}$ & -& $-0.02, 0.79$    
\enddata
\tablecomments{ Spearman test results ($r_s,p_s$) of EW variability versus different-velocity \mgii\ BALs and different-ion BALs at the same velocity from different subsets. $|\Delta \rm EW|_{\mgii}$2 and $|\Delta \rm EW|$/$\langle \rm EW \rangle_{\mgii}$2 are derived from those quasars having more than two different-velocity \mgii\ BALs. Highly significant correlations ($r_s>0.5,p_s>10^{-3}$) are highlighted by bold fonts.}
\label{tab_3disap_trans3}
\end{deluxetable*}

\section{Discussion}
\label{var_timescales}
We here discuss the implications of the results throughout Section \ref{stat_view} and propose a model to further address the time-dependent asymmetric variability (the preference for the EW to weaken over time rather than strengthen) observed in the \mgii-BAL sample. 
However, an exhaustive examination of potential models is beyond the scope of this work, so we note that while our proposed model is consistent with our data, there may be additional models that well describe the data.

\subsection{Asymmetry in EW variability}
\label{time_dependent_aysm}

In this subsection, we systematically analyze the asymmetric EW variability, one of the most dramatic differences between LoBAL and HiBAL variability. 
Previous variability studies of the full HiBAL population did not find any significantly asymmetric distributions with respect to weakening and strengthening of BAL troughs (e.g., \citealp{Gibson08,Filizak13,Wang15}), although we note that the asymmetric EW variations for the HiBAL sample of \citet{Filizak13} may start to appear at $\Delta t>2.5$ yr (see Figure 16 in their work).

The observed preference for negative changes in EW are unlikely due to biased measurements, as the distributions of $\Delta \rm{EW}$ derived with the reddened power-law and non-BAL template fits are consistent  (see Figure \ref{civEWvar_byTPfits}). Additionally, they are not biased by cases where both increasing and decreasing portions occur in the same BAL trough since these tentative variable troughs show similar distributions (see Figure \ref{ewave_vs_dew_tspan}).  
Note that there are $\sim$40 QSOs in the sample with apparent \feii\ absorption troughs. We checked the distributions for asymmetry after excluding these FeLoBAL QSOs and found that the distribution of EW variations is consistent with the entire sample via a two-sample K-S test. This result is somewhat as expected since we already exclude most of the \feii\ troughs during the identification of \mgii-BAL troughs; particularly, for these troughs mixed with overlapping \feii\ troughs, we only keep ``clean'' \mgii\ components (see Section \ref{ew_definition} for details). In addition, there is a high fraction ($\sim$23\%) of radio-loud LoBAL QSOs in the sample; we also checked the distributions for asymmetry after excluding them. Again, the observed asymmetry is still present.

One possible explanation for asymmetric EW variability may be associated with selection effects. 
Note that our initial threshold (BI$_{M,0}>10$ \kms) for selecting bona fide \mgii-BAL QSOs misses those variability cases where non-BAL/HiBAL QSOs change to LoBAL QSOs and hence may increase the fraction of positive EW variations in the sample. Therefore, we cannot rule out the possibility of such a selection effect, as the investigation of the emergence rate from non-BAL/HiBAL QSOs to LoBAL QSOs is beyond the scope of this work. However, it is more likely that the selection effect would only lead to a slight (or negligible) asymmetric distribution of EW variations, as BAL emergence/disappearance events are rare in BAL variability and the HiBAL emergence rate is broadly consistent with the HiBAL-disappearance rate (e.g., \citealp{Rogerson18}).

Another possibility for the asymmetric EW variability is that the lifetimes of \mgii\ BAL episodes may be short enough that over long rest-frame timescales ($>5$ yr), more objects will be observed gradually moving toward a non-LoBAL phase than increasing the strength of their LoBAL phase. 
In our sample, we found that the majority of quasars with more than three epoch spectra did show a similar long-term declining and short-term rising trend in BAL EW; in particular, we observed one BAL to disappear over a long timescale ($>5$ yr) and appear on a much shorter timescale ($<1$ yr; \citealp{Yi19}). Similarly, \citet{Trevese13} reported that the EW slowly declines accompanied by a sharp increase in a \civ\ BAL.

The most likely cause for our observed asymmetric EW variability, however, is that many of \mgii-BAL absorbers have relatively high optical depths under partial covering. 
We assess this possibility and test the effect in Section \ref{saturation_effect} and \ref{main_driver}.

\subsection{ EW variations of a \mgii\ trough in the ionization-change scenario}  \label{saturation_effect}

\begin{figure}[h]
\center{}
 \includegraphics[height=6cm,width=8.7cm,  angle=0]{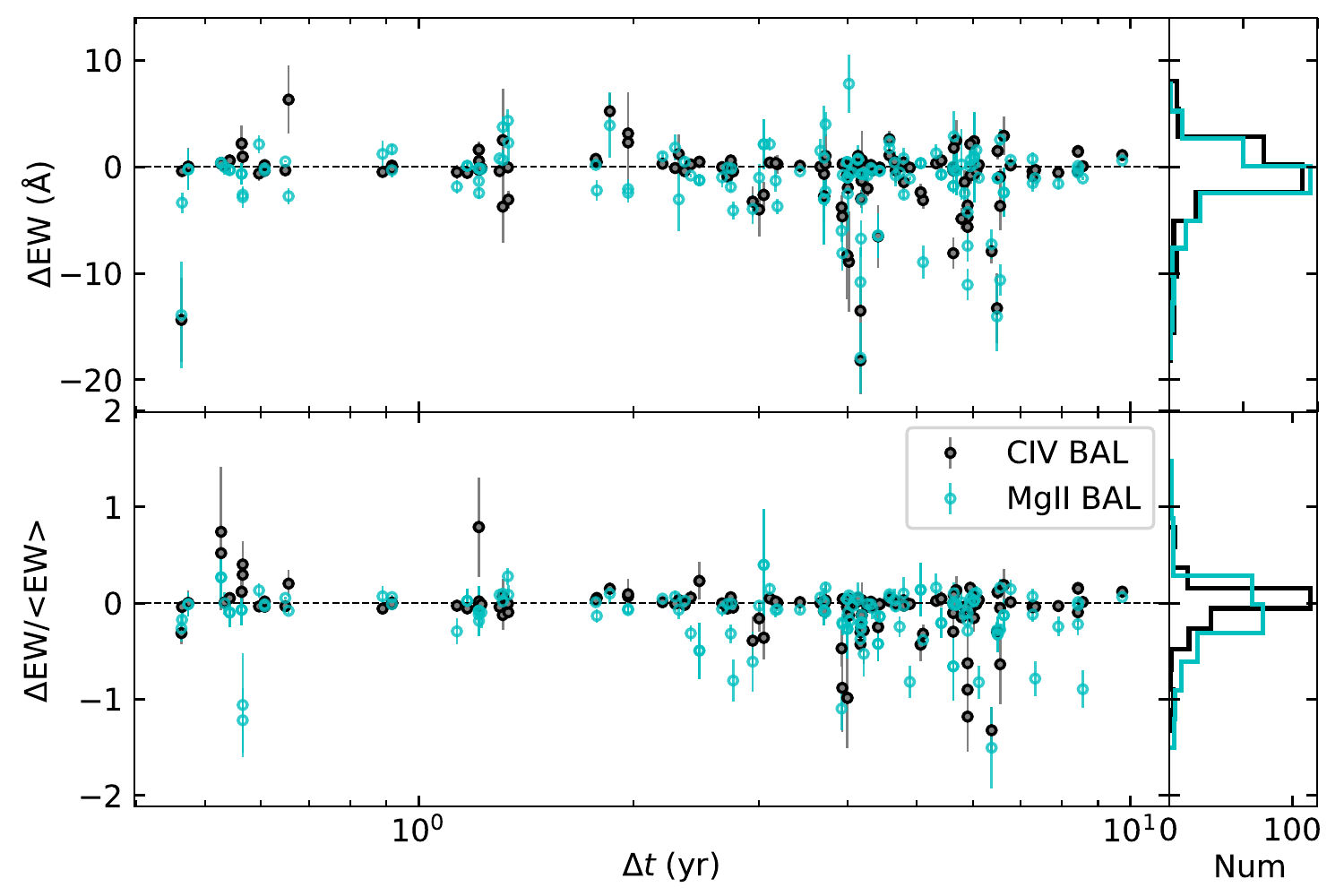} 
      \caption{Distributions of rest-frame timescale versus $\Delta \rm{EW}$ and $\Delta \rm{EW}/\langle \rm{EW} \rangle$ for \civ\ and \mgii\ BALs, both of which show a similar asymmetric distribution while \mgii\ BALs appear to have large EW variations.}  
      \label{civbal_dt_dew}
\end{figure}

A recent study based on a large sample of 6856 BAL quasars reveals that LoBAL QSOs are ubiquitously accompanied by \pv\ absorption \citep{Hamann18}, indicating high  saturation for the usual HiBALs like \nv\ and \civ. 
Indeed, observational evidence in support of saturation effects over \civ\ BALs was found throughout Section \ref{EWvar_mgii_civ} in our sample. 
\citet{Hamann18} also found that the \aliii\ doublet components appear to have trough-depth ratios consistent with $\sim1:1$ in most cases, suggesting saturation for low-ionization gas in LoBAL QSOs.

If saturation effects play a role in the \mgii-BAL troughs, it is natural to expect similar variability behavior in the \civ-BAL troughs at corresponding velocities. We choose QSOs from the subset showing both \civ\ and \mgii\ BALs in the same velocity range (see Section \ref{EWvar_mgii_civ}), which to some extent, can be used to investigate whether or not the \mgii\ BALs are saturated. 
Figure \ref{civbal_dt_dew} shows that we see a similar asymmetry in $\Delta \rm EW$ for \mgii, with a preference for negative $\Delta \rm EW$ as seen in \civ. 
Moreover, \civ-BAL variability is on average smaller than \mgii\ both in absolute and fractional EW variations,  suggesting that \civ\ BALs may be more saturated than \mgii\ BALs at the same velocity. Additional evidence supporting saturation effects among the \mgii\ BALs, as shown in Figure \ref{EW_matchciv_dEW}, is the similar ``flat'' distributions of $\Delta \rm EW$/$\langle \rm EW \rangle$ versus $\langle \rm EW \rangle$ between \mgii\ and matched \civ\ BAL troughs, in which $\Delta \rm EW$/$\langle \rm EW \rangle$ appears to level off at $\langle \rm EW \rangle >10$~\AA\ for both \civ\ and \mgii. Such a phenomenon is different from  \citet{Filizak13}, where the ``flat'' distribution appears at $\langle \rm EW \rangle >40$~\AA\ (see Figure 18 in their work). 
Most importantly, shallow \mgii\ BALs in our sample on average decrease their EWs in the later epoch (see Figure \ref{f_dew_width_depth}), which is opposite to the finding for shallow \civ\ BALs from \citet{Filizak13} (see Figure 17 in their work). 
 Such a remarkable difference could be due to the fact that \mgii-BAL absorbers in our sample tend to have relatively higher optical depths (if not saturated) and smaller LOS-covering factors than \civ-BAL absorbers from \citet{Filizak13}, as low-ionization gas is likely embedded inside high-ionization gas when they appear at the same velocity along our LOS (e.g., \citealp{Arav1999, Baskin2014a, Hamann18}). 
 
\begin{figure}[h]
\center{}
 \includegraphics[height=5.5cm,width=8.7cm,  angle=0]{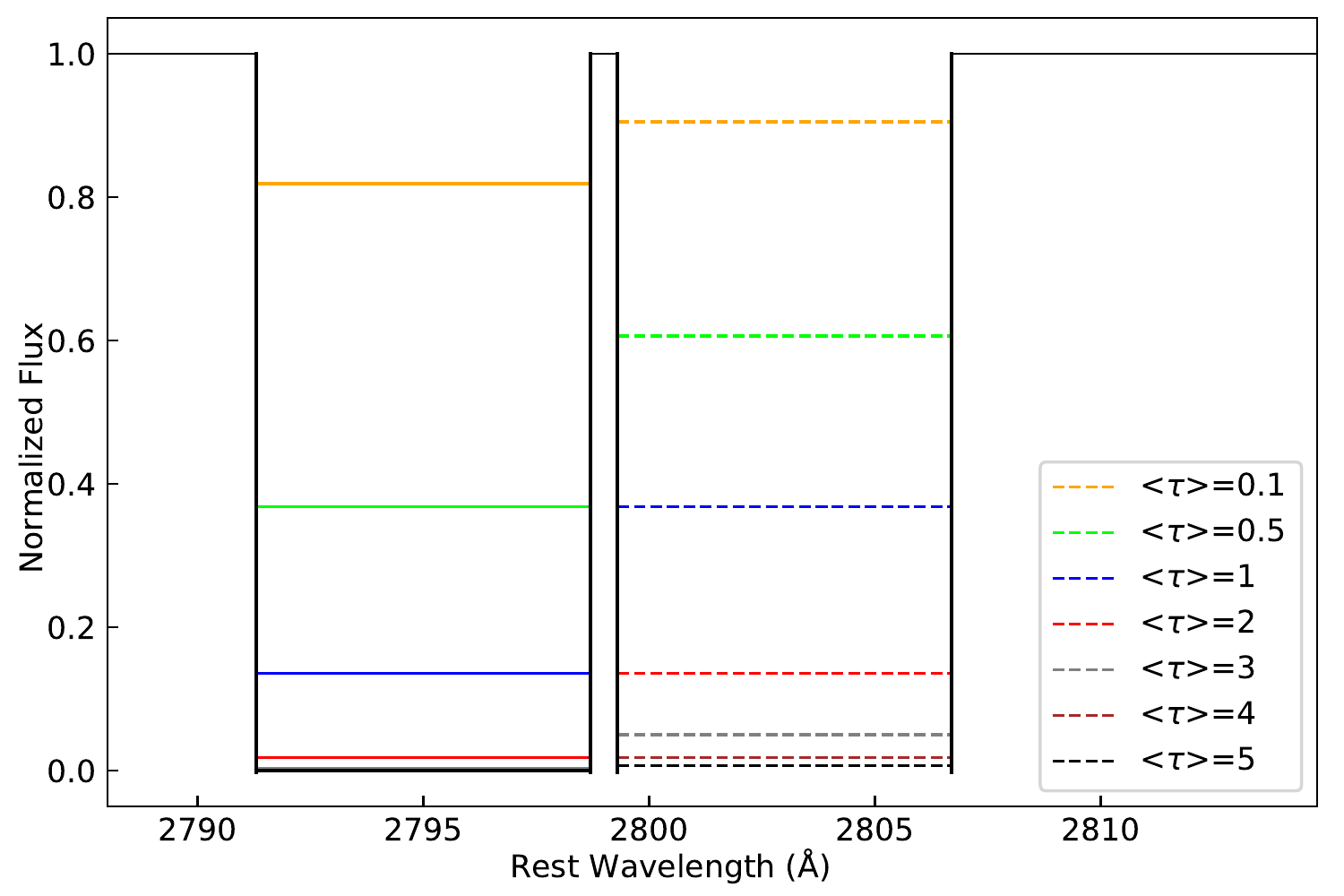} 
     \caption{Effective BAL profiles of a doublet at different mean optical depths (marked by different colors) under the full coverage condition. The strong (solid) and weak (dashed) components of the doublet have an inherent optical depth ratio of 2. }  
      \label{EW_meanAOD_demo}
\end{figure}

To better illustrate the detection of positive/negative EW variations of BAL troughs with relatively high optical depths and test the above argument,  we quantitatively examine EW variations of a doublet with an inherent optical depth ratio of 2 at different optical depths caused by ionization changes.  
In reality, the profile of a \mgii-BAL trough may have a wide range of shapes (see Figure \ref{mgiiBAL_diversity}). 
To demonstrate the final result of \mgii-BAL variability in this sample, we use effective troughs of the \mgii\ doublet at different mean optical depths (see Figure \ref{EW_meanAOD_demo}), from which we can visually quantify EW variations. 
In this demonstration, for simplicity, we ignore the EW error and assume $|\Delta {\rm EW_{thres}}|=1$ \AA\ (corresponding to the area below the dashed/solid red lines) is the threshold to distinguish non-variable and variable troughs.

If the weak component (dashed) of the doublet has an initial $\langle \tau \rangle=2$, then EW ($\approx$ 14 \AA) of the doublet is equal to the area above the dashed/solid red lines (the total EW of the two rectangle troughs is 15 \AA). As $\langle \tau \rangle$ randomly varies between 0 and 5 due to ionization changes, we cannot detect EW variability in all cases where $|\Delta {\rm EW}|<1$ \AA. Therefore, for this doublet, we can detect only weakening troughs with $|\Delta {\rm EW}|>1$~\AA\ as $\langle \tau \rangle$ randomly varies between 0 and 5 in the later epoch. The example demonstrated here is an analog for all high-saturation ($\langle \tau \rangle>5$) and partial-covering cases, where this effect would become more evident.

Similarly, when performing the same analysis for shallow, unsaturated (e.g., $\langle \tau \rangle<0.5$) BAL troughs, we found that we can detect more strengthening troughs than weakening troughs, which naturally explains the finding from \citet{Filizak13} that shallow \civ\ troughs tend to increase their EWs among HiBAL QSOs.

This also explains the systematic difference that the fraction of variable \mgii\ BALs ($\sim$37\%) found in this work is significantly lower than those of variable HiBALs (50\%--60\%) found from previous studies, as we cannot detect all strengthening \mgii-BAL troughs below their corresponding detection thresholds.

\subsection{Investigating BAL variability in a combined transverse-motion/ionization-change scenario} \label{main_driver}

The observation that only portions of BAL troughs show variations (see Figure \ref{width_VRs} and \ref{vcomJ112736}), the lack of significant correlation in the absolute EW variations over different-velocity \mgii\ troughs, the presence of strong correlations in BAL variability between \mgii\ and \aliii\ while no correlations in BAL variability between \mgii\ and \civ, and the observed asymmetry in EW variations together indicate that the ionization-change mechanism alone seems incapable to explain our observed BAL variability. 
Therefore, ionization change and transverse motion may be jointly taken into account to interpret BAL variability for the sample, as we have found strong observational evidence in support of the argument that \mgii-BAL absorbers in our sample tend to have relatively high optical depths (if not saturated) and small LOS-covering factors (see Section \ref{v_width_depth} and \ref{saturation_effect} for details).

Combining all the observational results throughout Section \ref{stat_view} and our analyses of the \mgii-BAL  EW variations in the ionization-change scenario, we provide a model for LoBAL variability in the sample that can further address the observed time-dependent asymmetry in EW variation.

Our observational results indicate that \mgii\ BALs in the sample have an average covering factor of 0.25 derived by the median depth of all \mgii-BAL troughs; in addition, these BALs likely have an initial mean optical depth $\langle \tau \rangle>1$ at the first epoch. Adopting this covering factor of 0.25 for a BAL absorber with an initial $\langle \tau \rangle=1$ (corresponding to EW $\approx$ 3 \AA\ and the maximum positive $\Delta {\rm EW}=$ 0.9~\AA\ by setting the maximum trough depth to 0.25 in Figure \ref{EW_meanAOD_demo}), we examine EW variability in a combined transverse-motion/ionization-change scenario for an individual quasar.  

\begin{figure}[h]
\center{}
 \includegraphics[height=5cm,width=8.7cm,  angle=0]{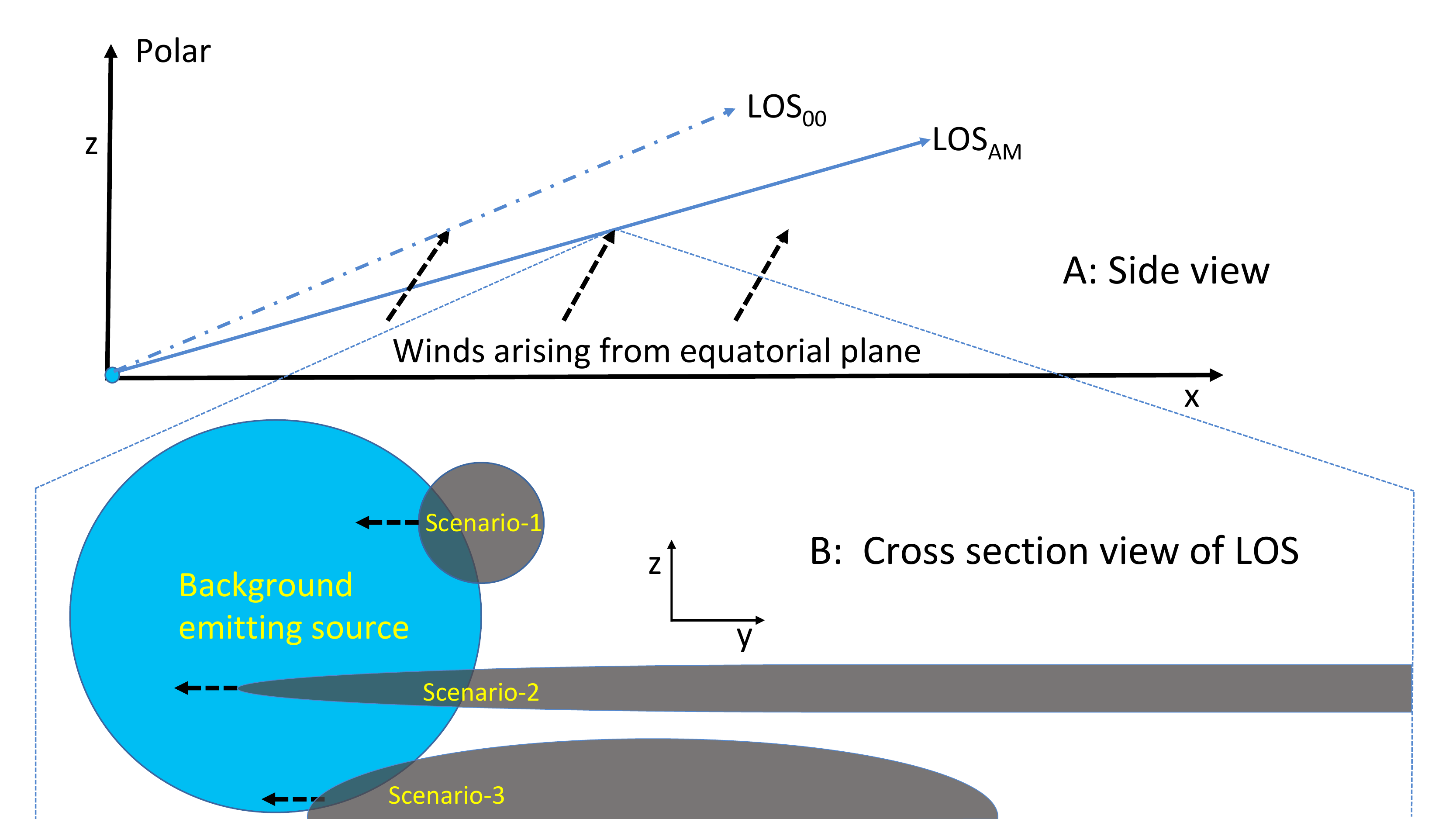} 
      \caption{Schematic illustration of BAL absorbers moving across our LOS. Upper : LOS$_{00}$ and LOS$_{\rm SA}$ represent HiBAL and LoBAL sightlines from a side view, respectively. Bottom: cross section of (zoom in) LOS$_{\rm SA}$, where scenario-1, scenario-2, and scenario-3 represent three effective-shape BAL absorbers crossing the LOS$_{\rm SA}$. The LOS$_{\rm SA}$ is closer to the equatorial plane and has a higher optical depth compared to the LOS$_{00}$. }
      \label{sketch_BAL_absorber}
\end{figure}

We use Figure \ref{EW_meanAOD_demo} and Figure \ref{sketch_BAL_absorber} together to demonstrate the final result. 
Although the BAL-absorber density is most likely inhomogeneous (e.g., \citealp{Hamann18}), we adopt an effective shape with a uniform density (same to the background continuum-source emission) to demonstrate the final result;  this result is the same if the ``clouds'' shown in Figure \ref{sketch_BAL_absorber} actually consist of numerous subunits. 
The effective projected shape of a \mgii-BAL absorber consisting of numerous subunits, in general, could be scenario-1 in Figure \ref{sketch_BAL_absorber} (a small cloud), or scenario-2 (a long flow tube), or scenario-3 (a large cloud), depending on the projected-size ratio between the BAL absorber and its background emitting area. 
Thus, scenario-1 represents all cases where the BAL-absorber size is less than that of the background source 
while scenario-2 and scenario-3 depict the opposite cases. 
However, our assessment holds regardless of whether the BAL-absorber structure is scenario-1, scenario-2, or scenario-3. As we adopt 0.25 as the maximum partial coverage for the BAL absorber, we need to constrain that the large cloud cannot cover its background source beyond 25\% at any time in the scenario-3 scenario. Therefore, scenario-1 and \break scenario-2 can be applied to the whole sample provided they are uniformly distributed, while scenario-3 may work only for individual objects.

During the analysis, we do not constrain the projected shape of a BAL absorber, as long as it meets the two conditions as required by the observations: 
(1) the absorber has a relatively high optical depth (e.g., $\langle \tau \rangle>1$) under partial covering, and 
(2) the timescale ($\Delta t_{1}$) from the point where a BAL absorber starts to increase the LOS coverage to the point of reaching the maximum LOS coverage is less than the timescale ($\Delta t_{2}$) over which the absorber moves across the LOS after reaching the maximum LOS coverage. In short, $\Delta t_{1}$ and $\Delta t_{2}$ represent intrinsic strengthening and weakening timescales in the LOS coverage,  respectively, which are independent of detection and projected BAL-absorber shape.

Specifically, we consider the case where $\Delta t_{2}=4\times \Delta t_{1}$ and assume $\Delta t_{1}=1$ yr in the scenario-1. 
Adopting the same threshold ($| \Delta {\rm EW_{thres}} |=1$ \AA) for the detection of EW variability (see Section \ref{saturation_effect}), we now examine the EW variability in a combined (transverse motion and ionization change) scenario with a maximum LOS coverage of 0.25.  
In this case, we can detect an increase in EW (due to increase of LOS coverage) within the one year from the start (EW = 0) to the point of reaching the maximum LOS coverage (EW $\approx$ 3 \AA); after that point, the BAL absorber will take four years to move across our LOS. During the four years, we cannot detect strengthening troughs because the maximum positive EW variation of the doublet ($\Delta {\rm EW}=$ 0.9 \AA) is below the detection threshold of 1~\AA, but we can detect weakening cases where $\Delta {\rm EW}<-1$~\AA\ due to ionization changes and/or other mechanisms. 
The example demonstrated here is an analog for all other $\Delta t_{2}>\Delta t_{1}$ and partial covering  cases, where this effect would become more evident with increasing $\Delta t_{2}$ when $\Delta t_{1}$ remains constant (e.g., the entire BAL outflow is much larger in size than the BAL structure observed in  scenario-2).

This model is supported by the observational evidence that the strengthening and weakening rates in EW variability are approximately equal to each other at $\Delta t <1$ yr but appears to be increasingly different at $\Delta t>1$ yr (see Figure \ref{ewave_vs_dew_tspan} and \ref{dew_sml_tspan}).  
In this picture, the detected BAL weakening rate in an individual object is higher than the BAL strengthening rate, leading to an overall decrease in EW as the sampling timescale increases.

This model also explains BAL-variability results reported from HiBAL QSO samples (e.g., \citealp{Filizak13,Wang15}), provided the majority of HiBAL absorbers in HiBAL QSOs have relatively low optical depths at the first epoch of observations. In that case, we will detect similar weakening/strengthening rates in EW variability in an ionization-change, transverse-motion, or mixed scenario. Furthermore, if the majority of HiBAL QSOs have $\Delta t_{2}>\Delta t_{1}$, ionization change would become the main driver of observed BAL variability, as reported by \citet{Wang15}.

\section{Summary and future work} \label{sum_sec}

We have studied BAL variability based on multi-epoch spectra in a sample of 134 \mgii-BAL quasars. Our major findings from the sample-based study are as follows: 

\begin{enumerate}

\item
We identify variable \mgii-BAL troughs using three metrics and find that the fraction ( $\sim$ 37\%, see Section \ref{variable_region_section}) of variable \mgii-BAL QSOs is significantly lower than that from previous HiBAL variability studies ($\sim$ 50\%--60\%). 
The result can be explained by the fact that \mgii-BAL absorbers tend to have relatively high optical depths under partial covering (see Section \ref{var_timescales}). 

\item
We found that larger fractional EW variations tend to occur in weak troughs with smaller EWs, particularly in shallower and higher velocity troughs. We do not find any significant correlations of $\Delta$EW or $\Delta \rm{EW}/\langle \rm{EW} \rangle$ 
with the QSO luminosity, redshift, radio loudness and Eddington ratio (see Section \ref{v_width_depth} and Section \ref{mag_rd_rl_Edd}). 

\item
We found that \mgii-BAL variations usually occur in relatively narrow portions of BAL troughs, consistent with HiBAL variability studies (e.g., \citealp{Filizak13}). The majority of variable regions have a width of 5 pixels ($\sim$ 275 \kms~), close to the lower limit set for searching variable regions (see Section \ref{variable_region_section}). 
 
\item
The absolute EW variation shows an increasing trend from short to intermediate timescales \break  ($\Delta t<5$ yr), but appears to level off with longer timescales ($\Delta t>5$ yr); however, the fractional EW variation shows an overall increasing trend at $\Delta t>0.01$ yr (see Section \ref{ew_variability}). 

\item
We investigate the behavior of \mgii\ BALs at different velocities in a small subsample of BALs that host multiple \mgii\ BALs. There is a significant correlation in the distribution of $\Delta \rm{EW}/\langle \rm{EW} \rangle$ with coordinated variability among \mgii-BAL troughs at different velocities. However, we find no correlation in $\Delta$EW, notably different from previous BAL variability studies based on HiBAL QSO samples  (see Section \ref{coordinate_var}).  

\item
By analyzing simultaneous BAL variability among quasars with \mgii\ (with an ionization potential of 15.0 eV), \aliii\ (28.5 eV) and \civ\ (64.5 eV) BALs, we find that two species with a smaller difference in their ionization potentials show significantly stronger correlations both in $\Delta$EW and $\Delta \rm{EW}/\langle \rm{EW} \rangle$. One possible cause of the correlations is the difference in ionization potentials  (see Section \ref{EWvar_mgii_civ}). 

\item
The absolute/fractional EW variations for \mgii-BAL troughs show remarkable asymmetric distributions, in that we see more troughs weakening ($\rm \Delta EW<0$) than strengthening ($\rm \Delta EW>~0$) and the asymmetry increases on longer timescales. 
We attribute these results to the possibility that transverse motions likely dominate the strengthening BAL troughs while ionization changes and/or other mechanisms dominate the weakening BAL troughs (see Section \ref{main_driver}). 

\end{enumerate}

The systematic comparisons and analyses in this work reveal some significant differences between HiBAL and LoBAL variability. 
Long-term sampling timescales from a large LoBAL sample act as powerful tools that allows us, for the first time, to pin down the role of main drivers in BAL variability (transverse-motion and ionization-change mechanisms dominate the strengthening and weakening BAL troughs, respectively). 
These findings highlight the importance of future investigations into BAL variability either from theoretical modeling or observational studies based on larger samples as well as investigations of individual objects, particularly BAL disappearance/emergence events. 
Compared with previous HiBAL variability studies, the significant lower fraction of variable \mgii\ BALs in our sample provides insight into the intrinsic properties of LoBALs that are likely associated with relatively high optical depths under partial covering.

For individual objects, a few quasars exhibit both remarkably stable and significantly variable \mgii-BAL troughs in EW;   several candidates show BAL disappearance; some quasars show an apparent monolithic shift in a single or multiple BAL troughs, possibly indicative of accelerating outflows.  These objects, in conjunction with large blueshift of broad/narrow emission  lines,  are of great interest for follow-up observations aimed at probing QSO feedback via multi-phase outflows.

In the future, a large-scale investigation of BAL versus broad emission line properties should give additional insights into the quasar-wind contribution to the formation of  high/low-ionization BALs. With the aid of multi-wavelength data, more comprehensive and further understanding of BAL variability can be gained. 
Furthermore, additional spectroscopic epochs can improve constraints upon the fraction of \mgii-BAL QSOs showing variability, BAL lifetime, BAL acceleration events, coordinated \mgii-BAL variability and correlated BAL variability from low- to high-ionization species. 
Follow-up data covering longer timescales will allow further assessment of the evolution of diverse \mgii-BAL troughs, the properties of \feii\ BALs, and perhaps establish the location of BAL outflows. 
As  dedicated campaigns such as TDSS continue to accumulate data, a wide variety of studies will be enabled, which will allow more systematic investigations into the nature of ionized outflows in the circumnuclear region and their effects to QSO feedback.

\acknowledgements{ACKNOWLEDGMENTS}

We thank Tinggui Wang for helpful discussions. 
We also acknowledge Robin Ciardullo and Mike Eracleous for assistance in the observations by the Hobby-Eberly Telescope.

W. Yi thanks the financial support from the program of China Scholarships Council (No. 201604910001) for his postdoctoral study at the Pennsylvania State University. 
W. Yi also thanks the support from the National Science Foundation of China (NSFC-11703076) and the West Light Foundation of the Chinese Academy of Sciences (Y6XB016001). This work is also supported by the Joint Research Fund in Astronomy (U1631127) under cooperative agreement between the National Science Foundation of China and Chinese Academy of Sciences. 
MV and WNB acknowledge support from NSF grant AST-1516784. 
PH acknowledges support from the Natural Sciences and Engineering Research Council of Canada (NSERC), funding reference number 2017-05983, and from the National Research Council Canada during his sabbatical at NRC Herzberg Astronomy \& Astrophysics. 
NFA thanks TUBITAK (115F037) for financial support. 

This research uses data obtained through the Telescope Access Program (TAP), which has been funded by the National Astronomical Observatories of China, the Chinese Academy of Sciences (the Strategic Priority Research Program "The Emergence of Cosmological Structures" Grant No. XDB09000000), and the Special Fund for Astronomy from the Ministry of Finance. Observations obtained with the Hale Telescope at Palomar Observatory were obtained as part of an agreement between the National Astronomical Observatories, Chinese Academy of Sciences, and the California Institute of Technology. 

Funding for SDSS-III has been provided by the Alfred P. Sloan Foundation, the Participating Institutions, the National Science Foundation, and the U.S. Department of Energy Office of Science. SDSS-IV acknowledges support and resources from the Center for High-Performance Computing at the University of Utah. The Hobby-Eberly Telescope (HET) is a joint project of the University of Texas at Austin, the Pennsylvania State University, Ludwig-Maximillians-Universit\"{a}t M\"{u}nchen, and Georg-August-Universit\"{a}t G\"{o}ttingen. The Hobby-Eberly 
Telescope is named in honour of its principal benefactors, William P. Hobby and Robert E. Eberly. We also acknowledge the support of the staff of the Lijiang 2.4m telescope (LJT). Funding for the telescope has been provided by CAS and the People's Government of Yunnan Province.


\begin{thebibliography}

\bibitem[Arav et al.(1999)]{Arav1999}  Arav, N.; Korista, K. T.; de Kool, M., et al. 1999, \apj, 516, 27
\bibitem[Arav et al.(2012)]{Arav12}Arav, N.; Edmonds, D.; Borguet, B., 2012, A\&A, 544, 33 
\bibitem[Arav et al.(2018)]{Arav18}Arav, N.; Liu, G.; Xu, X., et al. 2018, \apj, 857, 60 
\bibitem[Baskin et al.(2014)]{Baskin2014a} Baskin, A., Laor, A., \& Stern, J.\ 2014, \mnras, 445, 3025
\bibitem[Becker et al.(2000)]{Becker00}Becker, R. H., White, R. L. et al. 2000, \apj, 538, 72

\bibitem[Boroson \& Meyers (1992)]{Boroson92}Boroson, Todd A.; Meyers, Karie A, 1992, \apj, 397, 442
\bibitem[Borguet et al.(2013)]{Borguet13} Borguet, B. C. J.; Arav, N.; Edmonds, D.; Chamberlain, C.; Benn, C. 2013, \apj, 762, 49

\bibitem[Crenshaw \& Kraemer (2012)]{Crenshaw12} Crenshaw, D. M.; Kraemer, S. B. 2012, \apj, 753, 75

\bibitem[Capellupo et al.(2011)]{Cap11} Capellupo, D.~M., Hamann, F., Shields, J.~C., Rodr{\'{\i}}guez~Hidalgo, P., \& Barlow, T.~A. 2011, \mnras, 413, 908
\bibitem[Capellupo et al.(2012)]{Cap12} Capellupo, D. M.; Hamann, F. et al. 2012, \mnras,  422, 3249

\bibitem[Chonis et al.(2016)]{Chonis16}   Chonis, T. S.; Hill, G. J.; Lee, H. et al., 2016, SPIE, 9908, 99084C
\bibitem[De Cicco et al.(2018)]{DeCicco18} De Cicco, D.; Brandt, W. N. et al. 2018, A\&A, 616, 114

\bibitem[Di Matteo et al.(2005)]{Di05}Di Matteo T., Springel V., Hernquist L., 2005, Nature, 433, 604
\bibitem[Eisenstein et al.(2011)]{Eisenstein11} Eisenstein, D. J.; Weinberg, D. H. et al. 2011, \aj, 142, 72
\bibitem[Fabian et al.(2012)]{Fabian12}Fabian, A.C., 2012, ARA\&A, 50, 455
\bibitem[Fan et al.(2015)]{Fan15}Fan, Y. F.; Bai, J. M.; Zhang, J. J. et al. 2015, RAA, 15, 918
\bibitem[Filiz Ak et al.(2012)]{Filizak12} Filiz Ak, N.; Brandt, W. N.; Hall, P. B. et al, 2012, \apj, 757, 114
\bibitem[Filiz Ak et al.(2013)]{Filizak13} Filiz Ak, N.; Brandt, W. N.; Hall, P. B. et al, 2013, \apj, 777, 168
\bibitem[Filiz Ak et al.(2014)]{Filizak14} Filiz Ak, N.; Brandt, W. N.; Hall, P. B. et al, 2014, \apj, 791, 88

\bibitem[Gehrels et al.(1986)]{Gehrels1986} Gehrels, N., 1986, \apj, 303, 336
\bibitem[Grier et al.(2015)]{Grier15} Grier, C. J., et al.\ 2015, \apj, 806, 111
\bibitem[Grier et al.(2016)]{Grier16}Grier, C. J.; Brandt, W. N.; Hall, P. B. et al, 2016, \apj, 824, 130
\bibitem[Gibson et al.(2008)]{Gibson08} Gibson, R.~R., Brandt, W.~N., Schneider, D.~P., \& Gallagher, S.~C. 2008, \apj, 675, 985
\bibitem[Gibson et al.(2009)]{Gibson09}Gibson, R. R., Jiang, L., Brandt, W. N., et al. 2009, \apj, 692, 758
\bibitem[Gibson et al.(2010)]{Gibson10}Gibson, R. R., Brandt, W. N., Gallagher, S. C. et al. 2010, \apj, 713, 220
\bibitem[Gunn et al.(2006)]{Gunn06} Gunn, J. E.; Siegmund, W. A.; Mannery, E. J. et al. 2006, \aj, 131, 2332
\bibitem[Hall et al.(2002)]{Hall02}Hall, P. B.; Anderson, S. F.; Strauss, M. A. et al. 2002, \apjs, 141, 267

\bibitem[Hall et al.(2011)]{Hall2011}Hall, P. B.; Anosov, K.  et al., 2011, MNRAS, 411, 2653
\bibitem[Hall et al.(2013)]{Hall13} Hall, P. B.; Brandt, W. N.; Petitjean, P. et al. 2013, \mnras, 434, 222

\bibitem[Hamann \& Sabra (2004)]{Hamann04} Hamann, F.; Sabra, B., 2004, ASP Conf. Ser., 311, 203
\bibitem[Hamann et al.(2008)]{Hamann08} Hamann, F.; Kaplan, K. F. et al. 2008, \mnras, 391, 39
\bibitem[Hamann et al.(2011)]{Hamann11} Hamann, F.; Kanekar, N.; Prochaska, J. X., et al.  2011, \mnras, 410, 1957 
\bibitem[Hamann et al.(2013)]{Hamann13}Hamann, F. et al. 2013, MNRAS, 435, 133
\bibitem[Hamann et al.(2018)]{Hamann18} Hamann, F. et al. 2018, submitted, arXiv:1810.03686
\bibitem[He et al.(2017)]{He17}  He, Z.; Wang, T.; Zhou, H. et al. 2017, \apjs, 229, 22
\bibitem[Hewett \& Wild (2010)]{Hewett10}Hewett, Paul C.; Wild, Vivienne, 2010, MNRAS, 405, 2302
\bibitem[Junkkarinen et al.(2001)]{Junkkarinen2001}Junkkarinen, V.; Shields, A., et al., 2001, \apj, 155, 159
\bibitem[Lundgren et al.(2007)]{lundgren07} Lundgren, B.~F., Wilhite, B.~C., Brunner, R.~J., Hall, P.~B., Schneider, D.~P., York, D.~G., Vanden Berk, D.~E., \& Brinkmann, J. 2007, \apj, 656, 73
\bibitem[McGraw et al.(2015)]{McGraw2015}McGraw, S. M.; Shields, J. C.; Hamann, F. W. et al. 2015, \mnras, 453, 1379
\bibitem[McGraw et al.(2017)]{McGraw2017}McGraw, S. M.; Brandt, W. N.; Grier, C. J., et al. 2017, \mnras, 469, 3163


\bibitem[P\^{a}ris et al.(2017)]{Paris17} P\^{a}ris, I.; Petitjean, P.; Ross, N. et al. 2017, A\&A, 597, 79
\bibitem[Pei (1992)]{Pei92}Pei, Yichuan C., 1992, \apj, 395, 130
\bibitem[Randles et al.(1980)]{Randles80} Randles, R. H., Fligner, M. A., Policello, G. E., \& Wolfe, D. A. 1980, Journal of American Statistic Association, 75, 168

\bibitem[Ramsey et al.(1998)]{Ramsey98} Ramsey, L. W., et al. 1998, Proc. SPIE, 3352, 34
\bibitem[Rafiee et al.(2016)]{Rafiee2016}Rafiee, A.; Pirkola, P.; Hall, P. B., et al. 2016, \mnras, 459, 2472
\bibitem[Rogerson et al.(2018)]{Rogerson18} Rogerson, J. A.; Hall, P. B. et al. 2018, \apj, 862, 22
\bibitem[Salviander et al.(2007)]{Salviander07}Salviander, S.; Shields, G. A.; Gebhardt, K.; Bonning, E. W., 2007, \apj, 662, 131
\bibitem[Smee et al.(2013)]{Smee13} Smee, S. A.; Gunn, J. E.; Uomoto, A. et al. 2013, \apj, 146, 32
\bibitem[Schneider et al.(2010)]{Schneider10} Schneider, D.~P.,  et al.\ 2010, \aj, 139, 2360
\bibitem[Shen et al.(2011)]{Shen11} Shen, Y., Richards, G. T., et al., 2011,  \apjs, 194, 45
\bibitem[Shen et al.(2016)]{Shen16}Shen, Y.; Brandt, W. N.; Richards, G. et al. 2016, \apj, 831, 7
\bibitem[Schulze et al.(2017)]{Schulze17} Schulze, A.; Schramm, M.; Zuo, W. et al. 2017, \apj, 848, 104
\bibitem[Stern et al.(2017)]{Stern2017}Stern, D. et al., 2017, \apj, 839, 106
\bibitem[Stocke et al.(1992)]{Stocke92} Stocke, J. T.; Morris, S. L.; Weymann, R. J.; Foltz, C. B., 1992, \apj, 396, 487
\bibitem[Sun et al.(2015)]{Sun15} Sun, M.; Trump, J. R.; Shen, Y. et al.\ 2015,\apj, 811, 42

\bibitem[Trevese et al.(2013)]{Trevese13} Trevese, D.; Saturni, F.G.; Vagnetti, F. et al., 2013, \aa, 557, 91
\bibitem[Trump et al.(2006)]{Trump06}Trump, J. R.; Hall, P. B.; Reichard, T. et al. 2006, \apjs, 165, 1
\bibitem[Tsuzuki et al.(2006)]{Tsuzuki06}Tsuzuki, Y.; Kawara, K.; Yoshii, Y., et al. 2006, \apj, 650, 57
\bibitem[Vanden Berk et al.(2001)]{VandenBerk01}Vanden Berk, D. E.; Richards, G. T.; Bauer, A. et al. 2001, \aj, 122, 549
\bibitem[Vivek et al.(2012a)]{Vivek2012a}Vivek, M.; Srianand, R.; Mahabal, A.; Kuriakose, V. C., et al. 2012, \mnras, 421, 107
\bibitem[Vivek et al.(2012b)]{Vivek2012b}Vivek, M.; Srianand, R.; Petitjean, P., et al. 2012, MNRAS, 423, 2879
\bibitem[Vivek et al.(2014)]{Vivek14}Vivek, M.; Srianand, R., et al. 2014, MNRAS, 440, 799
\bibitem[Vivek et al.(2016)]{Vivek16}Vivek, M.; Srianand, R.; Gupta, N. 2016, MNRAS, 455, 136
\bibitem[Wang et al.(2015)]{Wang15}Wang, T.; Yang, C.; Wang, H.; Ferland, G., 2015, \apj, 814, 150
\bibitem[Welling et al.(2014)]{Welling14}Welling, C. A.; Miller, B. P.; Brandt, W. N.; Capellupo, D. M.; Gibson, R. R., 2009, \mnras, 440,, 2474
\bibitem[Weymann et al.(1991)]{Weymann91} Weymann, R. J.; Morris, S. L. et al. 1991, \apj, 373, 23 
\bibitem[Yi et al.(2017)]{Yi17}Yi, W.; Green, R. F.; Bai, J.-M. et al. 2017, \apj, 838, 135
\bibitem[Yi et al.(2019)]{Yi19}Yi, W.; Vivek, M.; Brandt, W. N., et al. 2019, \apjl, 870, 25
\bibitem[York et al.(2000)]{York00}York, D. J. et al. 2000, \aj, 120, 1579
\bibitem[Zhang et al.(2010)]{Zhang10}Zhang, S.; Wang, T.-G.; Wang, H. et al. 2010, \apj, 714, 367
\bibitem[Zhang et al.(2015)]{Zhang2015}Zhang, S.; Zhou, H.; Wang, T., et al. 2015, \apj, 803, 58
\bibitem[Wilson et al.(2004)]{Wilson04} Wilson, J.~C., Henderson, C.~P., Herter, T.~L., et al.\ 2004, \procspie, 5492, 1295
\end{thebibliography}
\end{document}